\documentclass[a4paper,fleqn,usenatbib]{mnras}

\usepackage[T1]{fontenc}
\usepackage{ae,aecompl}

\usepackage{graphicx}	
\usepackage{amsmath}	
\usepackage{amssymb}	
\usepackage{array}
\usepackage{booktabs}
\usepackage{pdflscape}
\usepackage{soul,color}

\newcommand\ubvri{\mbox{$U\!BV\!RI$}}   
\newcommand\bvri{\mbox{$BV\!RI$}}   
\newcommand\bvgri{\mbox{$BV\!gri$}}   
\newcommand\ugriz{\mbox{$ugriz$}}   
\newcommand\griz{\mbox{$griz$}}   
\newcommand\gri{\mbox{$gri$}}   
\newcommand\jhk{\mbox{$J\!H\!K$}}   
\newcommand\bv{\mbox{$B\!-\!V$}}   
\newcommand{\kms}{km\,s$^{-1}$}
\newcommand{\hnotunits}{\,km\,s$^{-1}$\,Mpc$^{-1}$} 
\newcommand{\isofe}{$^{56}$Fe}
\newcommand{\isoco}{$^{56}$Co}
\newcommand{\isoni}{$^{56}$Ni}

\newcommand{\naid}{\ion{Na}{I}\,D}
\newcommand{\lumunit}{erg\,s$^{-1}$}
\newcommand{\feii}{\ion{Fe}{II}~$\lambda5018$}
\newcommand{\msun}{M$_{\odot}$}
\newcommand{\EGBV}{$E_{B\!-\!V}^\mathrm{MW}$}
\newcommand{\EhBV}{$E_{B\!-\!V}^\mathrm{host}$}
\newcommand{\sunit}{mag\,(100\,d)$^{-1}$}

\title[Luminous SNe~II for their low expansion velocities]{Luminous Type II supernovae for their low expansion velocities}

\author[\'{O}. Rodr\'{i}guez et al.]{\'{O}. Rodr\'{i}guez,$^{1,2,3}$\thanks{E-mail: olrodrig@gmail.com}
G. Pignata,$^{1,2}$
J. P. Anderson,$^{4}$
T. J. Moriya,$^{5,6}$
A. Clocchiatti,$^{7,2}$
\newauthor
F. F\"{o}rster,$^{8,2}$
J. L. Prieto,$^{9,2}$
M. M. Phillips,$^{10}$
C. R. Burns,$^{11}$
C. Contreras,$^{10}$
\newauthor
G. Folatelli,$^{12,13,14}$
C. P. Guti\'errez,$^{15}$
M. Hamuy,$^{16,2}$
N. I. Morrell,$^{10}$
M. D. Stritzinger,$^{17}$
\newauthor
N. B. Suntzeff,$^{18}$
S. Benetti,$^{19}$
E. Cappellaro,$^{19}$
N. Elias-Rosa,$^{19}$
A. Pastorello,$^{19}$
\newauthor
M. Turatto,$^{19}$
J. Maza,$^{16}$
R. Antezana,$^{16}$
R. Cartier,$^{20}$
L. Gonz\'alez,$^{16}$
\newauthor
J. B. Haislip,$^{21}$
V. Kouprianov,$^{21}$
P. L\'opez,$^{14}$
S. Marchi-Lasch$^{16}$
and D. Reichart$^{21}$
\\
$^{1}$Departamento de Ciencias Fisicas, Universidad Andres Bello, Avda. Republica 252, Santiago, Chile\\
$^{2}$Millennium Institute of Astrophysics (MAS), Nuncio Monse\~nor Sotero Sanz 100, Providencia, Santiago, Chile\\
$^{3}$School of Physics and Astronomy, Tel Aviv University, Tel Aviv 69978, Israel\\
$^{4}$European Southern Observatory, Alonso de C\'ordova 3107, Casilla 19, Santiago, Chile\\
$^{5}$Division of Science, National Astronomical Observatory of Japan, National Institutes of Natural Sciences, 2-21-1 Osawa, Mitaka, Tokyo 181-8588, Japan\\
$^{6}$School of Physics and Astronomy, Faculty of Science, Monash University, Clayton, VIC 3800, Australia\\
$^{7}$Instituto de Astrof\'isica, Pontificia Universidad Cat\'olica de Chile, Av. Vicu\~na Mackenna 4860, 782-0436 Macul, Santiago, Chile\\
$^{8}$Center for Mathematical Modeling, University of Chile, Chile\\
$^{9}$N\'ucleo de Astronom\'ia de la Facultad de Ingenier\'ia y Ciencias, Universidad Diego Portales, Av. Ej\'ercito 441, Santiago, Chile\\
$^{10}$Carnegie Observatories, Las Campanas Observatory, Casilla 60, La Serena, Chile\\
$^{11}$Observatories of the Carnegie Institution for Science, 813 Santa Barbara Street, Pasadena, CA 91101, USA\\
$^{12}$Instituto de Astrof\'isica de La Plata (IALP), CONICET, Paseo del Bosque S/N, B1900FWA La Plata, Argentina\\
$^{13}$Facultad de Ciencias Astron\'omicas y Geof\'isicas, Universidad Nacional de La Plata, Paseo del Bosque, B1900FWA, La Plata, Argentina\\
$^{14}$Kavli Institute for the Physics and Mathematics of the Universe, Todai Institutes for Advanced Study, University of Tokyo, 5-1-5 Kashiwanoha, Kashiwa, Chiba 277-8583, Japan\\
$^{15}$Department of Physics and Astronomy, University of Southampton, Southampton, SO17 1BJ, UK\\
$^{16}$Departamento de Astronom\'ia, Universidad de Chile, Casilla 36-D, Santiago, Chile\\
$^{17}$Department of Physics and Astronomy, Aarhus University, Ny Munkegade 120, DK-8000 Aarhus C, Denmark\\
$^{18}$George P. and Cynthia Woods Mitchell Institute for Fundamental Physics and Astronomy, Texas A\&M University, College Station, TX 77843, USA\\
$^{19}$INAF -- Osservatorio Astronomico di Padova, Vicolo dell'Osservatorio 5, 35122 Padova, Italy\\
$^{20}$Cerro Tololo Inter-American Observatory, National Optical Astronomy Observatory, Casilla 603, La Serena, Chile\\
$^{21}$Department of Physics and Astronomy, University of North Carolina at Chapel Hill, Campus Box 325, Chapel Hill, NC 27599, USA
}

\date{Accepted XXX. Received YYY; in original form ZZZ}

\pubyear{2020}

\begin{document}
\label{firstpage}
\pagerange{\pageref{firstpage}--\pageref{lastpage}}
\maketitle

\begin{abstract} 
We present optical and near-IR data of three Type II supernovae (SNe~II), SN~2008bm, SN~2009aj, and SN~2009au. These SNe display the following common characteristics: signs of early interaction of the ejecta with circumstellar material (CSM), blue \bv\ colours, weakness of metal lines, low expansion velocities, and $V$-band absolute magnitudes 2{\textendash}3~mag brighter than those expected for normal SNe~II based on their expansion velocities. Two more SNe reported in the literature (SN~1983K and LSQ13fn) share properties similar to our sample. Analysing this set of five SNe~II, which are luminous for their low expansion velocities (LLEV), we find that their properties can be reproduced assuming ejecta-CSM interaction that lasts between 4{\textendash}11 weeks post explosion. The contribution of this interaction to the radiation field seems to be the dominant component determining the observed weakness of metal lines in the spectra rather than the progenitor metallicity. Based on hydrodynamic simulations, we find that the interaction of the ejecta with a CSM of $\sim3.6$~M$_\odot$ can reproduce the light curves and expansion velocities of SN~2009aj. Using data collected by the Chilean Automatic Supernova Search, we estimate an upper limit for the LLEV SNe~II fraction to be 2{\textendash}4 per cent of all normal SNe~II. With the current data-set, it is not clear whether the LLEV events are a separated class of SNe~II with a different progenitor system, or whether they are the extreme of a continuum mediated by CSM interaction with the rest of the normal SN~II population.
\end{abstract}




\section{Introduction}\label{sec:introduction}

Type II supernovae \citep[SNe~II;][]{Minkowski1941} are the final stage of the evolution of stars with an initial mass $>8$~\msun, that retain a significant fraction of hydrogen in their envelopes at the moment of the collapse of their iron cores. Among SNe~II, three special sub-types have been identified based on their photometric and spectroscopic characteristics: those showing hydrogen in early spectra that soon disappear \citep[SNe~IIb;][]{Woosley_etal1987,Filippenko1988}, those having light curves similar to SN~1987A \citep[1987A-like SNe, e.g.,][]{Hamuy_etal1988_SN1987A}, and those showing narrow hydrogen emission lines in the spectra due to the interaction of the ejecta with a circumstellar material (CSM) \citep[SNe~IIn;][]{Schlegel1990_SNeIIn}. For the rest of SNe~II \citep[77 per cent;][which show a range in luminosity decline rates]{Shivvers_etal2017} we will refer as normal SNe~II.

Among SNe~IIn, there are some events that after 25{\textendash}50~d past explosion start to look similar to normal SNe~II with broad P-Cygni profiles, but with bluer colours (hereafter SNe~IIn/II, e.g., SN~1979C, \citealt{Balinskaia_etal1980_SN1979C}, \citealt{Branch_etal1981_SN1979C}, \citealt{deVaucouleurs_etal1981_SN1979C}; SN~1998S, \citealt{Fassia_etal2000_SN1998S,Fassia_etal2001_SN1998S}; SN~2007pk, \citealt{Inserra_etal2013}; SN~2008fq, \citealt{Taddia_etal2013_CSP_SNeIIn}, \citealt{Faran_etal2014_IIL}; PTF11iqb, \citealt{Smith_etal2015}; SN~2013fc, \citealt{Kangas_etal2016_SN2013fc}). On the other hand, \citet{Smith_etal2015} suggested that some normal SNe~II could be classified as SNe~IIn if the classification spectra are taken at epochs early enough to detect the interaction of the ejecta with the CSM generated by progenitors with dense winds.

Recently, it has been shown that the presence of CSM around progenitors of normal SNe~II seems to be ubiquitous \citep{Forster_etal2018}. In fact, the ejecta-CSM interaction in normal SNe~II, previously detected only in few cases \citep[e.g., SN~2006bp;][]{Quimby_etal2007}, now is confirmed in many cases by the detection of narrow hydrogen emission lines in very early-time spectroscopy \citep[also called ``flash spectroscopy'',][]{Khazov_etal2016,Yaron_etal2017}. The effect of this early ejecta-CSM interaction over the radiation field is also inferred from the light curve modelling \citep{Moriya_etal2011,Morozova_etal2017,Morozova_etal2018,Das_Ray2017,Forster_etal2018} and, more recently, from the simultaneous modelling of light curves and spectra \citep{Hillier_Dessart2019}.

With the increasing number of discovered SNe~II, more rare events are being revealed. An example of this is LSQ13fn \citep{Polshaw_etal2016}, an SN~II that shares some of the properties seen on SNe~IIn/II (e.g., early ejecta-CSM interaction, blue colours, and high luminosities) but with two additional characteristics: (1) a weakness of metal lines in the spectra, and (2) low expansion velocities, with values between the subluminous SN~II~2005cs \citep{Pastorello_etal2006,Pastorello_etal2009} and the archetypal SN~II~1999em \citep{Hamuy_etal2001,Leonard_etal2002_99em,Elmhamdi_etal2003}, but having a luminosity comparable to the moderately luminous SN~II~2004et \citep{Sahu_etal2006,Maguire_etal2010_04et}. In addition to LSQ13fn, SN~1983K was also reported showing some of the aforementioned characteristics \citep{Niemela_etal1985_SN1983K,Phillips_etal1990_SN1983K}. SN~1983K and LSQ13fn, being more luminous than the expected from their expansion velocities, do not fall on the luminosity-expansion velocity relation found by \citet{Hamuy_Pinto2002} for normal SNe~II. \citet{Polshaw_etal2016} suggested a combined effect of a residual thermal energy from the early ejecta-CSM interaction, a low metallicity, and a large radius of the progenitor to explain the peculiarities seen in LSQ13fn. In this work we present optical and near-IR data of SN~2008bm, SN~2009aj, and SN~2009au, which show similar properties to those seen in SN~1983K and LSQ13fn. Throughout this paper we identify this sample of SNe~II using the acronym Luminous for their Low Expansion Velocities (LLEV) SNe~II.

The paper is organized as follows. In Section~\ref{sec:SNe_and_host_galaxies} we present the relevant information on our SN data-set and their host galaxies. Photometry along with the description of the data reduction is presented in Section~\ref{sec:observational_data}. In Section~\ref{sec:photometric_properties} and~\ref{sec:spectroscopic_properties}, we contrast the properties of the LLEV SNe~II with those of other SNe~II. A possible scenario that could explain the peculiar characteristics seen in the LLEV SNe~II is discussed in Section~\ref{sec:discussion}. We also present an estimation of the fraction of LLEV SNe~II with respect to normal SNe~II in Section~\ref{sec:SN_rate} and their impact on the use of normal SNe~II as distance indicators in Section~\ref{sec:SNeII_as_distance_indicators}. Finally, our conclusions are summarized in Section~\ref{sec:conclusions}.

\section{SUPERNOVAE AND HOST GALAXIES}\label{sec:SNe_and_host_galaxies}

In Table~\ref{table:SN_sample} we list the main parameters of the LLEV SNe~II in our set and their host galaxies. Throughout this work we assume the extinction curve given by \citet{Fitzpatrick1999} with $R_V=3.1$. To compute distances from recessional redshifts, we assume $H_0=73$~\hnotunits, $\Omega_m=0.27$, and $\Omega_\Lambda=0.73$. To estimate the error in distances due to peculiar velocities, we include a velocity dispersion of 382~\kms\ \citep{Wang_etal2006}.

\begin{table*}
\caption{SN and host galaxy parameters of the LLEV SN~II sample.}
\label{table:SN_sample}
\begin{tabular}{l c c c c c}
\hline
 SN data                                 & 1983K          & 2008bm             & 2009aj             & 2009au             & LSQ13fn\\
\hline
 RA (J2000.0)                            &12:46:36.41     & 13:02:58.78        & 13:56:45.33        &12:59:46.00         &11:51:17.29           \\
 DEC (J2000.0)                           &$-$8:21:21.9    & +10:30:27.0        & $-$48:29:36.2      &$-$29:36:07.5       &$-$29:36:41.10        \\
 Host galaxy                             &NGC~4699        &CGCG~71--101        &ESO~221--G18        &ESO~443--G21        &LEDA~727284           \\
 Host type$^a$                           &SAB(rs)b        & Sc                 & Sa?                &Scd                 &Sa$^e$                \\
 Distance modulus (mag)                  &$31.50\pm0.35$  &$35.66\pm0.08$      &$33.89\pm0.20$      &$33.31\pm0.22$      &$37.21\pm0.05$        \\
 \EGBV\ (mag)$^b$                        &$0.015\pm0.002$ &$0.022\pm0.003$     &$0.124\pm0.020$     &$0.079\pm0.013$     &$0.054\pm0.009$       \\
 \EhBV\ (mag)                            &$0.0^d$         &$0.00\pm0.03$       &$0.00\pm0.02$       &$0.35\pm0.17$       &$0.0^e$               \\
 Explosion epoch (MJD)                   &$45490.1\pm1.0$ &$54486.5\pm10.0$    &$54879.8\pm6.5$     &$54897.2\pm4.0$     &$56299.2\pm1.0$       \\
 SN heliocentric velocity (\kms)         &$1394\pm162$    &$9628\pm25$         &$2883\pm162$        &$2875\pm30$         &$18900\pm300^e$       \\
 12+log(O/H) (dex)$^c$                   &$8.55^\dagger$  &$8.33\pm0.02^\star$ &$8.29\pm0.03^\star$ &$8.56\pm0.03^\star$ &$<8.50\pm0.16^\otimes$\\
 $s_2$ (\sunit)                          &$0.73\pm0.10$   &$2.56\pm0.21$       &$0.80\pm0.02$       &$3.06\pm0.02$       &$0.86\pm0.10$         \\
 $M_V^{\mathrm{max}}$ (mag)              &$-18.83\pm0.36$ &$-18.40\pm0.18$     &$-18.81\pm0.21$     &$-17.55\pm0.58$     &$-17.95\pm0.21$       \\
 $M_V^{50\mathrm{d}}$ (mag)              &$-18.43\pm0.35$ &$-18.33\pm0.15$     &$-17.81\pm0.22$     &$-16.54\pm0.59$     &$-17.33\pm0.07$       \\
 \isoni\ mass (\msun)                    &$0.056\pm0.018$ &$>0.015\pm0.005$    &$>0.043\pm0.010$    & --                 &$>0.018\pm0.005$      \\
 (\bv)$_{50\mathrm{d}}$ (mag)            & $0.30\pm0.03$  & $<0.12\pm0.03$     &$0.41\pm0.05$       &$0.48\pm0.15$       &$0.34\pm0.11$         \\
 pEW(\feii)$_{50\mathrm{d}}$ (\AA)       & $3.46\pm0.22$  & $<8.00\pm3.00$     &$9.02\pm0.90$       &$14.65\pm2.63$      &$9.00\pm2.20$         \\
 $v_\mathrm{FeII}^{50\mathrm{d}}$ (\kms) &$2331\pm189$    &$1684\pm110$        &$2336\pm181$        &$1549\pm40$         &$2320\pm345$          \\
\hline
\multicolumn{6}{l}{$^a$ From NED, unless otherwise noted.}\\
\multicolumn{6}{m{0.9\linewidth}}{$^b$ Galactic colour excesses from \citet{Schlafly_Finkbeiner2011}, with a statistical error of 16 per cent \citep{Schlegel_etal1998}.}\\
\multicolumn{6}{m{0.9\linewidth}}{$^c$ Oxygen abundances, in the N2 calibration of \citet{Marino_etal2013}, measured by \citet{Kuncarayakti_etal2018} ($\dagger$), \citet{Anderson_etal2016} ($\star$), or \citet{Polshaw_etal2016} ($\otimes$). The latter is a recalibration of the original value reported in the N2 calibration of \citet{Pettini_Pagel2004}.}\\
\multicolumn{6}{l}{$^d$ Value from \citet{Niemela_etal1985_SN1983K}.}\\
\multicolumn{6}{l}{$^e$ Value from \citet{Polshaw_etal2016}.}
\end{tabular}
\end{table*}

\subsection{SN~2008bm}\label{sec:2008bm_info}
SN~2008bm was discovered in the galaxy CGCG~71{\textendash}101 on 2008 March 29.3~UT \citep{Drake_etal2008_SN2008bm_discovery} by the Catalina Real-time Transient Survey \citep[CRTS;][]{Drake_etal2009_CRTS}. The SN is also present on images taken on January 31.5~UT. The last non-detection was on January 11.5~UT, therefore the explosion epoch is constrained to occur at MJD~$54486.5\pm10.0$\footnote{We assume the explosion epoch as the midpoint of the range between the last non-detection and the first detection of the SN, with the error (not normal but uniform) being half the range.}, which is 67.8~d before the discovery. The event was classified as an SN~IIn a couple of months past explosion based on a spectrum obtained on April 7.1~UT. \citet{Stritzinger_Morrell2008_SN2008bm_classification} initially classified SN~2008bm as an SN~IIn, while latter spectra reported by \citet[][hereafter G17]{Gutierrez_etal2017_I} show clear Balmer absorption lines, typically seen in normal SNe~II. No radio detection was obtained in the 8.46~GHz band at the SN position on December 07.7~UT \citep{Chandra_Soderberg2008}.

CGCG~71{\textendash}101 has a recessional velocity of 9875~\kms\ (NED\footnote{NASA/IPAC Extragalactic Database (\url{https://ned.ipac.caltech.edu/}).}), which translates into a distance modulus of $35.66\pm0.08$~mag. No \naid\ absorption at the redshift of the host galaxy was detected in the SN spectra, indicating a negligible host galaxy reddening ($E_{B\!-\!V}^\mathrm{host}=0.00\pm0.03$~mag, \citealt{Anderson_etal2014_V_LC}, hereafter A14). SN~2008bm is located at a projected distance of $9.3\pm0.3$~kpc from the center of the apparently nearly face-on host galaxy, which is consistent with the low reddening scenario.

\subsection{SN~2009aj}\label{sec:2009aj_info}
SN~2009aj was discovered in the galaxy ESO~221{\textendash}G18 on 2009 February 24.3~UT \citep{Pignata_etal2009_SN2009aj_discovery} during the Chilean Automatic Supernova Search \citep[CHASE;][]{Pignata_etal2009_CHASE}. Nothing was visible at the SN position on February 11.2~UT, therefore the explosion is constrained to occur at MJD~$54879.8\pm6.5$ (i.e., $6.5$~d before the discovery). The event was classified by \citet{Stritzinger_etal2009_SN2009aj_SN2009au_classification} as an SN~II around maximum and reminiscent of SN~1983K.

We do not detect the presence of \naid\ at the redshift of the host galaxy in the SN spectra. In this case, we assume zero reddening with an error corresponding to the $3\sigma$ upper limit of the \naid\ pseudo equivalent width (pEW) non-detection. Using the relation of \citet{Poznanski_etal2012_NaID}\footnote{Where, as noted by \citet{Phillips_etal2013}, the error around the $\log(E_{B\!-\!V})$ relation is of 0.30~dex.} we obtain $E_{B\!-\!V}^\mathrm{host}=0.00\pm0.02$~mag, which we adopt as the host galaxy colour excess. ESO~221{\textendash}G18 has a recessional velocity corrected for the infall of the Local Group toward the Virgo cluster and the Great Attractor of $4380\pm112$~\kms\ (NED), which translates into a distance modulus of $33.89\pm0.20$~mag.

\subsection{SN~2009au}\label{sec:2009au_info}
SN~2009au was discovered in the galaxy ESO~443{\textendash}G21 on 2009 March 11.2~UT \citep{Pignata_etal2009_SN2009au_discovery} by the CHASE survey. Nothing was visible at the SN position on March 3.2~UT, therefore the explosion epoch is constrained to occur at MJD~$54897.2\pm4.0$ (i.e., 4~d before the discovery). The event appeared to be a young SN~IIn soon after explosion \citep{Stritzinger_etal2009_SN2009aj_SN2009au_classification}, however later spectra (see G17) show clear Balmer absorption features. No radio detection was obtained at the SN position in the 8.46~GHz band on December 13.6~UT \citep{Chandra_Soderberg2009}.

The distance modulus to ESO~443{\textendash}G21 is estimated using the Tully-Fisher relation to be $33.44\pm0.45$~mag \citep{Tully_etal2016}. In addition, ESO~443{\textendash}G21 is a member of the galaxy group HDCE~754 \citep{Crook_etal2007}, which has a recessional velocity corrected for Virgo infall of 3287~\kms\ (NED), corresponding to a distance modulus of $33.27\pm0.25$~mag. We adopt the weighted mean of these values ($\mu=33.31\pm0.22$~mag) as the distance to ESO~443{\textendash}G21.

SN~2009au is located at a projected distance of $1.7\pm0.1$~kpc from the center of its edge-on host galaxy, so the SN could be 
affected by a high amount of extinction. We measured a \naid~pEW of $1.33\pm0.21$~\AA\ at the redshift of the host galaxy in the SN spectra. However, the \naid~pEW becomes insensitive to estimate reddening for $\mathrm{pEW}>1.0$~\AA\ \citep{Phillips_etal2013}. Despite the above, we can estimate a lower limit for the \EhBV\ value as the saturation point ($\mathrm{pEW}=1.0$~\AA) in the \citet{Poznanski_etal2012_NaID} relation ($E_{B\!-\!V}^\mathrm{host}>0.21\pm0.14$~mag). On the other hand, matching the \bv\ colour curve of SN~2009au to the rest of LLEV SNe~II we obtain $E_{B\!-\!V}^{\mathrm{host}}=0.35\pm0.17$~mag, which is consistent with the previous lower limit. We adopt the colour excess from the \bv\ colour match as the \EhBV\ value for SN~2009au.

\section{OBSERVATIONAL MATERIAL}\label{sec:observational_data}

\subsection{Photometric data}
Optical and near-IR images of SN~2008bm, SN~2009aj, and SN~2009au were obtained over the course of the \textit{Carnegie Supernova Project}~I \citep[CSP-I;][]{Hamuy_etal2006}. Johnson $BV$ and Sloan $ugri$ images were obtained mostly with the 1~m Swope telescope at Las Campanas Observatory (LCO), while near-IR $Y\!JH$ images were obtained with both the 1~m Swope and the LCO 2.5~m du Pont telescope. The CSP-I data reduction is described in \citet{Contreras_etal2010} and \citet{Krisciunas_etal2017_CSP_3DR}.

In addition, Johnson-Kron-Cousins \ubvri\ and Sloan \griz\ images of SN~2009aj and SN~2009au were also obtained with the 41~cm Panchromatic Robotic Optical Monitoring and Polarimetry Telescopes \citep[PROMPT;][]{Reichart_etal2005} at Cerro Tololo inter-American Observatory (CTIO), as part of the CHASE follow-up program. We also include \bvri\ images of SN~2009au obtained with the SMARTS 1.3~m telescope at CTIO, equipped with the A Novel Dual Imaging Camera (ANDICAM)\footnote{Images are available on the NOAO science archive (\url{http://archive1.dm.noao.edu/}).}. Data reduction includes bias subtraction, overscan correction, flat-field correction, cosmic ray rejection, and image combination.

As part of the CSP-I and CHASE programs, images of the SNe host galaxies 1.5{\textendash}4.0 years after the SN explosion were obtained in order to remove the host galaxy contamination. In the case of the ANDICAM images of SN~2009au, to construct a template for each filter, we selected the earliest images of the SN field obtained under good observing conditions with such camera. Then, using the SNOoPY\footnote{SNOoPy is a package for SN photometry using PSF fitting and/or template subtraction developed by E. Cappellaro. A package description can be found at \url{http://sngroup.oapd.inaf.it/snoopy.html}} package, we removed the SN flux. Such templates were then subtracted from the rest of the images acquired with ANDICAM. Instrumental magnitudes were measured using the point spread function (PSF) technique. Final magnitudes were computed with respect to a local sequence of stars (Tables~\ref{table:SN2008bm_CSP_griz_sequence}{\textendash}\ref{table:SN2009au_CSP_YJH_sequence}), which are calibrated using \citet{Landolt1992_UBVRI} \ubvri, \citet{Smith_etal2002_ugriz} \ugriz, and \citet{Persson_etal1998} \jhk\ standard stars \citep[$J$- and $K$-band observations were used to derive $Y$-band magnitudes of the standard stars, see][]{Hamuy_etal2006}. 

\begin{figure}
\includegraphics[width=0.99\columnwidth]{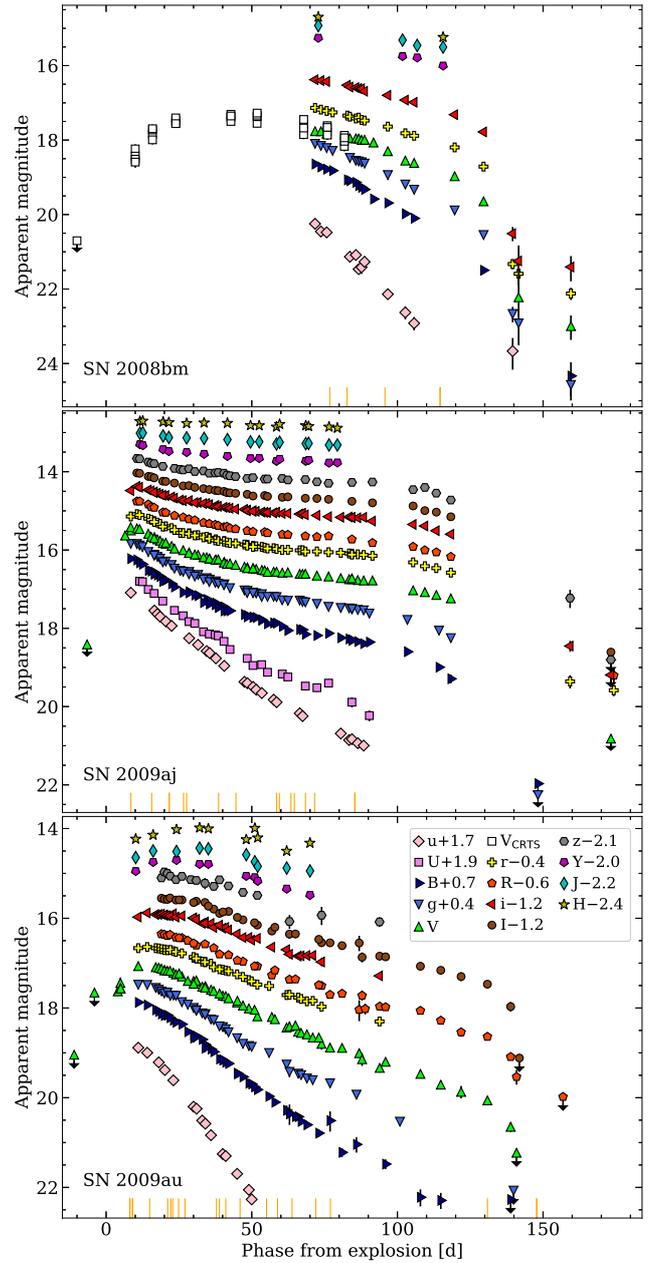}
\caption{Light curves of SN~2008bm (top), SN~2009aj (middle), and SN~2009au (bottom). Bands and magnitude shifts with respect to the original values are quoted in the legend. Arrows indicate upper limits. Orange ticks mark the epochs of the spectroscopy. For SN~2008bm we also plot the CRTS photometry. }
\label{fig:light_curves}
\end{figure}

Tables~\ref{table:SN2008bm_photometry}{\textendash}\ref{table:SN2009au_photometry} list the standard system photometry of SN~2008bm, SN~2009aj, and SN~2009au, while Fig.~\ref{fig:light_curves} shows their optical and near-IR light curves. Systematic differences between the CSP-I and CHASE \bvgri photometry, CSP-I and ANDICAM $BV$ photometry, and CHASE and ANDICAM $RI$ photometry are, on average, lower than 0.02~mag. We stress the fact that the photometry obtained with the SMARTS 1.3~m and PROMPT telescopes was calibrated using the magnitudes of the CSP-I local sequences.

In the analysis we include the unfiltered photometry transformed to $V$-band magnitudes ($V_\mathrm{CRTS}$) of SN~2008bm obtained by the CRTS\footnote{Available on \url{http://nesssi.cacr.caltech.edu/DataRelease/CRTS-I_transients.html}}. We note that the $V_\mathrm{CRTS}$ photometry lies very close to our $r$-band photometry in the phase interval where the light curves overlap. In order to reach a better agreement between the $V_\mathrm{CRTS}$ and the filtered $V$-band photometry, we add 0.26 mag to the $V_\mathrm{CRTS}$ magnitudes, which is the average $V\!-r$ colour at 50~d since explosion we obtained for SN~2009aj and LSQ13fn. 

\subsection{Spectroscopic data}
Optical spectra of SN~2009au were obtained with telescopes and instruments listed in Table~\ref{table:SN09au_spectra}. Data reduction includes bias subtraction, flat-field correction, wavelength calibration, one-dimensional spectrum extraction and sky subtraction, and flux calibration. The left part of Fig.~\ref{fig:SN09au_spectra} shows the optical spectral evolution of SN~2009au.

We also obtained near-IR spectra of SN~2009aj with telescopes and instruments listed in Table~\ref{table:SN09aj_spectra}. Data reduction includes the subtraction of the pairs of images to remove the sky background, images combination, wavelength calibration, one-dimentional spectrum extraction, telluric correction, and flux calibration. The right part of Fig.~\ref{fig:SN09au_spectra} shows the near-IR spectral evolution of SN~2009aj.

In the analysis we include the optical spectroscopy for SN~2008bm, SN~2009aj, and SN~2009au, which was obtained by the CSP-I and already published in G17.

\subsection{Sample of supernovae}
In addition to SN~2008bm, SN~2009aj, and SN~2009au, we include SN~1983K \citep{Niemela_etal1985_SN1983K,Phillips_etal1990_SN1983K} and LSQ13fn \citep{Polshaw_etal2016} into the analysis, given the similarity of their photometric and spectral properties (see Section~\ref{sec:photometric_properties} and \ref{sec:spectroscopic_properties}). SN~1983K was discovered in NGC~4699 on 1983 June 6.1~UT \citep{Maza_etal1983_SN1983K} at $17.1$~mag \citep{Phillips_etal1990_SN1983K}. There is no information about the last non-detection. However, the SN was at $13.3$~mag on June 10.1~UT \citep{Phillips_etal1990_SN1983K}, which indicates that the SN was discovered close to the explosion. Fitting a quadratic polynomial to the $B$-band rise photometry, we obtain the explosion epoch to be MJD~$45490.1\pm1.0$ (i.e., 1~d before the discovery). The distance modulus to NGC~4699 is estimated with the Tully-Fisher relation to be $31.45\pm0.45$~mag \citep{Tully_etal2016}. In addition, NGC~4699 is a member of the galaxy group HDCE~740 \citep{Crook_etal2007}, which has a recessional velocity corrected for Virgo infall of 1506~\kms\ (NED), corresponding to a distance modulus of $31.57\pm0.55$~mag. We adopt the weighted mean ($\mu=31.50\pm0.35$~mag) as the distance to NGC~4699. LSQ13fn was discovered in LEDA~727284 on 2013 January 10.2~UT by the La Silla-Quest Variability Survey \citep{Baltay_etal2013_LSQ}, which also obtained early-time photometry. The first detection was on January 8.2~UT, while the last non detection was on January 6.2~UT, so the explosion is constrained to occur at MJD~$56299.2\pm1.0$ (i.e., $3$~d before the discovery). Using the redshift given in \citet{Polshaw_etal2016} and the cosmic microwave background dipole model of \citet{Fixsen_etal1996}, we compute a recessional velocity of $19225\pm300$~\kms\ for LEDA~727284, corresponding to a distance modulus of $37.21\pm0.05$~mag. The main parameters for SN~1983K, LSQ13fn, and their host galaxies are summarized in Table \ref{table:SN_sample}.

\section{PHOTOMETRIC PROPERTIES}\label{sec:photometric_properties}

In this section we compare photometric properties of the LLEV SNe~II with those of other SNe~II. 

\subsection{$V$-band light curves}\label{sec:MVmax_vs_s2}

\begin{figure}
\includegraphics[width=1.0\columnwidth]{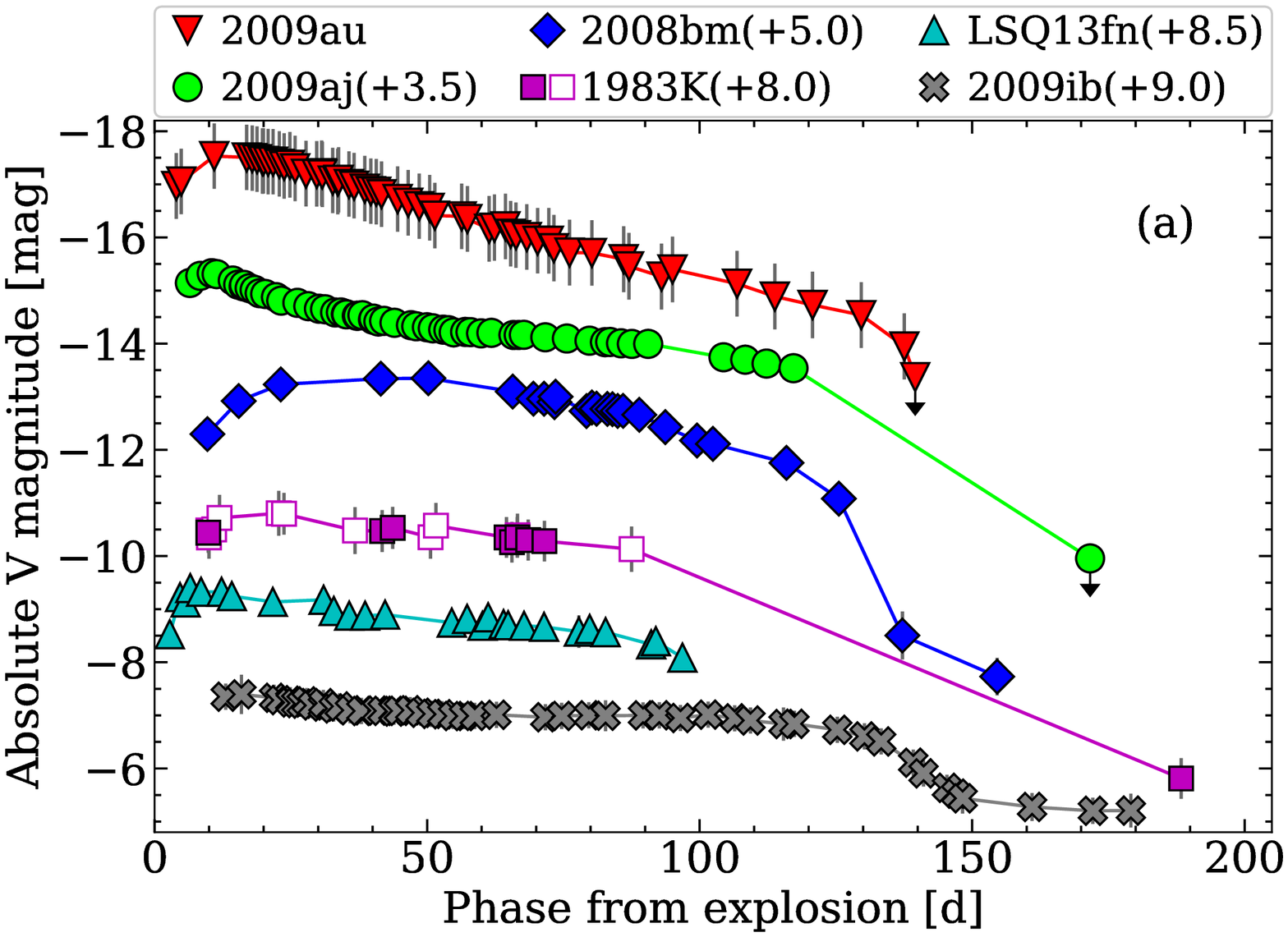}
\includegraphics[width=1.0\columnwidth]{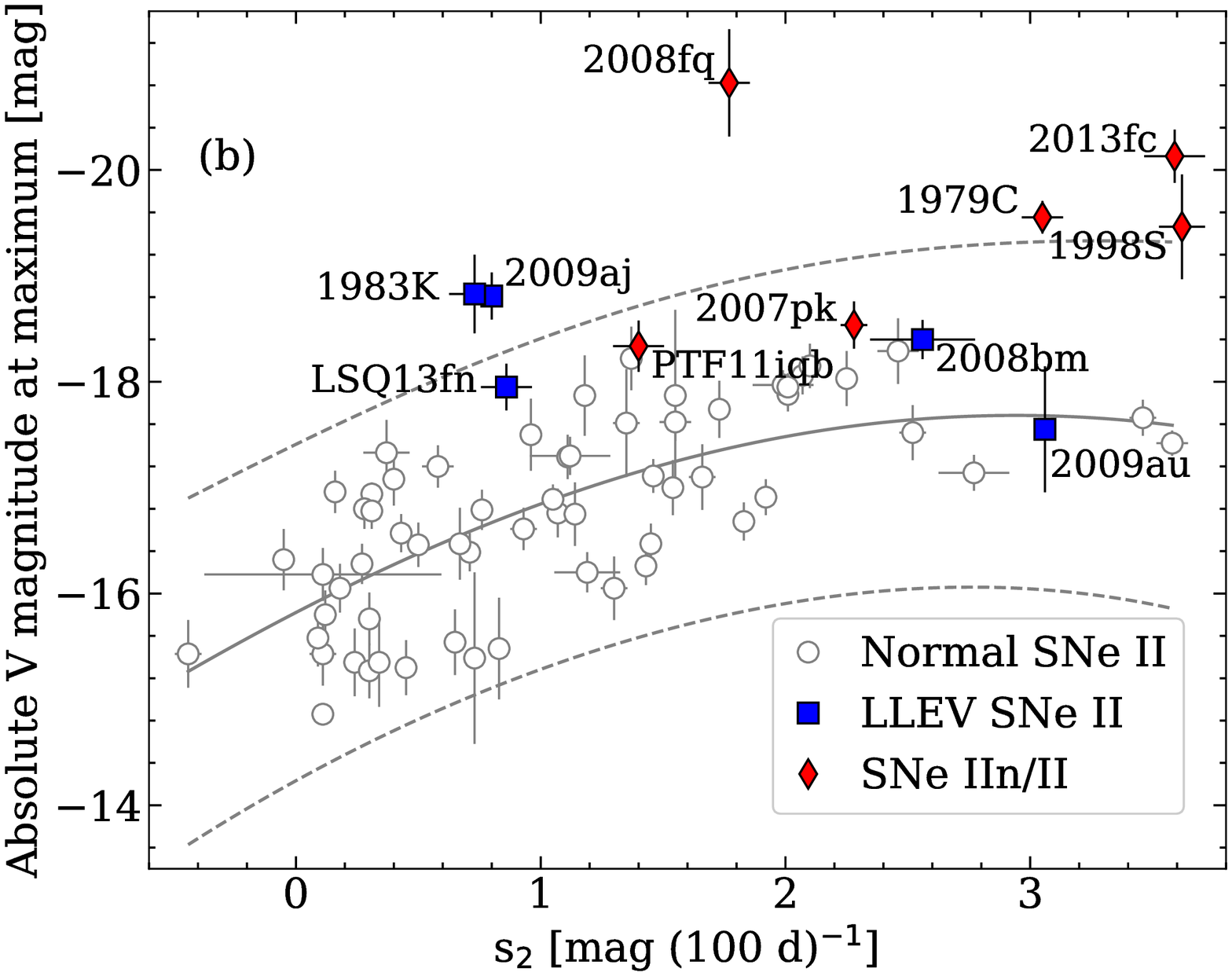}
\caption{(a) Absolute $V$-band light curves of the LLEV SNe~II in our set (coloured symbols), and of the normal SN~II 2009ib (gray crosses). For clarity we averaged photometric points into 2~hr bins, and applied and offset to magnitudes (indicated in parentheses). Arrows indicate upper limits. (b) Absolute $V$-band magnitude at maximum versus $V$-band decline rate during the plateau, showing the LLEV SNe~II in our set (blue squares), the normal SNe~II in the A14 sample corrected for \EhBV\ (empty circles), and the SNe~IIn/II from the literature (red diamonds). The solid line corresponds to the Gaussian process fit, where dashed lines indicate the 3$\sigma$ error around the fit.}
\label{fig:MVmax_vs_s2}
\end{figure}

Fig.~\ref{fig:MVmax_vs_s2}a shows the absolute $V$-band light curves of the LLEV SNe~II in our set. We note that the length of the photospheric phase (also called optically thick phase duration, OPTd) of SN~2008bm and SN~2009aj is about 120~d, and of about 130~d for SN~2009au. These values are comparable to that of SNe~II~2004er \citep[120~d;][]{Anderson_etal2014_V_LC} and 2009ib \citep[][gray crosses]{Takats_etal2015}, which are among the normal SNe~II with longest plateau.

Analysing a set of 116 SN~II $V$-band light curves, A14 found a correlation between the absolute $V$-band magnitude at maximum ($M_{V}^{\mathrm{max}}$) and the decline rate of the second, shallower slope in the light curve ($s_2$), which suggests a continuum in the normal SN~II population in this parameter space. Similar peak magnitude-decline rate correlations were also obtained by \citet{Sanders_etal2015}, \citet{Galbany_etal2016}, and \citet{Valenti_etal2016}. 

For SN~2008bm, SN~2009aj, SN~2009au, and LSQ13fn we have $V$-band photometry around the maximum light. Therefore we can check if such correlation holds also for LLEV SNe~II. In the case of SN~1983K, the $V$-band light curve (purple filled squares) is less sampled than the $B$-band one \citep[see Fig.~3 of][]{Phillips_etal1990_SN1983K}, and the maximum light (observed in $B$-band) in the $V$-band is missed. To estimate the $V$-band photometry around the maximum, we interpolate the \bv\ colour (which is monotonically increasing, see Section~\ref{sec:colour_evolution}) to the epochs of the $B$-band photometry without $V$-band measurements (${t^*}$), and then we compute $V_{t^*}=B_{t^*}-(B\!-\!V)_{t^*}$ (purple empty squares). To estimate the $V$-band maximum of our LLEV SNe~II, we fit a fourth order polynomial to the photometry close in time to the brightest point. To estimate $s_2$, we fit a straight line to the $V$-band light curve during the plateau phase (for more details, see A14). Values of $M_V^{\mathrm{max}}$ and $s_2$ for the LLEV SNe~II set are listed in Table~\ref{table:SN_sample}.

Fig.~\ref{fig:MVmax_vs_s2}b shows the normal SNe~II in the A14 sample\footnote{We remove SN~2008bm and SN~2009au from the A14 sample since they are in our LLEV SN~II sample.} (empty circles; where values are from Table~6 of A14) in the $M_V^{\mathrm{max}}$ versus $s_2$ space, where we remove those SNe without estimation of \EhBV. To characterize the distribution of the A14 sample in this space, we perform a Gaussian process fit (solid line), where the dashed lines indicate the $\pm3\sigma$ limits. In the figure we plot the LLEV SNe~II as blue squares. We see that they have $M_V^{\mathrm{max}}\lesssim-$17.5~mag, where SN~2008bm and SN~2009au are fast decliners, while for SN~1983K, SN~2009aj, and LSQ13fn the decline rate is lower, indicating that LLEV SNe~II seems not to be connected to a particular light curve decline rate. In addition, we see that SN~2008bm, SN~2009au, and LSQ13fn are within 3$\sigma$ limit, while SN~1983K and SN~2009aj are outliers in the distribution, i.e., they are significantly brighter than what would be implied from their $s_2$ decline rates. In the figure we also plot the SNe~IIn/II (red diamonds) that we found in the literature (the set and main properties are listed in Table~\ref{table:SN_IInII_sample}). We see that the LLEV SNe~II tend to have lower decline rates than SNe~IIn/II. 

The case of SN~2009au is special. Its OPTd and $s_2$ values are not common in normal SNe~II, where faster decliners have also shorter OPTd (see for example, Fig.12 in A14). Performing radiation hydrodynamics simulations, \citet{Hillier_Dessart2019} showed that the ejecta-CSM interaction scenario could produce fast decliners with an OPTd similar to slow decliners. Based on the latter, we suggest that the characteristics of the $V$-band light curve of SN~2009au are consequence of the interaction of its ejecta with a CSM.

\subsection{Pseudo-bolometric light curves}
To compute pseudo-bolometric light curves, we proceed as follows: (1) We convert broad-band magnitudes into monochromatic fluxes ($f_x$), associated to their respective effective wavelengths ($l_x$). For a fair comparison with other SNe~II in the literature, we use \bvri\ and \gri\ magnitudes. (2) For each photometric epoch, we perform linear interpolations between the ($l_x$, $f_x$) points, which we adopt as the spectral energy distribution (SED). (3) We correct the SED for redshift and colour excess, and then we integrate it from 4200 to 7500~\AA\ (which is a wavelength range covered by the filters). (4) We convert the integrated flux into luminosity using the corresponding SN distance.

\begin{figure}
\includegraphics[width=1.0\columnwidth]{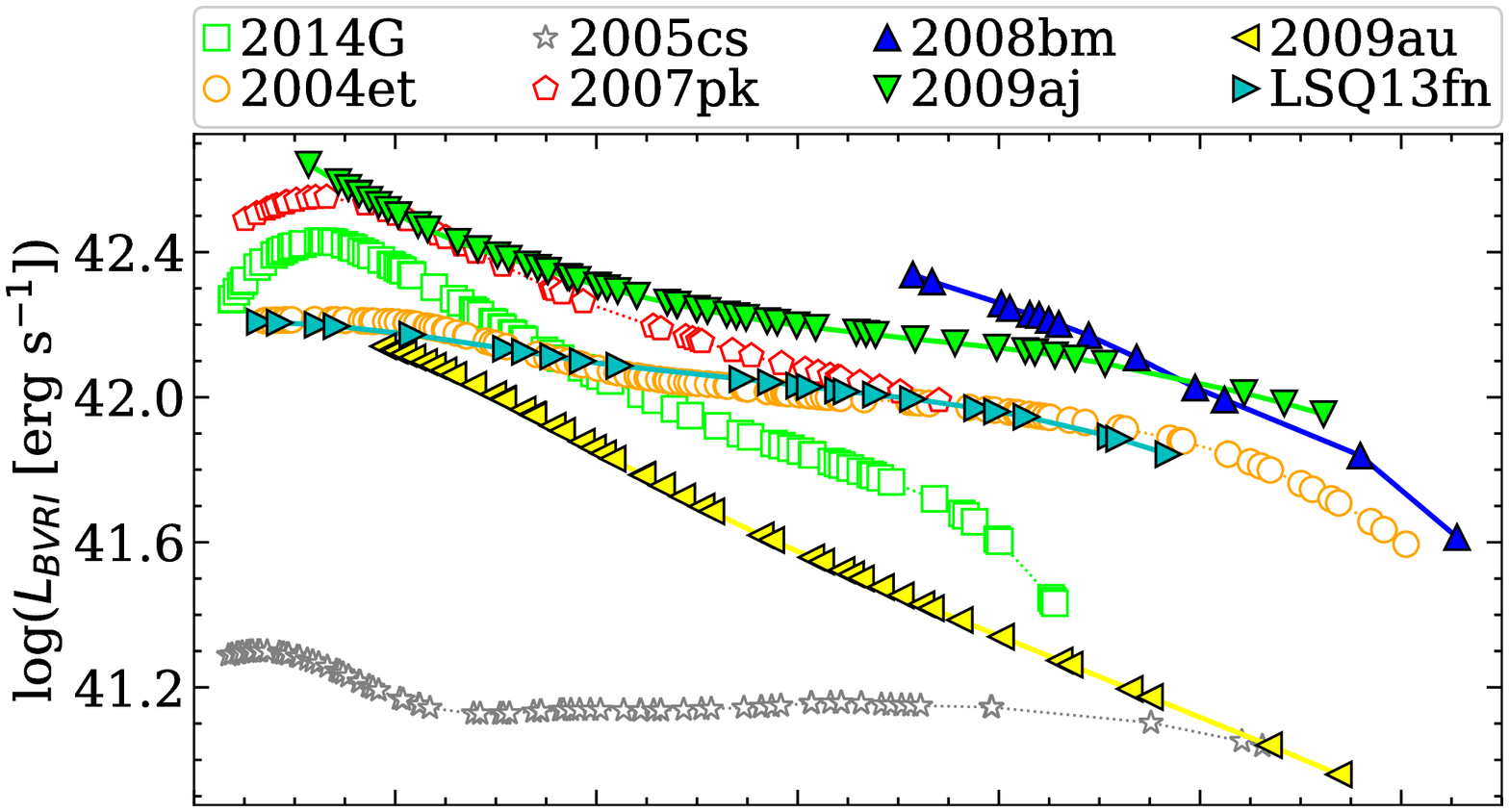}
\includegraphics[width=1.0\columnwidth]{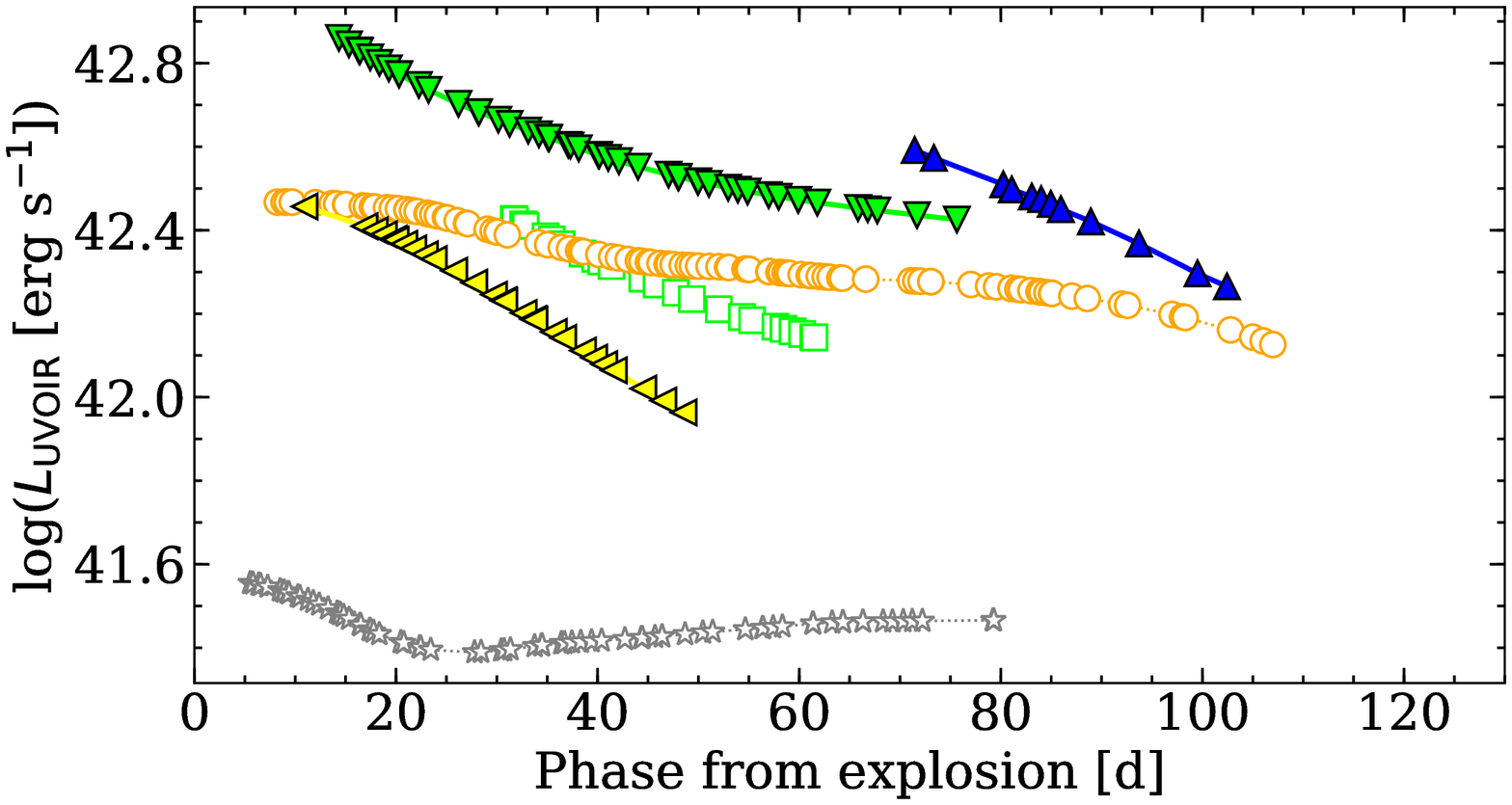}
\caption{\bvri\ (top) and UVOIR (bottom) pseudo-bolometric light curves of the LLEV SNe~II in our set (filled symbols) and a selection of SNe~II that share some of the properties seen in LLEV SNe~II (empty symbols).}
\label{fig:Lbol}
\end{figure}

The top part of Fig.~\ref{fig:Lbol} shows the \bvri\ pseudo-bolometric light curves of the LLEV SNe~II in our sample, except SN~1983K 
for which the photometry in the necessary bands is not available. For comparison, we also compute the \bvri\ pseudo-bolometric light curves for a selection of SNe~II from the literature that share some of the properties observed in our LLEV SN~II sample: normal SNe~II~2014G \citep{Bose_etal2016,Terreran_etal2016} and SN~2004et, which have decline rates similar to SN~2008bm and SN~2009au, and to SN~2009aj and LSQ13fn, respectively; the SN~IIn/II~2007pk; and the subluminous SN~II~2005cs, which has low expansion velocities (see Section~\ref{sec:expansion_velocities}). We see that SN~2008bm and SN~2009aj have luminosities higher than SNe~II with a similar decline rate, confirming the result obtained in Fig.~\ref{fig:MVmax_vs_s2}b, where the luminosity of SN~2009aj is comparable to those of the SN~2007pk at epochs earlier than 40~d since explosion. After that, the evolution of the SN~2009aj luminosity is similar to that of SN~2004et, though brighter. SN~2009au is less luminous than SN~2014G in all the phase range reported in the plot, with a luminosity at maximum light similar to SN~2004et, and similar to SN~2005cs at 100~d. LSQ13fn, as reported by \citet{Polshaw_etal2016}, has a luminosity similar to SN~2004et. We also compute the pseudo-bolometric light curve using ultraviolet, optical, and near-IR (UVOIR) photometry, from $u$/$U$- up to $H$-band (3600{\textendash}16500~\AA, bottom part of Fig.~\ref{fig:Lbol}). We find that SN~2008bm and SN~2009aj are not only brighter in the optical (top part of Fig.~\ref{fig:Lbol}), but also in the UVOIR wavelength range.

\subsection{Colour evolution}\label{sec:colour_evolution}
Normal SNe~II are found to form a continuum population in observed colours, showing a large diversity at all epochs. While there are some (red) normal SNe~II that clearly show the effects of strong host galaxy reddening, most of the colour dispersion apparently arises from intrinsic colour differences between normal SNe~II (\citealt{deJaeger_etal2018_SNII_colors}, hereafter D18).

\begin{figure}
\includegraphics[width=1.0\columnwidth]{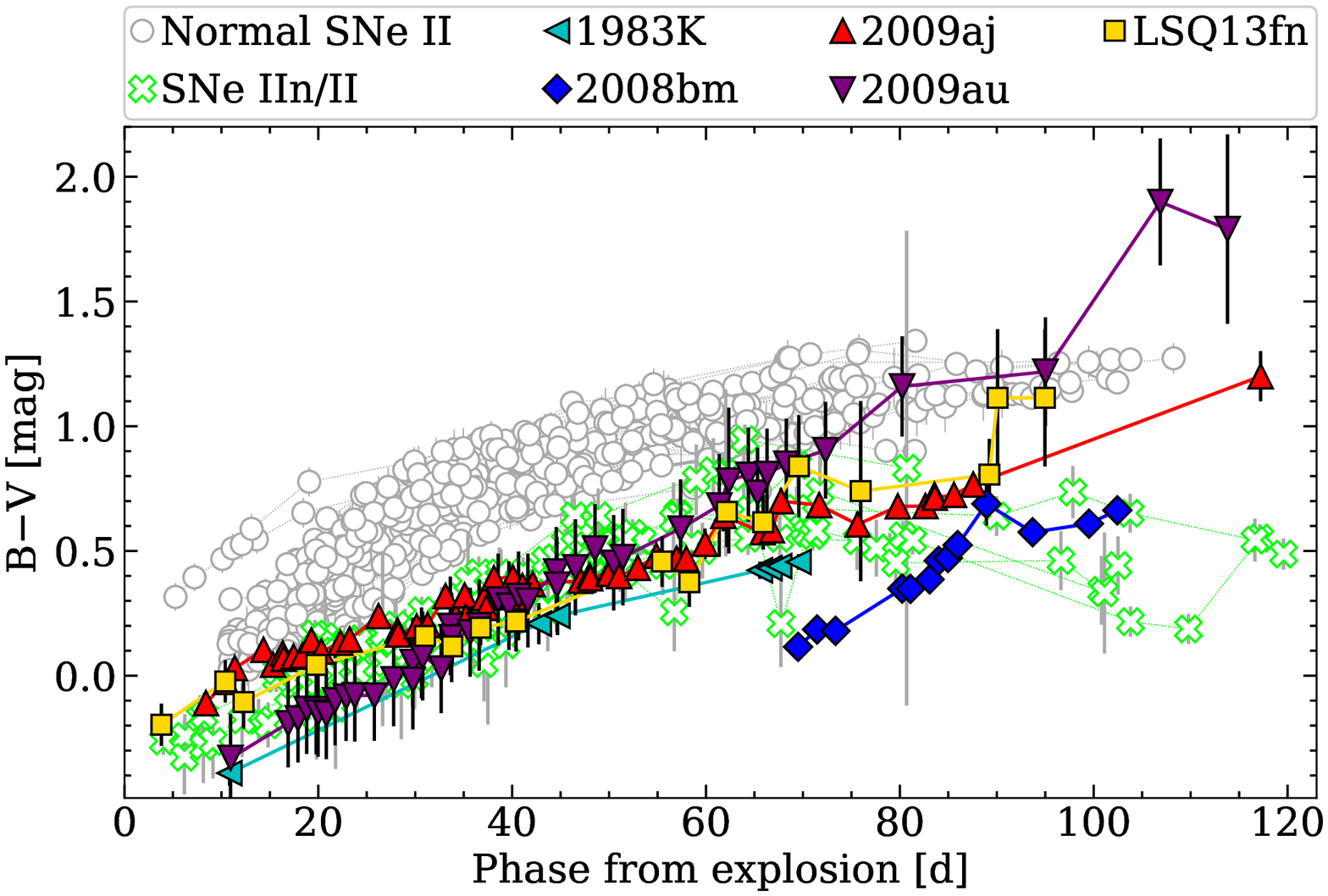}
\includegraphics[width=1.0\columnwidth]{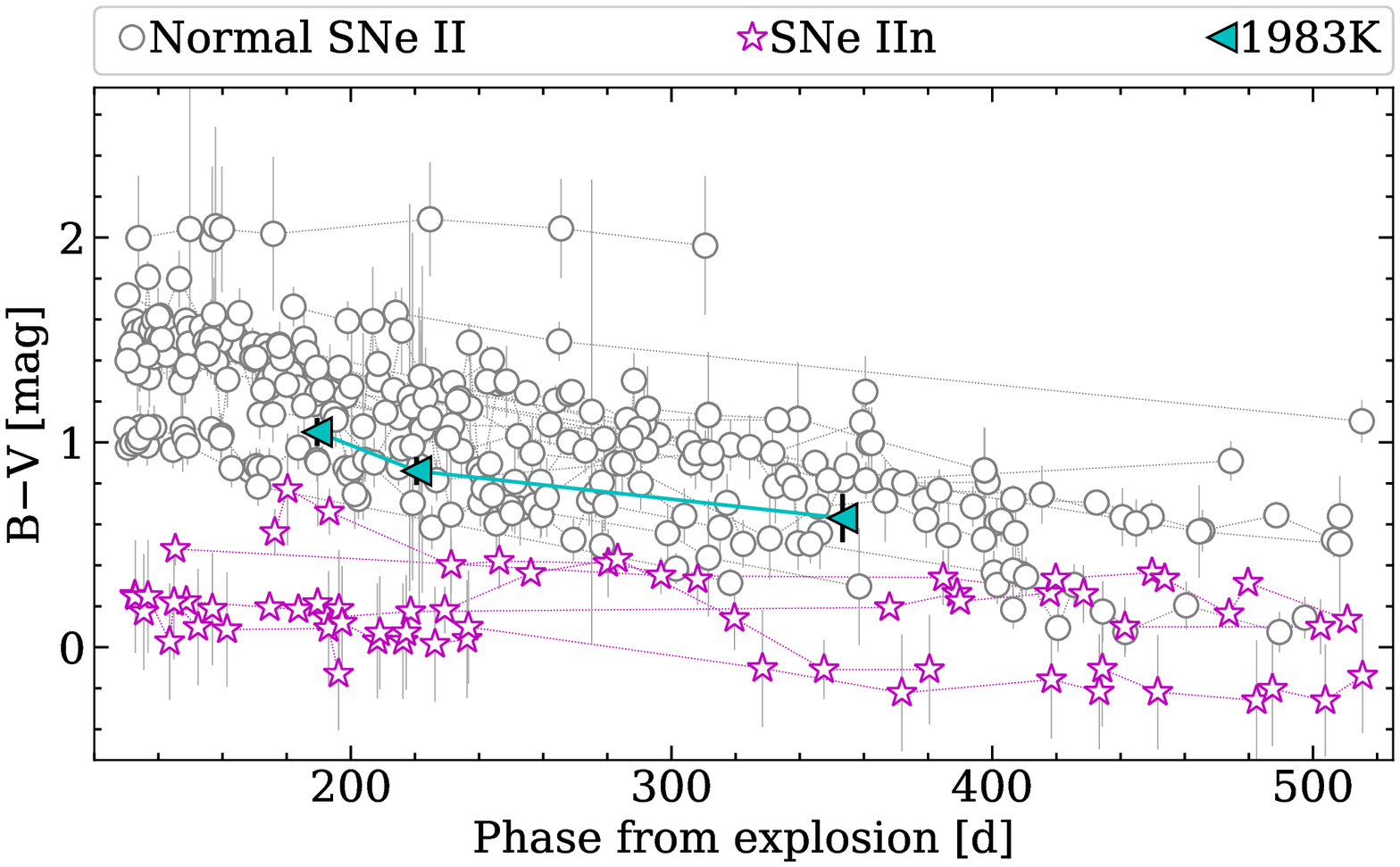}
\caption{Evolution of the \bv\ colour, corrected for \EhBV. Top: during the first 120~d since explosion of the LLEV SNe~II in our set (filled symbols) compared with the normal SNe~II in the D18 sample (empty circles). Bottom: during the late-time phase of SN~1983K (filled triangle) and of normal SNe~II from the literature with $B$- and $V$-band photometry during 190{\textendash}350~d since explosion (empty circles). We also include SNe~IIn/II (top, green crosses) and SNe~IIn (bottom, magenta stars) to depict the effect of the ejecta-CSM interaction over the \bv\ colour.}
\label{fig:B-V}
\end{figure}

The top part of Fig.~\ref{fig:B-V} shows the \bv\ colour curves of the normal SNe~II analysed in D18\footnote{We remove SN~2009au from the D18 sample since it is in our LLEV SN~II sample.} (empty circles) and corrected for \EhBV\ (using values of A14), along with the LLEV SNe~II in our set. We note that the LLEV SNe~II are systematically bluer than normal SNe~II in all the phase range reported in the plot\footnote{In the case of SN~2009au, its \bv\ colour curve was matched to the rest of LLEV SNe~II in order to estimate \EhBV\ (see Section~\ref{sec:2009au_info}).}. The latter could be due to low metallicity progenitors \citep[where the lower the progenitor metallicity, the bluer the colour;][]{Dessart_etal2014}, or a consequence of an ejecta-CSM interaction, which makes colours bluer than normal SNe~II \citep[e.g.,][]{Hillier_Dessart2019} as in the case of SNe~IIn/II, which are included in Fig.~\ref{fig:B-V} for comparison.

The bottom part of Fig.~\ref{fig:B-V} shows the evolution of the late-time \bv\ colour of SN~1983K, compared with normal SNe~II that we found in the literature with $B$- and $V$-band photometry during 190{\textendash}350~d since explosion, i.e., the phase range covered by the late-time \bv\ colour of SN~1983K. This sample is listed in Table~\ref{table:normal_SNII_sample}. We see that the \bv\ colour evolution of SN~1983K is consistent with the rest of normal SNe~II in the plot. We also include a set of SNe~IIn (sample listed in Table~\ref{table:SN_IIn_sample}) to depict the effect of a long lasting ejecta-CSM interaction over the \bv\ colours. We see that the ejecta-CSM interaction on SNe~IIn makes their colour bluer than normal SNe~II. The latter suggests that for SN~1983K, the late-time flux is dominated by the decay of \isoco\ to \isofe.

\subsection{Nickel mass}
Assuming that the observed flux during the radioactive tail is due to the decay of \isoco\ to \isofe, and that all the $\gamma$-rays from that decay are thermalized, the \isoni\ mass synthesized during a SN~II explosion can be estimated as
\begin{equation}\label{eq:Ni_mass}
\frac{M(^{56}\mathrm{Ni})}{\mathrm{M}_\odot} = \left(\frac{L_t}{L_*}\right) \mathrm{exp}\left[\frac{(t-t_0)/(1+z)-6.1\mathrm{\,d}}{111.26\mathrm{\,d}}\right]
\end{equation}
\citep[e.g.,][]{Hamuy2003}, where $L_t$ is the bolometric luminosity measured at epoch $t$, $L_*=1.271\times10^{43}$~\lumunit, and $t_0$ and $z$ are the explosion epoch and heliocentric redshift of the SN, respectively.

To estimate $L_t$ for the LLEV SNe~II in our set, we use
\begin{equation}
\log\left(L_t/L_*\right) = (\mu-V_{0,t}-\mathrm{BC})/2.5 -7.38,
\end{equation} 
where $V_{0,t}$ is the $V$-band magnitude at epoch $t$ during the radioactive tail corrected by extinction, $\mu$ is the SN distance modulus, and BC is the bolometric correction. For the latter, we assume the same BC for SNe~II during the radioactive tail \citep[$\mathrm{BC}=0.26\pm0.06$~mag;][]{Hamuy2001}. For SN~2009aj we convert the last $Rr$ photometry (at 173~d since explosion) to $V$-band magnitude using $V\!-\!R$ and $V\!-\!r$ colours of normal SNe~II nearly at the same epoch of the $Rr$ photometry, obtaining $V=20.60\pm0.13$~mag. For LSQ13fn, we compute $V=26.24\pm0.30$~mag from its nebular spectrum (at 335~d since explosion). For SN~1983K we measure a $V$-band radioactive tail slope ($s_3$) of $1.01\pm0.02$~\sunit, which is consistent with the slope of 0.98~\sunit\ expected for the complete $\gamma$-ray trapping scenario. For SN~2008bm, SN~2009aj, and LSQ13fn, there are not enough photometric data during the nebular phase to estimate $s_3$. Since we cannot check the complete $\gamma$-ray trapping scenario for those SNe, we adopt their \isoni\ mass estimations as lower limits.  \isoni\ mass values are listed in Table~\ref{table:SN_sample}

\begin{figure}
\includegraphics[width=1.0\columnwidth]{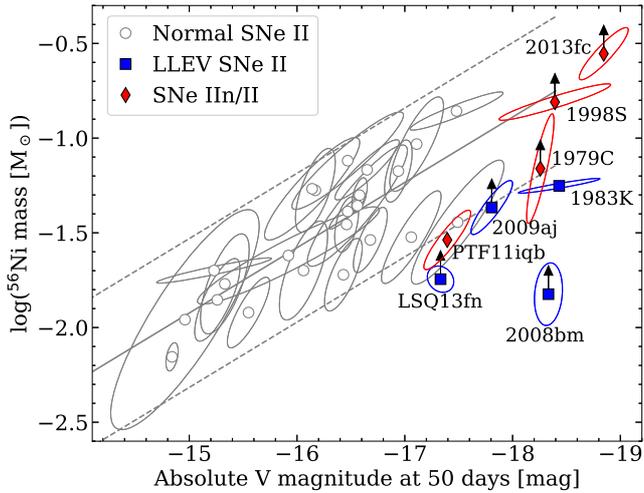}
\caption{\isoni\ mass versus absolute $V$-band magnitude at 50~d since explosion, showing the normal SNe~II (empty circles), the SNe~IIn/II from the literature (red diamonds), and the LLEV SNe~II in our set (blue squares). The ellipses indicate the $1\sigma$ statistical error. The solid line corresponds to a linear fit, where dashed lines indicate the $2\sigma$ error around the fit. Arrows indicate lower limits.}
\label{fig:Ni_mass_vs_MV50d}
\end{figure}

Fig.~\ref{fig:Ni_mass_vs_MV50d} shows the location of the LLEV SNe~II (blue squares) in the log \isoni\ mass versus the absolute $V$-band magnitude at 50~d since explosion ($M_V^{50\mathrm{d}}$) space\footnote{Since the errors in the $\log(M$(\isoni)) versus $M_V^{50\mathrm{d}}$ space are dominated by errors in reddenings and/or distances, the confidence regions are elongated ellipsoids \citep[e.g.,][]{Pejcha_Prieto2015b}.}. For comparison, we include normal SNe~II (empty circles) in the A14 sample, which have \isoni\ mass values estimated in the same manner than our LLEV SNe~II, and the normal SNe~II in the \citet{Hamuy2003} sample, where we recompute the \isoni\ masses using new estimations for the distance and host galaxy reddening (the SN set and parameters are listed in Table~\ref{table:new_Ni_masses}). We also include the SN~IIn/II set (red diamonds). As noted by \citet{Hamuy2003} and subsequently by other authors \citep[e.g.,][]{Pejcha_Prieto2015b,Valenti_etal2016,Muller_etal2017_SNII_nickel_mass}, and as visible in Fig.~\ref{fig:Ni_mass_vs_MV50d}, for normal SNe~II there is a correlation between the \isoni\ mass and $M_V^{50\mathrm{d}}$. In order to characterize the distribution of normal SNe~II, we fit a straight line (solid line), where the dashed lines indicate the $\pm2\sigma$ limits. We note that SN~1983K is below the $-2\sigma$ limit, i.e., at 50~d since explosion it is brighter than those explosions producing the same amount of \isoni. The latter, along with the colour evolution of SN~1983K (Fig.~\ref{fig:B-V}), indicates that for that SN there is a source of photons, still relevant at 50~d since explosion, which becomes negligible at the radioactive tail. We speculate that the source of these photons is the early interaction between the SN ejecta and the CSM.

\section{SPECTROSCOPIC PROPERTIES}\label{sec:spectroscopic_properties}

\begin{figure*}
\includegraphics[width=1.0\textwidth]{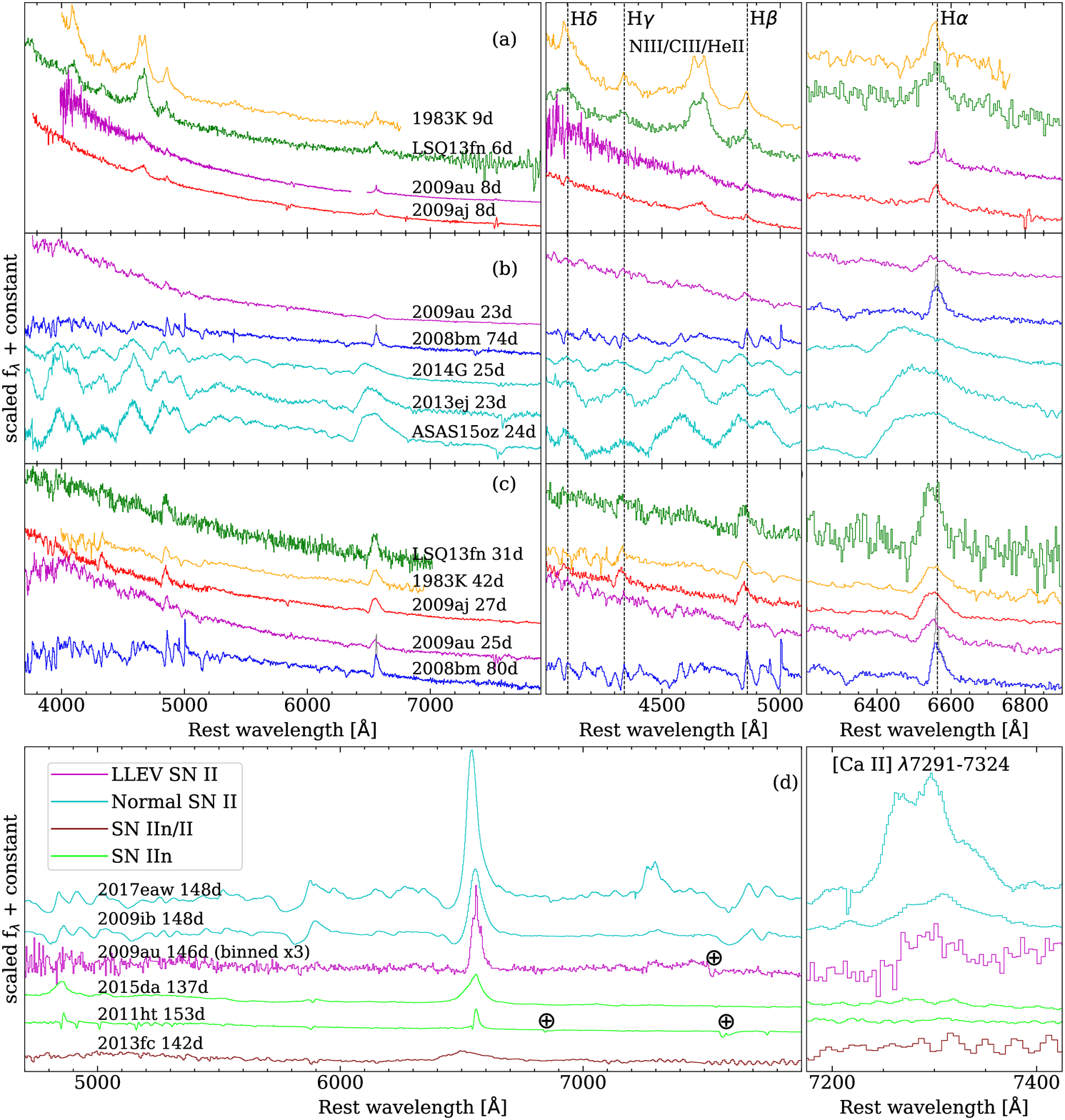}
\caption{Panels (a), (b), and (c) show the early spectra, the last spectra still affected by the ejecta-CSM interaction, and the first spectra showing Balmer absorptions for our LLEV SN~II, respectively, and zooms around H$\delta$, H$\gamma$, and H$\beta$ (middle column) and H$\alpha$ (right column). In Panel (b) we also plot normal SNe~II (cyan) for comparison. Panel (d): latest spectrum of SN~2009au compared with spectra of SNe~IIn, SNe~IIn/II, and normal SNe~II at similar epochs, and a zoom around $[$\ion{Ca}{II}$]$~$\lambda7291$-$7324$ (right). Circled crosses indicate the presence of telluric lines. Epochs are since the explosion. Spectra were corrected for redshift and reddening.}
\label{fig:csm_interaction}
\end{figure*}

\begin{figure*}
\includegraphics[width=1.0\textwidth]{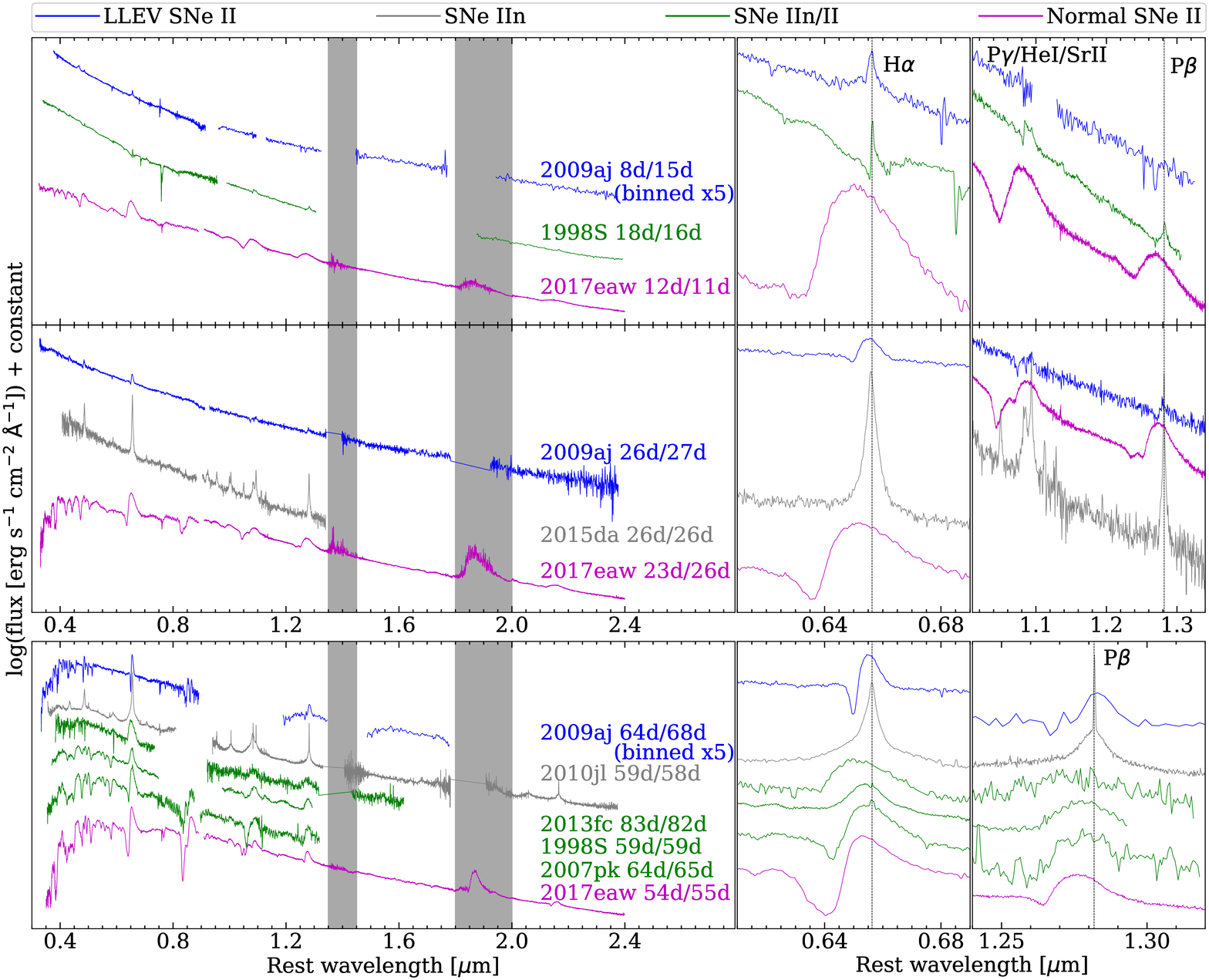}
\caption{Left column: Near-IR spectra of SN~2009aj combined with the closest in time optical spectra (blue), compared with the combined optical and near-IR spectra of SNe~IIn (gray), SNe~IIn/II (green), and normal SN~II~2017eaw (magenta) at a similar epochs. Middle column: zooms around H$\alpha$. Right column: zooms around the P$\gamma$/\ion{He}{I}/\ion{Sr}{II} and P$\beta$ profiles.}
\label{fig:IR_spectra}
\end{figure*}

In the previous section, we found evidence in favour of an early ejecta-CSM interaction scenario for LLEV SNe~II. In this section, analysing the spectroscopic data, we present further evidence that reinforces this hypothesis.

\subsection{Spectral evolution}\label{sec:early_ejecta_CSM_interaction}
Fig.~\ref{fig:csm_interaction}a shows the early spectra of the LLEV SNe~II (left), along with zooms around H$\delta$, H$\gamma$ and H$\beta$ (middle) and H$\alpha$ (right). We can see that those SNe show broad H$\alpha$ and H$\beta$ in emission, centred in the laboratory wavelengths, and with no prominent absorption features. The latter is characteristic of an SN~II ejecta interacting with a CSM, where the line broadening mechanism is dominated by electron scattering \citep{Chugai2001_SN1998_electron_scattering}. The spectral feature located between H$\gamma$ and H$\beta$ (prominent in SN~1983K and LSQ13fn, but weaker in SN~2009aj and SN~2009au spectra) corresponds to \ion{He}{II}~$\lambda$4686 possibly blended with \ion{N}{III}~$\lambda\lambda$4634-40-42 \citep{Niemela_etal1985_SN1983K} and \ion{C}{III}~$\lambda$4648 \citep{Polshaw_etal2016}. The high temperature needed to produce the \ion{N}{III}/\ion{C}{III}/\ion{He}{II} feature could come from the conversion of kinetic energy into thermal energy during an ejecta-CSM interaction, which favours this scenario.

Fig.~\ref{fig:csm_interaction}b shows the spectra of SN~2009au and SN~2008bm at 23 and 74~d since explosion, respectively. Since the H$\alpha$ profile in the SN~2008bm spectrum is highly contaminated by \ion{H}{II} region lines (depicted as gray lines), we remove that contamination modelling the H$\alpha$ profile as a mixture of three Gaussians (SN H$\alpha$ absorption and emission, along with the H$\alpha$ emission from the \ion{H}{II} region). In the middle row of the right column of Fig.~\ref{fig:csm_interaction} we see that the absorption part of the H$\alpha$ P-Cygni profile is not strong. The latter could arise from the fact that SN~2008bm and SN~2009au are fast decliners, which typically have weaker H$\alpha$ P-Cygni profile absorptions \citep[e.g.,][]{Gutierrez_etal2017_II}, but also could indicate that the ejecta-CSM interaction is still ongoing. In Fig.~\ref{fig:csm_interaction}b we also plot spectra of the fast decliners SN~2014G, SN~2013ej \citep[e.g.,][]{Bose_etal2015_SN2013ej,Huang_etal2015_SN2013ej}, and ASASSN-15oz \citep{Bostroem_etal2019_ASAS15oz} at epochs close in time to the SN~2009au spectrum. We see that the H$\alpha$ P-Cygni absorption component of SN~2008bm and SN~2009au is weaker (and narrower) than those of the other fast decliners, which indicates that the weakness of the H$\alpha$ P-Cygni absorption component of SN~2008bm and SN~2009au is not only related to their high photometric decline rates but also due to still ongoing ejecta-CSM interaction.

Fig.~\ref{fig:csm_interaction}c shows the spectra of LSQ13fn, SN~1983K, SN~2009aj, SN~2009au, and SN~2008bm at 31, 42, 27, 27, and 80~d since explosion, respectively. At these epochs, the LLEV SNe~II in our set start to show the absorption part of the H$\alpha$ P-Cygni profile, and the emission peaks appear blueshifted, where the broader the emission the more blueshifted the peak. The latter characteristic is common in normal SNe~II \citep{Anderson_etal2014_blueshifted_emission}, indicating that the effect of the ejecta-CSM interaction decreases substantially after 4{\textendash}11 weeks since explosion.

Fig.~\ref{fig:csm_interaction}d shows the latest spectrum of SN~2009au at 146~d since explosion, compared to spectra at similar epochs of SNe~IIn (green) and SNe~IIn/II (brown) that we found in the literature, and some normal SNe~II (cyan). The most prominent spectral feature in the SN~2009au spectrum is the H$\alpha$ profile in emission, while the rest of the spectrum seems featureless compared to normal SNe~II, as in the case of the SNe~IIn and SNe~IIn/II in the plot. However, we detect $[$\ion{Ca}{II}$]$~$\lambda7291$-$7324$ which is present in normal SNe~II but absent in the SN~IIn and SN~IIn/II spectra in the plot.

The left part of Fig.~\ref{fig:IR_spectra} shows the near-IR spectra of SN~2009aj combined with the closest (in time) optical spectra, along with zooms around H$\alpha$ (middle column), P$\beta$ and the spectral feature corresponding to P$\gamma$ blended with \ion{He}{I}~$\lambda10830$ and \ion{Sr}{II}~$\lambda10920$ (right column). For comparison, we also include combined optical and near-IR spectra of the SNe~IIn (red) 2015da \citep{Tartaglia_etal2020_SN2015da} and 2010jl \citep{Zhang_etal2012_SN2010jl,Borish_etal2015_SN2010jl}, the SNe~IIn/II (green) 1998S, 2007pk, and 2013fc, and the normal SN~II~2017eaw \citep[e.g.,][]{Szalai_etal2019_SN2017eaw}. All the previous SNe~II were found having optical and near-IR spectra close in time to the combined spectra of SN~2009aj. At 1{\textendash}2 weeks since explosion, the spectrum of SN~2009aj does not show the P$\beta$ profile, being more similar to the spectrum of SN~IIn~1998S than to the normal SN~II~2017eaw. The absorption at 1.27~$\mu$m in the SN~2009aj spectrum is not followed by a redder emission, so this feature might not belong to the SN. At 3{\textendash}4 weeks since explosion, the spectrum of SN~2009aj is more similar to the normal SN~II~2017eaw than to the SN~IIn~2015da, but with H$\alpha$, P$\beta$, and P$\gamma$/\ion{He}{I}/\ion{Sr}{II} profiles weaker and narrower than the observed in SN~2017eaw. At 9{\textendash}10 weeks since explosion, the H$\alpha$ profile and the P$\beta$ emission component are more similar to SNe~IIn/II and SN~2017eaw, but still weaker.

The aforementioned characteristics of the near-IR spectral evolution of SN~2009aj are quite similar to those found in the optical spectral evolution of the LLEV~SNe~II.

\subsection{Weakness of metal lines}\label{sec:lack_of_metal_lines}
\begin{figure}
\includegraphics[width=1.0\columnwidth]{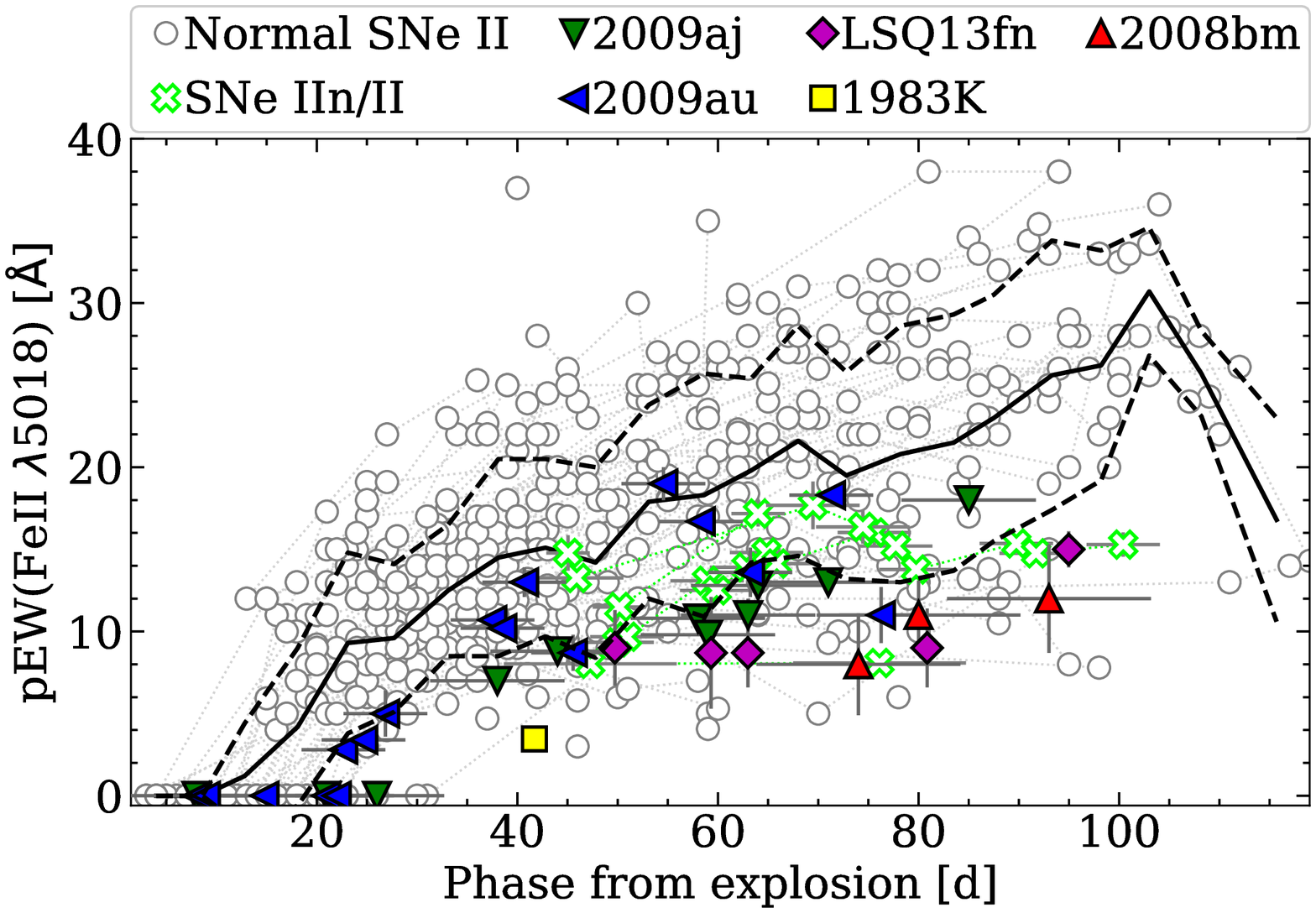}
\includegraphics[width=1.0\columnwidth]{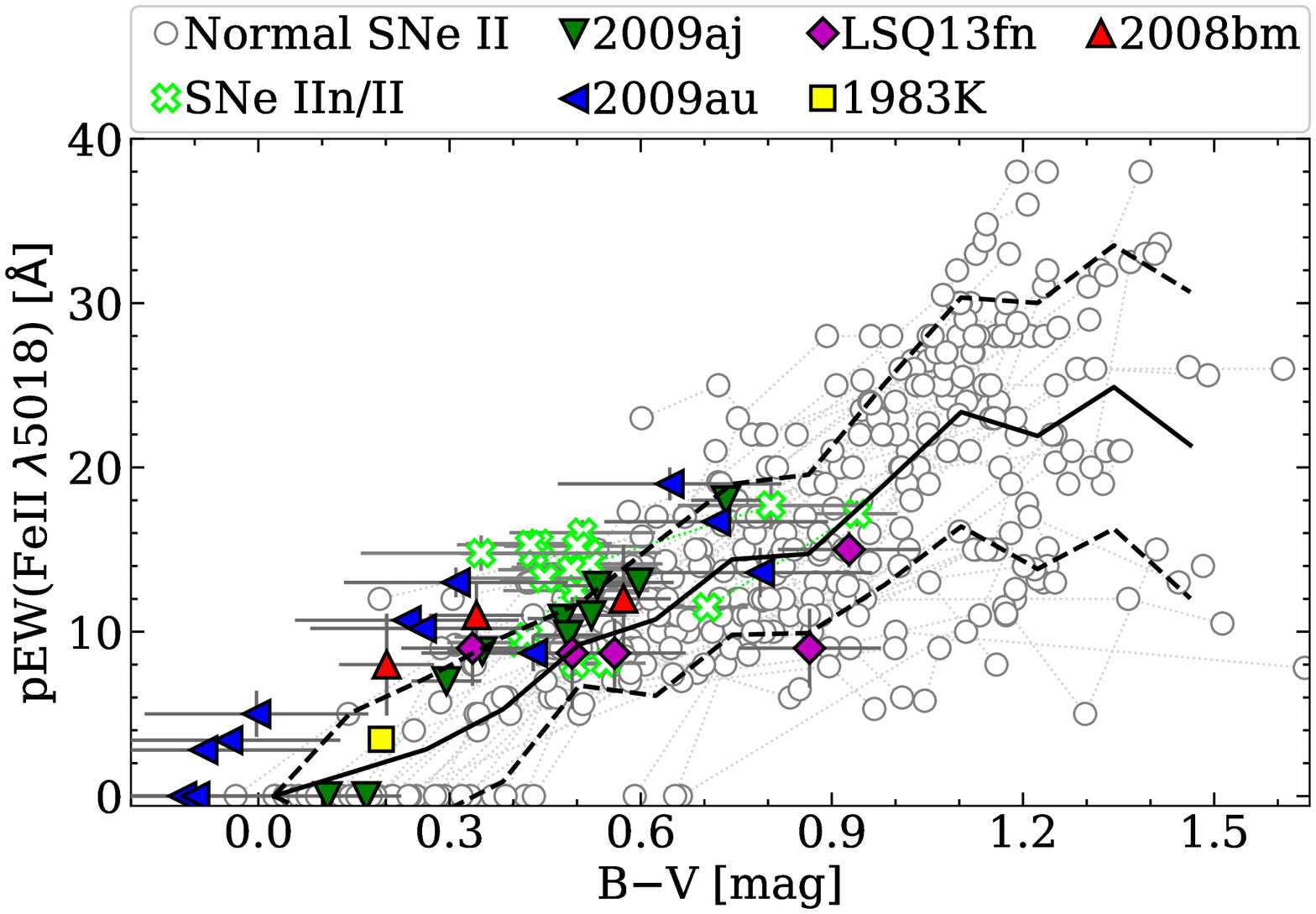}
\caption{Evolution of the \feii~pEW, as a function of the time since explosion (top) and the \bv\ colour (bottom), of the LLEV SNe~II (filled symbols), the SNe~IIn/II from the literature (empty crosses), and the normal SNe~II in the G17 sample (empty circles). Black solid and dashed lines correspond to the mean values and standard deviations, respectively.}
\label{fig:pEW_FeII5018}
\end{figure}

The top part of Fig.~\ref{fig:pEW_FeII5018} shows the evolution of the \feii~pEW of the LLEV SNe~II in our set (coloured symbols), compared to the normal SNe~II in the G17 sample (empty circles). We see that the LLEV SNe~II are mostly below the $-1\sigma$ limit (dashed line) of the G17 sample, which indicates the weakness of metal lines in the LLEV SNe~II spectra in all the phase range reported in the plot. 

\citet{Polshaw_etal2016} had already reported the weakness of metal lines in the spectra of LSQ13fn compared to normal SNe~II. In that work, one of the parameters that they suggested to explain the weakness of metal lines is a low metallicity of the SN progenitor. In fact, \citet{Dessart_etal2014} explored theoretically the dependence on the SN~II progenitor metallicity of some spectral features. The general behaviour they obtained is: the lower the progenitor metallicity, the weaker the metal lines. A lower metallicity also implies a lower line blanketing and therefore a colour bluer than other SNe~II with a higher metallicity, which in principle could explain the blue colour we observe in LLEV SNe~II.

Nevertheless, the strength of metal lines not only depends on the progenitor metallicity, but also on the temperature of the line formation region (which modifies the opacity) and, when the SN is surrounded by a CSM, on the contribution to the flux generated from the ejecta-CSM interaction \citep{Leloudas_etal2015}. Regarding the temperature, \citet[][hereafter A16]{Anderson_etal2016} show that the pEW of the \feii\ line in normal SN~II spectra increases with the $V\!-\!I$ and $V\!-i$ colour, which are used as a proxy for temperature.

\begin{figure}
\includegraphics[width=1.0\columnwidth]{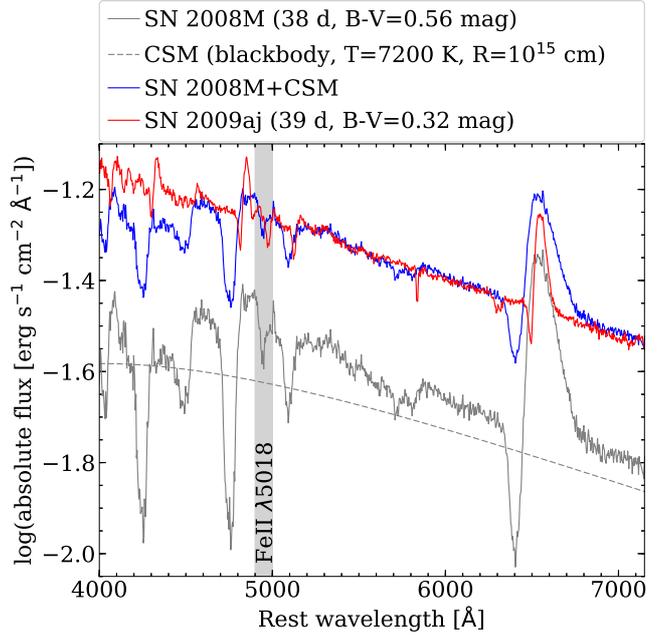}
\caption{Comparison of the spectrum of SN~2009aj at 39~d since explosion (red line) with a composite spectrum (blue line), which is the sum of the SN~2008M spectrum at 38~d since explosion (gray line) and a blackbody continuum (dashed line). Gray region indicates the location of the \feii\ line.}
\label{fig:CSM_veiling}
\end{figure}

The bottom part of Fig.~\ref{fig:pEW_FeII5018} shows the \feii~pEW as a function of the \bv\ colour. In the plot we see that LLEV SNe~II occupy a region similar to SNe~IIn/II with both groups scattering around the $+1\sigma$ limit of normal SNe~II. This indicates that in LLEV SNe~II the weakness of metal lines is at least partially due to the higher temperature in the line forming region, but also that the line dilution play a role. Regarding the latter, to quantify its effect to first order we select SN~2008M (A14, G17) because it is one of the bluest normal SNe~II in the D18 sample. With this, we can isolate as best as possible the effect of CSM dilution on the pEW of the metal lines from the similar effect of temperature we have analysed previously. Fig.~\ref{fig:CSM_veiling} shows the spectrum of SN~2009aj at 39~d since explosion (red line) and a spectrum of SN~2008M nearly at the same epoch and scaled to the SN~2009aj distance (gray line). Spectra of SN~2009aj and SN~2008M have \feii~pEW of 7.1 and 9.7~\AA, respectively. In order to include the contribution of a CSM to the flux of SN~2008M, we model the CSM as a blackbody. We choose the CSM parameters such that the continuum of SN~2008M plus the CSM matches to the continuum of SN~2009aj, which is reached using a radius of $10^{15}$~cm and a temperature of 7200~K. The \feii~pEW of the composite spectrum is of 5.7~\AA, which matches better to the value for SN~2009aj. Therefore, even when the SN~2008M spectrum is not as blue as the spectrum of SN~2009aj, and the oxygen abundance at the site of SN~2008M (8.43~dex; A16, which we use as a proxy for the progenitor metallicity) is higher than the value for SN~2009aj (8.29~dex; see Table~\ref{table:SN_sample}), the flux contribution due to the ejecta-CSM interaction seems to be an efficient driver of the observed weakness of metal lines in its spectra.

\begin{figure}
\includegraphics[width=1.0\columnwidth]{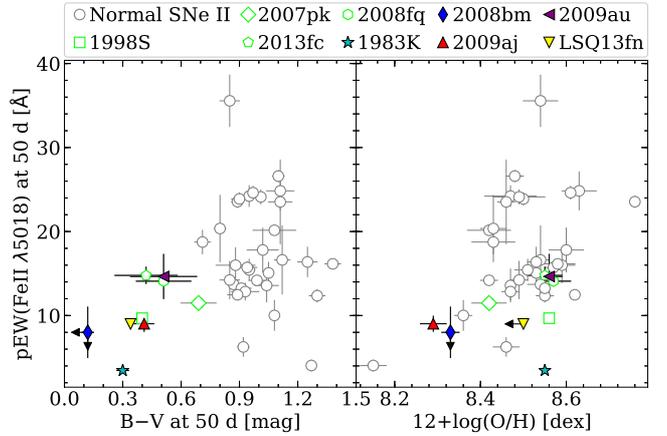}
\caption{pEW of the \feii\ line at 50~d since explosion as a function of the \bv\ colour (corrected for reddening) at the same epoch (left), and as a function of the oxygen abundance (right), showing the LLEV SNe~II in our set (coloured filled symbols), the SNe~IIn/II from the literature (green empty symbols) and the normal SNe~II in the A14 sample (gray empty circles). Arrows indicate upper limits.}
\label{fig:pEW_vs_BV_Z}
\end{figure}

Fig.~\ref{fig:pEW_vs_BV_Z} shows the \feii~pEW at 50~d since explosion, as a function of the \bv\ colour (corrected for \EhBV) at the same epoch (left), and as a function of the oxygen abundance in the N2 calibration of \citet{Marino_etal2013} (right) for the LLEV SNe~II in our set (coloured filled symbols). For comparison, we plot the normal SNe~II in the A16 sample\footnote{We remove SN~2008bm, SN~2009aj, and SN~2009au from the sample since they are in our LLEV SN~II sample.\label{footnote}} (gray empty circles, where we use the colour curves plotted in Fig.~\ref{fig:B-V} to compute \bv\ colours at 50~d since explosion), and the SNe~IIn/II (green empty symbols). In the left panel we see that SN~1998S, SN~2009aj, and LSQ13fn have similar \feii~pEW values and \bv\ colours, while in the right panel we see that the same SNe have different oxygen abundances. The latter indicates that (1) the metallicity is not a dominant component determining the \feii~pEW values as \bv\ colour is, or (2) SN~1998S and LSQ13fn suffer more line dilution compared to SN~2009aj. In Fig.~\ref{fig:pEW_vs_BV_Z} we also see SN~2013fc, LSQ13fn, and SN~1983K having similar \bv\ colour and oxygen abundances, but different \feii~pEW values. One explanation for this difference is that SN~1983K and LSQ13fn suffer more line dilution than SN~2013fc. Based on these findings, we suggest that the progenitor metallicity is not the main driver of the weakness of metal lines seen on LLEV SNe~II spectra, but a combination of higher temperatures at the line formation region and line dilution (both being consequence of an early ejecta-CSM interaction).

\begin{figure}
\includegraphics[width=1.0\columnwidth]{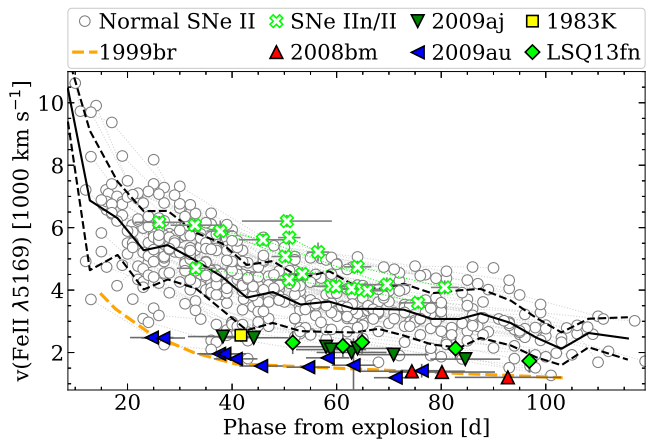}
\caption{Evolution of the expansion velocities, measured through the \ion{Fe}{II}~$\lambda$5169 minimum absorption lines, of the LLEV SNe~II (filled symbols), the SNe~IIn/II from the literature (empty crosses), and the normal SNe~II in the G17 sample (empty circles). Black solid and dashed lines correspond to the mean values and standard deviations, respectively, of the expansion velocities in the G17 sample. The orange dashed line corresponds to the expansion velocities for SN~1999br.}
\label{fig:FeII5169}
\end{figure}

\subsection{Expansion velocities}\label{sec:expansion_velocities}
Fig.~\ref{fig:FeII5169} shows the evolution of the expansion velocities, measured through the \ion{Fe}{II}~$\lambda$5169 minimum absorption lines ($v_\mathrm{FeII}$), of the LLEV SNe~II in our set (coloured symbols) and the normal SNe~II in the G17 sample\textsuperscript{\ref{footnote}} (empty circles). Expansion velocities for LLEV SNe~II are all below the $-1\sigma$ limit (dashed line) of the G17 sample (2820~\kms\ at 50~d since explosion). In particular, SN~2008bm and SN~2009au have expansion velocities similar to SN~1999br (orange dashed line), which is one of the SNe~II with the lowest expansion velocities in the G17 sample ($v_\mathrm{FeII}\approx1550$~\kms\ at 50~d since explosion). These low expansion velocities can be explained by the loss of kinetic energy by the ejecta due to their interaction with CSM.

\section{DISCUSSION}\label{sec:discussion}

\subsection{A break in the magnitude-velocity relation}
\begin{figure}
\includegraphics[width=1.0\columnwidth]{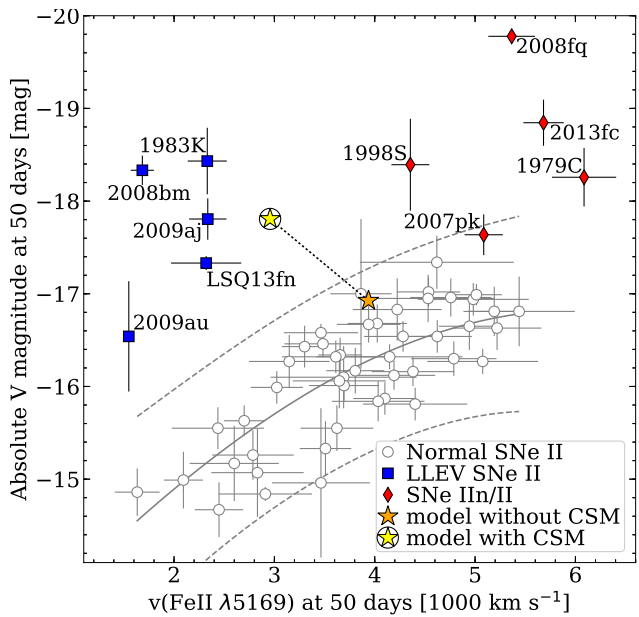}
\caption{Absolute $V$-band magnitude versus expansion velocity, both at 50~d since explosion, showing the LLEV SN~II set (blue squares), the normal SNe~II in the A14 sample (empty circles), and the SNe~IIn/II from the literature (red diamonds). The solid line corresponds to the Gaussian process fit, where dashed lines indicate the 3$\sigma$ error around the fit. We also plot the location of our model for SN~2009aj (see Section~\ref{SN2009aj_model}) without (orange star) and with (circled yellow star) CSM.}
\label{fig:MV50d_vs_v50d}
\end{figure}

As we have seen in Section~\ref{sec:MVmax_vs_s2} and \ref{sec:expansion_velocities}, LLEV SNe~II are characterized by having luminous peak magnitudes and low expansion velocities. This is not expected in a scenario where more energetic SN~II explosions that produce high luminosities also have high expansion velocities. This trend is shown in Fig.~\ref{fig:MV50d_vs_v50d}, where we plot the $V$-band absolute magnitude and the expansion velocity, both at 50~d post explosion, for the normal SNe~II in the A14 SN sample (gray circles, where expansion velocities are from Table~3 of \citealt{Gutierrez_etal2017_II}). To characterize the distribution of the A14 sample in this space, we perform a Gaussian process fit (solid line), where dashed lines indicate the $\pm3\sigma$ error around the fit. We can see that the LLEV SNe~II\footnote{For SN~1983K and SN~2008bm we extrapolate the expansion velocities at 50~d since explosion using the power-law given in \citet{Nugent_etal2006}.} (blue squares) populate a region where the only known member was LSQ13fn, marking a discrepancy of 2{\textendash}3~mag with respect to the trend obtained with the A14 sample. At this point, it is not clear whether the LLEV SNe~II can be considered as a separated class of SNe~II, or whether they are part of a continuum in the SN~II distribution induced by an increasing ejecta-CSM interaction.

To test the latter, we include in Fig.~\ref{fig:MV50d_vs_v50d} the SNe~IIn/II set. We see that SN~1998S, SN~2008fq, SN~2013fc, and, possibly, SN~1979C, are brighter than the 3$\sigma$ limit. However, they do not fill the gap between the distribution of normal SNe~II and the LLEV SNe~II.

\begin{figure}
\includegraphics[width=1.0\columnwidth]{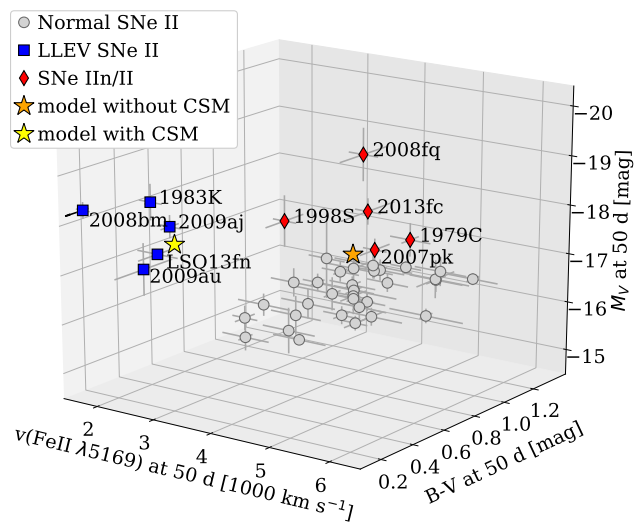}
\caption{Absolute $V$-band magnitude versus expansion velocity and \bv\ colour, all at 50~d since explosion, for the LLEV SNe~II in our set (blue squares), the normal SNe~II in the A14/D18 sample (gray circles), and the SNe~IIn/II from the literature (red diamonds). The \bv\ colour of SN~2008bm is an upper limit. We also plot the location of our model for SN~2009aj (see Section~\ref{SN2009aj_model}) without (orange star) and with (yellow star) CSM.}
\label{fig:Mv50d_vs_v50d_vs_BV50d}
\end{figure}

Fig.~\ref{fig:Mv50d_vs_v50d_vs_BV50d} shows the absolute magnitude as a function of the expansion velocity and the \bv\ colour, all at 50~d since explosion of the LLEV SNe~II in our set (blue squares). In this space we see that the LLEV SNe~II seem to form a separated set of objects from the rest of normal SNe~II (gray circles). With the aim to confirm this visual finding, we run the mean-shift algorithm to search for clusters in this space. We perform simulations where points are moved within their errors (assuming a normal distribution), and we find that in all of the realizations the LLEV SNe~II form a separated group with respect to the rest of SNe~II. Although, given the low number of LLEV SNe~II, this result has to be taken with caution, the fact that these SNe always cluster in a different parameter space region, indicates that these SNe could indeed be a new sub-type of SNe~II.

\subsection{Interaction of the ejecta with a massive CSM}\label{SN2009aj_model}
As mentioned in the previous paragraphs, a scenario where the ejecta of a normal SN~II interacts with a CSM medium could explain all the characteristics observed in LLEV SNe~II. In fact, during the ejecta-CSM interaction, part of the kinetic energy is converted into thermal energy and photons, which slows down the ejecta, increases the temperature and luminosity, and veils the spectral lines.

In order to test whether the ejecta-CSM interaction is responsible of converting a normal SN~II into a LLEV SN~II, we perform hydrodynamical simulations. We focus on the modelling of SN~2009aj here because its explosion epoch and host galaxy colour excess are better constrained than those for SN~2008bm and SN~2009au. We adopt the same numerical method as in \citet{2017MNRAS.469L.108M,2018MNRAS.476.2840M} and use the radiation hydrodynamics code \texttt{STELLA} \citep{1998ApJ...496..454B,2000ApJ...532.1132B,2006A&A...453..229B} for our numerical light curve modelling.

\begin{figure}
\includegraphics[width=1.0\columnwidth]{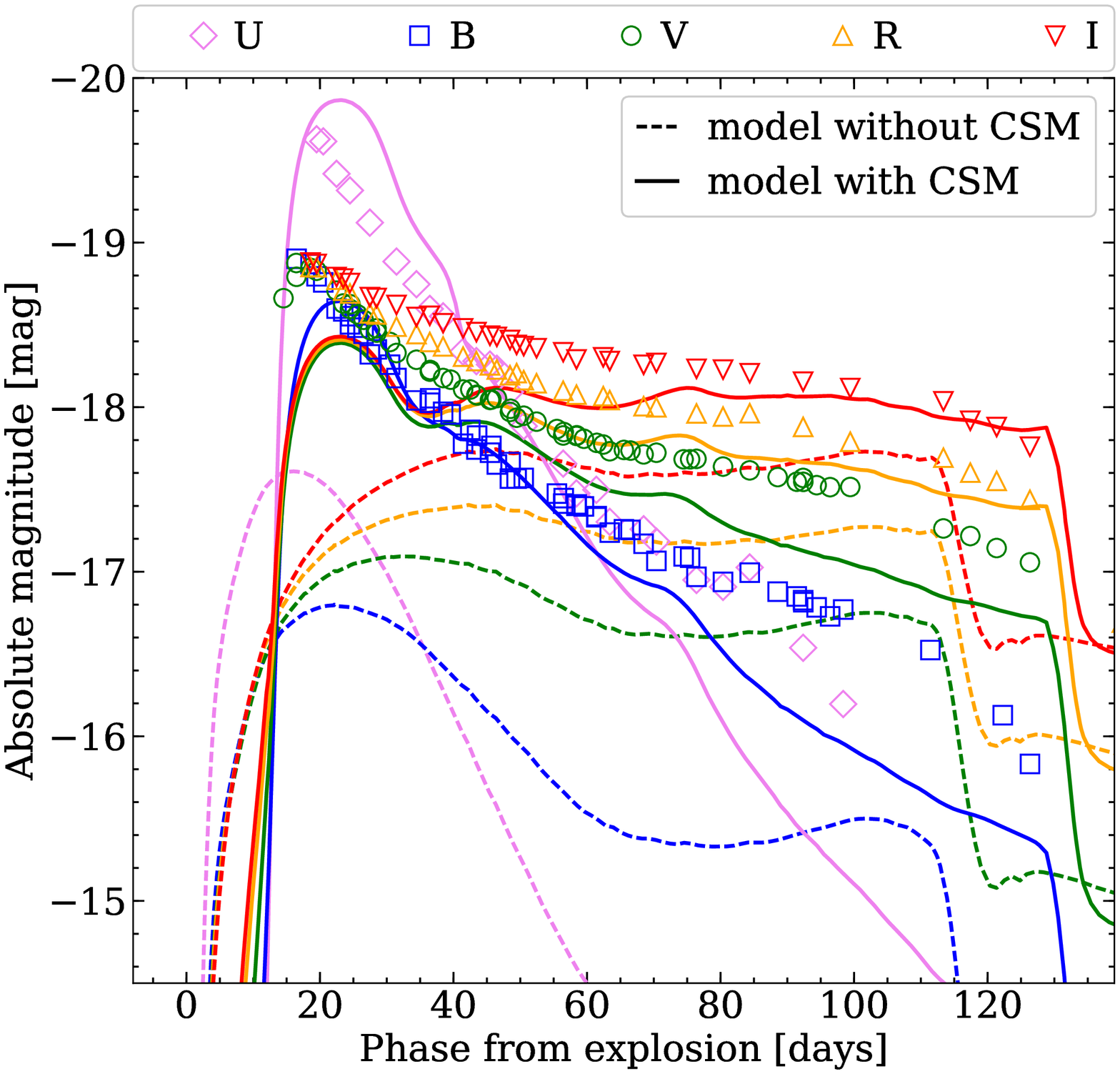}
\includegraphics[width=1.0\columnwidth]{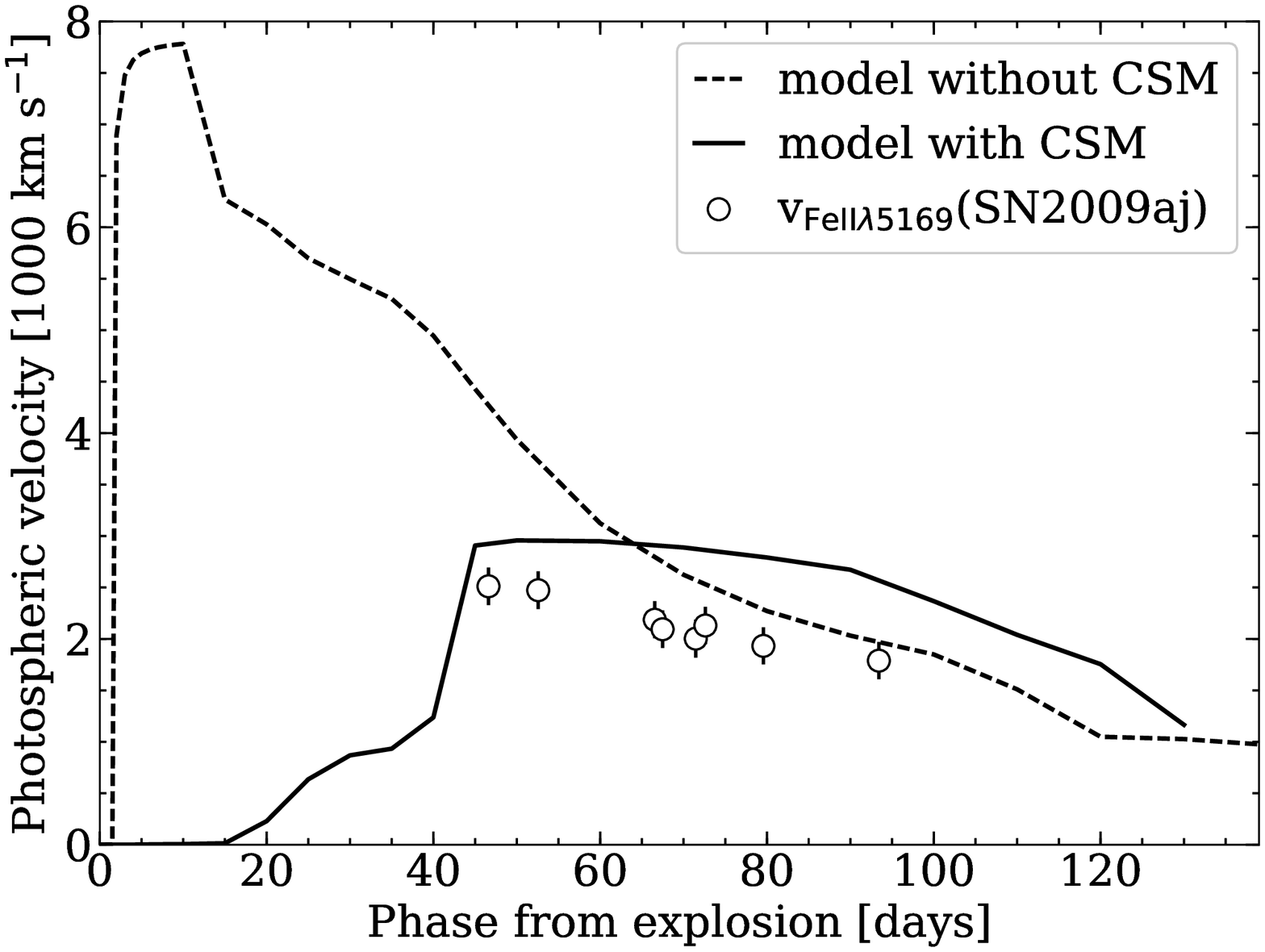}
\caption{Absolute \ubvri\ light curves (top) and expansion velocities (middle) of our model with CSM (solid lines) and of SN~2009aj (empty symbols). The same model but without CSM (dashed lines) is displayed for comparison.}
\label{fig:SN2009aj_model}
\end{figure}

We take the 14~\msun\ progenitor model in \citet{2018MNRAS.476.2840M} and attached a dense CSM above the progenitor by adopting the $\beta$ law wind velocity. We take one model with the mass-loss rate of $3\times 10^{-2}~\mathrm{M_\odot~yr^{-1}}$, the terminal wind velocity of $10~\mathrm{km~s^{-1}}$, $\beta=5$, and the dense CSM radius of $10^{15}~\mathrm{cm}$. This dense CSM has a mass of 3.6~\msun. When we explode the progenitor in this system with the explosion energy of $8\times 10^{50}~\mathrm{erg}$, we obtain the \ubvri\ absolute light curves, and expansion velocities as in Fig.~\ref{fig:SN2009aj_model} (solid lines). We can see that the observed absolute magnitudes and expansion velocities of SN~2009aj (empty symbols) match well to the numerical results, adopting an explosion epoch 1.5~d prior to the last nondetection\footnote{Given the limiting magnitude of the last non-detection, we cannot discard the presence of the SN at $>18.4$~mag.}.

The effect on the luminosity-velocity relation produced by the interaction of the ejecta with a massive CSM is depicted in Fig.~\ref{fig:MV50d_vs_v50d}. We can see that the SN~II model without CSM (orange star), consistent with the rest of the observed SNe~II, is translated outside of the $3\sigma$ limit in direction to the location of the LLEV SNe~II if we include a massive CSM (circled yellow star).

\subsection{The observed fraction of LLEV SNe~II}\label{sec:SN_rate}
In order to estimate a first approximation of the fraction of LLEV within the normal SN~II family, we use the monitoring campaign carried out by the CHASE survey. Since LLEV SNe~II are typically more luminous than normal SNe~II, and given that the CHASE is a magnitude-limit survey, the estimation of the fraction of LLEV SNe~II we will provided should be regarded as an upper limit.

During March~2008{\textendash}December~2014, CHASE reported the discovery of 46 normal SNe~II and 3 SNe~IIn (among them, SN~2009aj and SN~2009au). In addition, we include 58 normal SNe~II and 6 SNe~IIn discovered by other surveys during the same period, which were discovered by CHASE independently. In order to identify possible LLEV SNe~II in the CHASE normal SN~II sample, we check the classification reports (and classification spectra, if they are available). We select as possible LLEV SNe~II those SNe whose classification spectra best match to a subluminous SN~II (e.g., SN~2005cs), or have low expansion velocities (e.g., $\mathrm{H}_\alpha\lesssim5000$~\kms). From this list, we discard those SNe having absolute magnitude\footnote{Discovery and confirmation magnitudes reported by CHASE are obtained from unfiltered images, which are similar to $R$-band magnitudes.} (at the moment of the discovery or in posterior confirmation epochs) consistent with subluminous SNe~II \citep[$\gtrsim-16.0$~mag;][]{Lisakov_etal2018_LLSNeII}. From the CHASE SN~IIn sample we select as possible LLEV SN~II those SNe~IIn that spectroscopically do not evolve as an SN~IIn but as a normal SN~II with narrow P-Cygni profiles. For this, we use the information available in the literature.

Among the 104 normal SNe~II, 15 show spectral characteristics consistent with low expansion velocities, of which 7 are published subluminous SNe~II, and 6 have absolute magnitudes (corrected for \EhBV\ inferred from \naid\ at the redshift of the host galaxy) $>-15.5$~mag. Only SN~2009aj shows narrow P-Cygni profiles and high luminosities, while for SN~2010jc we do not have enough information to confirm or discard it. Among the 9 SNe~IIn, 7 of them evolve as SNe~IIn. Only SN~2009au shows a posterior evolution different than an SN~IIn, while for SN~2008gm there is not enough information to confirm or discard its LLEV nature. If we consider SN~2009aj and SN~2009au as the only LLEV SNe~II in the CHASE sample, then its fraction could be around 2 per cent. On the other hand, if we consider all the unconfirmed candidates (SN~2010jc and SN~2008gm) as LLEV SNe~II, then the upper limit increases to 4 per cent. The low fraction of LLEV SNe~II points toward an uncommon progenitor, which should have experienced a high mass loss rate close in time to core collapse in order to generate a sufficiently massive shell ($\sim4$~\msun) close to its surface to produce the spectroscopic and photometric characteristics that we observe in LLEV SNe~II.

\subsection{Impact on SNe~II as distance indicators}\label{sec:SNeII_as_distance_indicators}
Since the LLEV SNe~II do not follow the luminosity-velocity relation observed for normal SNe~II, we have to analyse their impact over the use of normal SNe~II as distance indicators.

As mentioned in \citet{Rodriguez_etal2019}, the SN~II distance precision using the Photospheric Magnitude Method \citep{Rodriguez_etal2014} could be up to 0.23~mag within a 99 per cent of confidence level. In this case, the location of the LLEV SNe~II ($\geq2$~mag brighter than the expected from their expansion velocities) in the Hubble diagram (HD) will be $\geq8.7\sigma$ below the Hubble law fit. Therefore these events can be easily discarded with a sufficient amount of data. 

Since the oxygen abundance of LLEV SNe~II is not necessarily low, we do not have evidence that those SNe are related to low metallicity galaxies, so we do not expect that the fraction of LLEV SNe~II will increase with redshift (e.g., $z>0.5$) as consequence of the evolution of the metallicity of the Universe. However, we expect to find a higher fraction of LLEV SNe~II due to the Malmquist bias. For example, for the Large Synoptic Survey Telescope (LSST) the $r$-band 5$\sigma$ limiting magnitude is estimated to be $\sim24.3$~mag, so for $z\geq0.5$ we expected to detect SNe~II with $M_r\lesssim-18.0$~mag. In this case, the observed fraction of LLEV SNe~II could be as high as 10{\textendash}18 per cent, so the LLEV SNe~II will be only $\geq2.9${\textendash}$3.9\sigma$ below the Hubble law, which makes it difficult to recognize them as outliers. This means that the existence of LLEV SNe~II has to be taken into account if normal SNe~II at high redshifts are used to derive cosmological parameters. Nevertheless, as shown in this work, the current sample have expansion velocities which are in the lower end of the normal SN~II distribution, therefore a cut on expansion velocities could greatly reduce their contamination.

\section{CONCLUSIONS}\label{sec:conclusions}

In this work we presented optical and near-IR data of SN~2008bm, SN~2009aj, and SN~2009au. From the analysis of these data together with already published data we found that they show similar characteristics with those of SN~1983K and LSQ13fn. In the luminosity-expansion velocity plane this possible family of SNe~II, that we call LLEV, forms a separate group which have $V$-band absolute magnitudes 2{\textendash}3~mag brighter that those expected from their expansion velocities by the luminosity-velocity relation observed for normal SNe~II. 

The ejecta-CSM interaction observed in LLEV SNe~II lasts up to 4{\textendash}11 weeks since the explosion. Subsequently, spectra show P-Cygni profiles characterized by low expansion velocities, and a weakness of metal lines. We found evidence that the metal line weakness seems not to be related to the metallicity of the progenitor but with a combined effect of the line dilution due to the contribution of the CSM to the flux, and the higher temperatures than normal SNe~II at similar epochs. Through hydrodynamic simulations, which consider a RSG progenitor of 14~\msun\ with an explosion energy of $8\times10^{50}$~erg, we found that the high luminosity and low expansion velocities seen on SN~2009aj can be explained if the ejecta interacts with a CSM of $\sim3.6$~\msun\ located very close to the progenitor. 

Based on the discoveries by the CHASE survey, we estimated an upper limit for the LLEV SNe~II fraction to be 2{\textendash}4 per cent of all normal SNe~II. This low fraction, together with the high CSM mass we obtain from the hydrodynamic simulations, may indicate an uncommon progenitor with a high mass loss rate close in time to core collapse in order to generate a sufficiently massive shell close to its surface. Based on the available data, it is unclear whether the LLEV SNe~II are a separated class of SNe~II with a different progenitor system, or there is a continuum of objects connecting them with the normal SNe~II. It is necessary to populate the luminosity-velocity space with more LLEV SNe~II in order to reveal the nature of these peculiar SNe~II. Finally, we showed that, based on the current sample, LLEV SNe~II should not represent a severe contaminant in the use of normal SNe~II as standardizable candles.

\section*{Acknowledgements}

We gratefully acknowledge support of the CSP-I by NSF under grants AST-0306969, AST-0607438, and AST-1008343. OR, GP, AC, FF, and JLP acknowledge support by the Ministry of Economy, Development, and Tourism's Millennium Science Initiative through grant IC120009, awarded to The Millennium Institute of Astrophysics, MAS. OR acknowledge support from CONICYT PAI/INDUSTRIA 79090016. TJM is supported by the Grants-in-Aid for Scientific Research of the Japan Society for the Promotion of Science (JP17H02864, JP18K13585) and by Japan Society for the Promotion of Science Open Partnership Bilateral Joint Research Project between Japan and Chile. Support for JLP is provided in part by FONDECYT through the grant 1191038. CPG acknowledges support from EU/FP7-ERC grant no. [615929]. MDS is funded by a project grant (8021-00170B) from the Independent Research Fund Denmark and a generous grant (13261) from VILLUM FONDEN. Numerical computations were in part carried out on PC cluster at Center for Computational Astrophysics, National Astronomical Observatory of Japan. NBS acknowledges support from the NSF through grant AST-1613455, and through the Texas A\&M University Mitchell/Heep/Munnerlyn Chair in Observational Astronomy. Additional support has been provided by the George P. and Cynthia Woods Mitchell Institute for Fundamental Physics and Astronomy. This work is partially based on observations collected at the European Southern Observatory under ESO programmes 082.A-0526 and 083.D-0970. This paper is based on observations obtained at the Gemini Observatory, Cerro Pachon, Chile (Gemini Program GS-2009A-Q-43). This research has made use of the NASA/IPAC Extragalactic Database (NED) which is operated by the Jet Propulsion Laboratory, California Institute of Technology, under contract with the National Aeronautics and Space Administration.




\bibliographystyle{mnras}
\bibliography{references}

\begin{thebibliography}{}
\makeatletter
\relax
\def\mn@urlcharsother{\let\do\@makeother \do\$\do\&\do\#\do\^\do\_\do\%\do\~}
\def\mn@doi{\begingroup\mn@urlcharsother \@ifnextchar [ {\mn@doi@}
  {\mn@doi@[]}}
\def\mn@doi@[#1]#2{\def\@tempa{#1}\ifx\@tempa\@empty \href
  {http://dx.doi.org/#2} {doi:#2}\else \href {http://dx.doi.org/#2} {#1}\fi
  \endgroup}
\def\mn@eprint#1#2{\mn@eprint@#1:#2::\@nil}
\def\mn@eprint@arXiv#1{\href {http://arxiv.org/abs/#1} {{\tt arXiv:#1}}}
\def\mn@eprint@dblp#1{\href {http://dblp.uni-trier.de/rec/bibtex/#1.xml}
  {dblp:#1}}
\def\mn@eprint@#1:#2:#3:#4\@nil{\def\@tempa {#1}\def\@tempb {#2}\def\@tempc
  {#3}\ifx \@tempc \@empty \let \@tempc \@tempb \let \@tempb \@tempa \fi \ifx
  \@tempb \@empty \def\@tempb {arXiv}\fi \@ifundefined
  {mn@eprint@\@tempb}{\@tempb:\@tempc}{\expandafter \expandafter \csname
  mn@eprint@\@tempb\endcsname \expandafter{\@tempc}}}

\bibitem[\protect\citeauthoryear{{Anderson} et~al.,}{{Anderson}
  et~al.}{2014a}]{Anderson_etal2014_blueshifted_emission}
{Anderson} J.~P.,  et~al., 2014a, \mn@doi [\mnras] {10.1093/mnras/stu610},
  \href {http://adsabs.harvard.edu/abs/2014MNRAS.441..671A} {441, 671}

\bibitem[\protect\citeauthoryear{{Anderson} et~al.,}{{Anderson}
  et~al.}{2014b}]{Anderson_etal2014_V_LC}
{Anderson} J.~P.,  et~al., 2014b, \mn@doi [\apj] {10.1088/0004-637X/786/1/67},
  \href {http://adsabs.harvard.edu/abs/2014ApJ...786...67A} {786, 67}

\bibitem[\protect\citeauthoryear{{Anderson} et~al.,}{{Anderson}
  et~al.}{2016}]{Anderson_etal2016}
{Anderson} J.~P.,  et~al., 2016, \mn@doi [\aap] {10.1051/0004-6361/201527691},
  \href {http://adsabs.harvard.edu/abs/2016A%26A...589A.110A} {589, A110}

\bibitem[\protect\citeauthoryear{{Andrews} et~al.,}{{Andrews}
  et~al.}{2011}]{Andrews_etal2011_SN2007it}
{Andrews} J.~E.,  et~al., 2011, \mn@doi [\apj] {10.1088/0004-637X/731/1/47},
  \href {https://ui.adsabs.harvard.edu/abs/2011ApJ...731...47A} {731, 47}

\bibitem[\protect\citeauthoryear{{Balinskaia}, {Bychkov}  \&
  {Neizvestnyi}}{{Balinskaia} et~al.}{1980}]{Balinskaia_etal1980_SN1979C}
{Balinskaia} I.~S.,  {Bychkov} K.~V.,   {Neizvestnyi} S.~I.,  1980, \aap, \href
  {https://ui.adsabs.harvard.edu/abs/1980A&A....85L..19B} {85, L19}

\bibitem[\protect\citeauthoryear{{Baltay} et~al.,}{{Baltay}
  et~al.}{2013}]{Baltay_etal2013_LSQ}
{Baltay} C.,  et~al., 2013, \mn@doi [\pasp] {10.1086/671198}, \href
  {http://adsabs.harvard.edu/abs/2013PASP..125..683B} {125, 683}

\bibitem[\protect\citeauthoryear{{Barbon}, {Ciatti}, {Rosino}, {Ortolani}  \&
  {Rafanelli}}{{Barbon} et~al.}{1982}]{Barbon_etal1982}
{Barbon} R.,  {Ciatti} F.,  {Rosino} L.,  {Ortolani} S.,   {Rafanelli} P.,
  1982, \aap, \href {https://ui.adsabs.harvard.edu/abs/1982A&A...116...43B}
  {116, 43}

\bibitem[\protect\citeauthoryear{{Benetti} et~al.,}{{Benetti}
  et~al.}{2016}]{Benetti_etal2016_SN1996al}
{Benetti} S.,  et~al., 2016, \mn@doi [\mnras] {10.1093/mnras/stv2811}, \href
  {https://ui.adsabs.harvard.edu/abs/2016MNRAS.456.3296B} {456, 3296}

\bibitem[\protect\citeauthoryear{{Blinnikov}, {Eastman}, {Bartunov},
  {Popolitov}  \& {Woosley}}{{Blinnikov} et~al.}{1998}]{1998ApJ...496..454B}
{Blinnikov} S.~I.,  {Eastman} R.,  {Bartunov} O.~S.,  {Popolitov} V.~A.,
  {Woosley} S.~E.,  1998, \mn@doi [\apj] {10.1086/305375}, \href
  {https://ui.adsabs.harvard.edu/abs/1998ApJ...496..454B} {496, 454}

\bibitem[\protect\citeauthoryear{{Blinnikov}, {Lundqvist}, {Bartunov}, {Nomoto}
   \& {Iwamoto}}{{Blinnikov} et~al.}{2000}]{2000ApJ...532.1132B}
{Blinnikov} S.,  {Lundqvist} P.,  {Bartunov} O.,  {Nomoto} K.,   {Iwamoto} K.,
  2000, \mn@doi [\apj] {10.1086/308588}, \href
  {https://ui.adsabs.harvard.edu/abs/2000ApJ...532.1132B} {532, 1132}

\bibitem[\protect\citeauthoryear{{Blinnikov}, {R{\"o}pke}, {Sorokina},
  {Gieseler}, {Reinecke}, {Travaglio}, {Hillebrand t}  \&
  {Stritzinger}}{{Blinnikov} et~al.}{2006}]{2006A&A...453..229B}
{Blinnikov} S.~I.,  {R{\"o}pke} F.~K.,  {Sorokina} E.~I.,  {Gieseler} M.,
  {Reinecke} M.,  {Travaglio} C.,  {Hillebrand t} W.,   {Stritzinger} M.,
  2006, \mn@doi [\aap] {10.1051/0004-6361:20054594}, \href
  {https://ui.adsabs.harvard.edu/abs/2006A&A...453..229B} {453, 229}

\bibitem[\protect\citeauthoryear{{Borish}, {Huang}, {Chevalier}, {Breslauer},
  {Kingery}  \& {Privon}}{{Borish} et~al.}{2015}]{Borish_etal2015_SN2010jl}
{Borish} H.~J.,  {Huang} C.,  {Chevalier} R.~A.,  {Breslauer} B.~M.,  {Kingery}
  A.~M.,   {Privon} G.~C.,  2015, \mn@doi [\apj] {10.1088/0004-637X/801/1/7},
  \href {https://ui.adsabs.harvard.edu/abs/2015ApJ...801....7B} {801, 7}

\bibitem[\protect\citeauthoryear{{Bose} et~al.,}{{Bose}
  et~al.}{2013}]{Bose_etal2013}
{Bose} S.,  et~al., 2013, \mn@doi [\mnras] {10.1093/mnras/stt864}, \href
  {http://adsabs.harvard.edu/abs/2013MNRAS.433.1871B} {433, 1871}

\bibitem[\protect\citeauthoryear{{Bose} et~al.,}{{Bose}
  et~al.}{2015}]{Bose_etal2015_SN2013ej}
{Bose} S.,  et~al., 2015, \mn@doi [\apj] {10.1088/0004-637X/806/2/160}, \href
  {https://ui.adsabs.harvard.edu/abs/2015ApJ...806..160B} {806, 160}

\bibitem[\protect\citeauthoryear{{Bose}, {Kumar}, {Misra}, {Matsumoto},
  {Kumar}, {Singh}, {Fukushima}  \& {Kawabata}}{{Bose}
  et~al.}{2016}]{Bose_etal2016}
{Bose} S.,  {Kumar} B.,  {Misra} K.,  {Matsumoto} K.,  {Kumar} B.,  {Singh} M.,
   {Fukushima} D.,   {Kawabata} M.,  2016, \mn@doi [\mnras]
  {10.1093/mnras/stv2351}, \href
  {https://ui.adsabs.harvard.edu/#abs/2016MNRAS.455.2712B} {455, 2712}

\bibitem[\protect\citeauthoryear{{Bostroem} et~al.,}{{Bostroem}
  et~al.}{2019}]{Bostroem_etal2019_ASAS15oz}
{Bostroem} K.~A.,  et~al., 2019, \mn@doi [\mnras] {10.1093/mnras/stz570}, \href
  {https://ui.adsabs.harvard.edu/abs/2019MNRAS.tmp..562B} {p.~562}

\bibitem[\protect\citeauthoryear{{Branch}, {Falk}, {McCall}, {Rybski}, {Uomoto}
   \& {Wills}}{{Branch} et~al.}{1981}]{Branch_etal1981_SN1979C}
{Branch} D.,  {Falk} S.~W.,  {McCall} M.~L.,  {Rybski} P.,  {Uomoto} A.~K.,
  {Wills} B.~J.,  1981, \mn@doi [\apj] {10.1086/158755}, \href
  {https://ui.adsabs.harvard.edu/\#abs/1981ApJ...244..780B} {244, 780}

\bibitem[\protect\citeauthoryear{{Chandra} \& {Soderberg}}{{Chandra} \&
  {Soderberg}}{2008}]{Chandra_Soderberg2008}
{Chandra} P.,  {Soderberg} A.,  2008, The Astronomer's Telegram, \href
  {https://ui.adsabs.harvard.edu/abs/2008ATel.1869....1C} {1869, 1}

\bibitem[\protect\citeauthoryear{{Chandra} \& {Soderberg}}{{Chandra} \&
  {Soderberg}}{2009}]{Chandra_Soderberg2009}
{Chandra} P.,  {Soderberg} A.,  2009, The Astronomer's Telegram, \href
  {https://ui.adsabs.harvard.edu/abs/2009ATel.2351....1C} {2351, 1}

\bibitem[\protect\citeauthoryear{{Chandra}, {Chevalier}, {Chugai}, {Fransson}
  \& {Soderberg}}{{Chandra} et~al.}{2015}]{Chandra_etal2015_SN2010jl}
{Chandra} P.,  {Chevalier} R.~A.,  {Chugai} N.,  {Fransson} C.,   {Soderberg}
  A.~M.,  2015, \mn@doi [\apj] {10.1088/0004-637X/810/1/32}, \href
  {https://ui.adsabs.harvard.edu/abs/2015ApJ...810...32C} {810, 32}

\bibitem[\protect\citeauthoryear{{Chugai}}{{Chugai}}{2001}]{Chugai2001_SN1998_electron_scattering}
{Chugai} N.~N.,  2001, \mn@doi [\mnras] {10.1111/j.1365-2966.2001.04717.x},
  \href {https://ui.adsabs.harvard.edu/abs/2001MNRAS.326.1448C} {326, 1448}

\bibitem[\protect\citeauthoryear{{Ciatti}, {Rosino}  \& {Bertola}}{{Ciatti}
  et~al.}{1971}]{Ciatti_etal1971_SN1969L}
{Ciatti} F.,  {Rosino} L.,   {Bertola} F.,  1971, \memsai, \href
  {https://ui.adsabs.harvard.edu/abs/1971MmSAI..42..163C} {42, 163}

\bibitem[\protect\citeauthoryear{{Clocchiatti} et~al.,}{{Clocchiatti}
  et~al.}{1996}]{Clocchiatti_etal1996_SN1992H}
{Clocchiatti} A.,  et~al., 1996, \mn@doi [\aj] {10.1086/117874}, \href
  {https://ui.adsabs.harvard.edu/abs/1996AJ....111.1286C} {111, 1286}

\bibitem[\protect\citeauthoryear{{Contreras} et~al.,}{{Contreras}
  et~al.}{2010}]{Contreras_etal2010}
{Contreras} C.,  et~al., 2010, \mn@doi [\aj] {10.1088/0004-6256/139/2/519},
  \href {http://adsabs.harvard.edu/abs/2010AJ....139..519C} {139, 519}

\bibitem[\protect\citeauthoryear{{Crook}, {Huchra}, {Martimbeau}, {Masters},
  {Jarrett}  \& {Macri}}{{Crook} et~al.}{2007}]{Crook_etal2007}
{Crook} A.~C.,  {Huchra} J.~P.,  {Martimbeau} N.,  {Masters} K.~L.,  {Jarrett}
  T.,   {Macri} L.~M.,  2007, \mn@doi [\apj] {10.1086/510201}, \href
  {https://ui.adsabs.harvard.edu/#abs/2007ApJ...655..790C} {655, 790}

\bibitem[\protect\citeauthoryear{{Dall'Ora} et~al.,}{{Dall'Ora}
  et~al.}{2014}]{DallOra_etal2014}
{Dall'Ora} M.,  et~al., 2014, \mn@doi [\apj] {10.1088/0004-637X/787/2/139},
  \href {http://adsabs.harvard.edu/abs/2014ApJ...787..139D} {787, 139}

\bibitem[\protect\citeauthoryear{{Das} \& {Ray}}{{Das} \&
  {Ray}}{2017}]{Das_Ray2017}
{Das} S.,  {Ray} A.,  2017, \mn@doi [\apj] {10.3847/1538-4357/aa97e1}, \href
  {https://ui.adsabs.harvard.edu/abs/2017ApJ...851..138D} {851, 138}

\bibitem[\protect\citeauthoryear{{Dessart} et~al.,}{{Dessart}
  et~al.}{2008}]{Dessart_etal2008}
{Dessart} L.,  et~al., 2008, \mn@doi [\apj] {10.1086/526451}, \href
  {http://adsabs.harvard.edu/abs/2008ApJ...675..644D} {675, 644}

\bibitem[\protect\citeauthoryear{{Dessart} et~al.,}{{Dessart}
  et~al.}{2014}]{Dessart_etal2014}
{Dessart} L.,  et~al., 2014, \mn@doi [\mnras] {10.1093/mnras/stu417}, \href
  {http://adsabs.harvard.edu/abs/2014MNRAS.440.1856D} {440, 1856}

\bibitem[\protect\citeauthoryear{{Drake}, {Djorgovski}, {Williams}, {Mahabal},
  {Graham}, {Beshore}, {Larson}  \& {Christensen}}{{Drake}
  et~al.}{2008}]{Drake_etal2008_SN2008bm_discovery}
{Drake} A.~J.,  {Djorgovski} S.~G.,  {Williams} R.,  {Mahabal} A.,  {Graham}
  M.~J.,  {Beshore} E.~C.,  {Larson} S.~M.,   {Christensen} E.,  2008, The
  Astronomer's Telegram, \href
  {https://ui.adsabs.harvard.edu/abs/2008ATel.1447....1D} {1447, 1}

\bibitem[\protect\citeauthoryear{{Drake} et~al.,}{{Drake}
  et~al.}{2009}]{Drake_etal2009_CRTS}
{Drake} A.~J.,  et~al., 2009, \mn@doi [\apj] {10.1088/0004-637X/696/1/870},
  \href {https://ui.adsabs.harvard.edu/abs/2009ApJ...696..870D} {696, 870}

\bibitem[\protect\citeauthoryear{{Elias-Rosa} et~al.,}{{Elias-Rosa}
  et~al.}{2016}]{EliasRosa_etal2016_SN2015bh}
{Elias-Rosa} N.,  et~al., 2016, \mn@doi [\mnras] {10.1093/mnras/stw2253}, \href
  {https://ui.adsabs.harvard.edu/abs/2016MNRAS.463.3894E} {463, 3894}

\bibitem[\protect\citeauthoryear{{Elmhamdi} et~al.,}{{Elmhamdi}
  et~al.}{2003}]{Elmhamdi_etal2003}
{Elmhamdi} A.,  et~al., 2003, \mn@doi [\mnras]
  {10.1046/j.1365-8711.2003.06150.x}, \href
  {http://adsabs.harvard.edu/abs/2003MNRAS.338..939E} {338, 939}

\bibitem[\protect\citeauthoryear{{Faran} et~al.,}{{Faran}
  et~al.}{2014}]{Faran_etal2014_IIL}
{Faran} T.,  et~al., 2014, \mn@doi [\mnras] {10.1093/mnras/stu1760}, \href
  {http://adsabs.harvard.edu/abs/2014MNRAS.445..554F} {445, 554}

\bibitem[\protect\citeauthoryear{{Fassia} et~al.,}{{Fassia}
  et~al.}{2000}]{Fassia_etal2000_SN1998S}
{Fassia} A.,  et~al., 2000, \mn@doi [\mnras]
  {10.1046/j.1365-8711.2000.03797.x}, \href
  {https://ui.adsabs.harvard.edu/\#abs/2000MNRAS.318.1093F} {318, 1093}

\bibitem[\protect\citeauthoryear{{Fassia} et~al.,}{{Fassia}
  et~al.}{2001}]{Fassia_etal2001_SN1998S}
{Fassia} A.,  et~al., 2001, \mn@doi [\mnras]
  {10.1046/j.1365-8711.2001.04282.x}, \href
  {https://ui.adsabs.harvard.edu/\#abs/2001MNRAS.325..907F} {325, 907}

\bibitem[\protect\citeauthoryear{{Filippenko}}{{Filippenko}}{1988}]{Filippenko1988}
{Filippenko} A.~V.,  1988, \mn@doi [\aj] {10.1086/114940}, \href
  {https://ui.adsabs.harvard.edu/\#abs/1988AJ.....96.1941F} {96, 1941}

\bibitem[\protect\citeauthoryear{{Fitzpatrick}}{{Fitzpatrick}}{1999}]{Fitzpatrick1999}
{Fitzpatrick} E.~L.,  1999, \mn@doi [\pasp] {10.1086/316293}, \href
  {http://adsabs.harvard.edu/abs/1999PASP..111...63F} {111, 63}

\bibitem[\protect\citeauthoryear{{Fixsen}, {Cheng}, {Gales}, {Mather}, {Shafer}
   \& {Wright}}{{Fixsen} et~al.}{1996}]{Fixsen_etal1996}
{Fixsen} D.~J.,  {Cheng} E.~S.,  {Gales} J.~M.,  {Mather} J.~C.,  {Shafer}
  R.~A.,   {Wright} E.~L.,  1996, \mn@doi [\apj] {10.1086/178173}, \href
  {http://adsabs.harvard.edu/abs/1996ApJ...473..576F} {473, 576}

\bibitem[\protect\citeauthoryear{{F{\"o}rster} et~al.,}{{F{\"o}rster}
  et~al.}{2018}]{Forster_etal2018}
{F{\"o}rster} F.,  et~al., 2018, \mn@doi [Nature Astronomy]
  {10.1038/s41550-018-0563-4}, \href
  {https://ui.adsabs.harvard.edu/abs/2018NatAs...2..808F} {p.~122}

\bibitem[\protect\citeauthoryear{{Fransson} et~al.,}{{Fransson}
  et~al.}{2014}]{Fransson_etal2014_SN2010jl}
{Fransson} C.,  et~al., 2014, \mn@doi [\apj] {10.1088/0004-637X/797/2/118},
  \href {https://ui.adsabs.harvard.edu/abs/2014ApJ...797..118F} {797, 118}

\bibitem[\protect\citeauthoryear{{Galbany} et~al.,}{{Galbany}
  et~al.}{2016}]{Galbany_etal2016}
{Galbany} L.,  et~al., 2016, \mn@doi [\aj] {10.3847/0004-6256/151/2/33}, \href
  {http://adsabs.harvard.edu/abs/2016AJ....151...33G} {151, 33}

\bibitem[\protect\citeauthoryear{{Guti{\'e}rrez} et~al.,}{{Guti{\'e}rrez}
  et~al.}{2017a}]{Gutierrez_etal2017_I}
{Guti{\'e}rrez} C.~P.,  et~al., 2017a, \mn@doi [\apj]
  {10.3847/1538-4357/aa8f52}, \href
  {http://adsabs.harvard.edu/abs/2017ApJ...850...89G} {850, 89}

\bibitem[\protect\citeauthoryear{{Guti{\'e}rrez} et~al.,}{{Guti{\'e}rrez}
  et~al.}{2017b}]{Gutierrez_etal2017_II}
{Guti{\'e}rrez} C.~P.,  et~al., 2017b, \mn@doi [\apj]
  {10.3847/1538-4357/aa8f42}, \href
  {https://ui.adsabs.harvard.edu/abs/2017ApJ...850...90G} {850, 90}

\bibitem[\protect\citeauthoryear{{Hamuy}}{{Hamuy}}{2001}]{Hamuy2001}
{Hamuy} M.~A.,  2001, PhD thesis, The University of Arizona

\bibitem[\protect\citeauthoryear{{Hamuy}}{{Hamuy}}{2003}]{Hamuy2003}
{Hamuy} M.,  2003, \mn@doi [\apj] {10.1086/344689}, \href
  {https://ui.adsabs.harvard.edu/#abs/2003ApJ...582..905H} {582, 905}

\bibitem[\protect\citeauthoryear{{Hamuy} \& {Pinto}}{{Hamuy} \&
  {Pinto}}{2002}]{Hamuy_Pinto2002}
{Hamuy} M.,  {Pinto} P.~A.,  2002, \mn@doi [\apjl] {10.1086/339676}, \href
  {http://adsabs.harvard.edu/abs/2002ApJ...566L..63H} {566, L63}

\bibitem[\protect\citeauthoryear{{Hamuy}, {Suntzeff}, {Gonzalez}  \&
  {Martin}}{{Hamuy} et~al.}{1988}]{Hamuy_etal1988_SN1987A}
{Hamuy} M.,  {Suntzeff} N.~B.,  {Gonzalez} R.,   {Martin} G.,  1988, \mn@doi
  [\aj] {10.1086/114613}, \href
  {https://ui.adsabs.harvard.edu/\#abs/1988AJ.....95...63H} {95, 63}

\bibitem[\protect\citeauthoryear{{Hamuy} et~al.,}{{Hamuy}
  et~al.}{2001}]{Hamuy_etal2001}
{Hamuy} M.,  et~al., 2001, \mn@doi [\apj] {10.1086/322450}, \href
  {http://adsabs.harvard.edu/abs/2001ApJ...558..615H} {558, 615}

\bibitem[\protect\citeauthoryear{{Hamuy} et~al.,}{{Hamuy}
  et~al.}{2006}]{Hamuy_etal2006}
{Hamuy} M.,  et~al., 2006, \mn@doi [\pasp] {10.1086/500228}, \href
  {http://adsabs.harvard.edu/abs/2006PASP..118....2H} {118, 2}

\bibitem[\protect\citeauthoryear{{Hicken} et~al.,}{{Hicken}
  et~al.}{2017}]{Hicken_etal2017}
{Hicken} M.,  et~al., 2017, \mn@doi [\apjs] {10.3847/1538-4365/aa8ef4}, \href
  {http://adsabs.harvard.edu/abs/2017ApJS..233....6H} {233, 6}

\bibitem[\protect\citeauthoryear{{Hillier} \& {Dessart}}{{Hillier} \&
  {Dessart}}{2019}]{Hillier_Dessart2019}
{Hillier} D.~J.,  {Dessart} L.,  2019, \mn@doi [\aap]
  {10.1051/0004-6361/201935100}, \href
  {https://ui.adsabs.harvard.edu/abs/2019A&A...631A...8H} {631, A8}

\bibitem[\protect\citeauthoryear{{Ho}, {Van Dyk}, {Peng}, {Filippenko},
  {Leonard}, {Matheson}, {Treffers}  \& {Richmond}}{{Ho}
  et~al.}{2001}]{Ho_etal2001_SN1994Y}
{Ho} W. C.~G.,  {Van Dyk} S.~D.,  {Peng} C.~Y.,  {Filippenko} A.~V.,  {Leonard}
  D.~C.,  {Matheson} T.,  {Treffers} R.~R.,   {Richmond} M.~W.,  2001, \mn@doi
  [\pasp] {10.1086/323970}, \href
  {https://ui.adsabs.harvard.edu/abs/2001PASP..113.1349H} {113, 1349}

\bibitem[\protect\citeauthoryear{{Huang} et~al.,}{{Huang}
  et~al.}{2015}]{Huang_etal2015_SN2013ej}
{Huang} F.,  et~al., 2015, \mn@doi [\apj] {10.1088/0004-637X/807/1/59}, \href
  {https://ui.adsabs.harvard.edu/abs/2015ApJ...807...59H} {807, 59}

\bibitem[\protect\citeauthoryear{{Huang} et~al.,}{{Huang}
  et~al.}{2016}]{Huang_etal2016_SN2014cx}
{Huang} F.,  et~al., 2016, \mn@doi [\apj] {10.3847/0004-637X/832/2/139}, \href
  {https://ui.adsabs.harvard.edu/abs/2016ApJ...832..139H} {832, 139}

\bibitem[\protect\citeauthoryear{{Humphreys}, {Davidson}, {Jones}, {Pogge},
  {Grammer}, {Prieto}  \& {Pritchard}}{{Humphreys}
  et~al.}{2012}]{Humphreys_etal2012_SN2011ht}
{Humphreys} R.~M.,  {Davidson} K.,  {Jones} T.~J.,  {Pogge} R.~W.,  {Grammer}
  S.~H.,  {Prieto} J.~L.,   {Pritchard} T.~A.,  2012, \mn@doi [\apj]
  {10.1088/0004-637X/760/1/93}, \href
  {https://ui.adsabs.harvard.edu/abs/2012ApJ...760...93H} {760, 93}

\bibitem[\protect\citeauthoryear{{Inserra} et~al.,}{{Inserra}
  et~al.}{2012}]{Inserra_etal2012_SN2009bw}
{Inserra} C.,  et~al., 2012, \mn@doi [\mnras]
  {10.1111/j.1365-2966.2012.20685.x}, \href
  {https://ui.adsabs.harvard.edu/abs/2012MNRAS.422.1122I} {422, 1122}

\bibitem[\protect\citeauthoryear{{Inserra} et~al.,}{{Inserra}
  et~al.}{2013}]{Inserra_etal2013}
{Inserra} C.,  et~al., 2013, \mn@doi [\aap] {10.1051/0004-6361/201220496},
  \href {http://adsabs.harvard.edu/abs/2013A%26A...555A.142I} {555, A142}

\bibitem[\protect\citeauthoryear{{Jang} \& {Lee}}{{Jang} \&
  {Lee}}{2017}]{Jang_Lee2017_V}
{Jang} I.~S.,  {Lee} M.~G.,  2017, \mn@doi [\apj] {10.3847/1538-4357/836/1/74},
  \href {https://ui.adsabs.harvard.edu/#abs/2017ApJ...836...74J} {836, 74}

\bibitem[\protect\citeauthoryear{{Kangas} et~al.,}{{Kangas}
  et~al.}{2016}]{Kangas_etal2016_SN2013fc}
{Kangas} T.,  et~al., 2016, \mn@doi [\mnras] {10.1093/mnras/stv2567}, \href
  {https://ui.adsabs.harvard.edu/abs/2016MNRAS.456..323K} {456, 323}

\bibitem[\protect\citeauthoryear{{Khazov} et~al.,}{{Khazov}
  et~al.}{2016}]{Khazov_etal2016}
{Khazov} D.,  et~al., 2016, \mn@doi [\apj] {10.3847/0004-637X/818/1/3}, \href
  {https://ui.adsabs.harvard.edu/abs/2016ApJ...818....3K} {818, 3}

\bibitem[\protect\citeauthoryear{{Krisciunas} et~al.,}{{Krisciunas}
  et~al.}{2017}]{Krisciunas_etal2017_CSP_3DR}
{Krisciunas} K.,  et~al., 2017, \mn@doi [\aj] {10.3847/1538-3881/aa8df0}, \href
  {https://ui.adsabs.harvard.edu/abs/2017AJ....154..211K} {154, 211}

\bibitem[\protect\citeauthoryear{{Kuncarayakti} et~al.,}{{Kuncarayakti}
  et~al.}{2018}]{Kuncarayakti_etal2018}
{Kuncarayakti} H.,  et~al., 2018, \mn@doi [\aap] {10.1051/0004-6361/201731923},
  \href {https://ui.adsabs.harvard.edu/abs/2018A&A...613A..35K} {613, A35}

\bibitem[\protect\citeauthoryear{{Landolt}}{{Landolt}}{1992}]{Landolt1992_UBVRI}
{Landolt} A.~U.,  1992, \mn@doi [\aj] {10.1086/116242}, \href
  {https://ui.adsabs.harvard.edu/\#abs/1992AJ....104..340L} {104, 340}

\bibitem[\protect\citeauthoryear{{Leloudas} et~al.,}{{Leloudas}
  et~al.}{2015}]{Leloudas_etal2015}
{Leloudas} G.,  et~al., 2015, \mn@doi [\aap] {10.1051/0004-6361/201322035},
  \href {https://ui.adsabs.harvard.edu/\#abs/2015A&A...574A..61L} {574, A61}

\bibitem[\protect\citeauthoryear{{Leonard} et~al.,}{{Leonard}
  et~al.}{2002}]{Leonard_etal2002_99em}
{Leonard} D.~C.,  et~al., 2002, \mn@doi [\pasp] {10.1086/324785}, \href
  {http://adsabs.harvard.edu/abs/2002PASP..114...35L} {114, 35}

\bibitem[\protect\citeauthoryear{{Lisakov}, {Dessart}, {Hillier}, {Waldman}  \&
  {Livne}}{{Lisakov} et~al.}{2018}]{Lisakov_etal2018_LLSNeII}
{Lisakov} S.~M.,  {Dessart} L.,  {Hillier} D.~J.,  {Waldman} R.,   {Livne} E.,
  2018, \mn@doi [\mnras] {10.1093/mnras/stx2521}, \href
  {https://ui.adsabs.harvard.edu/abs/2018MNRAS.473.3863L} {473, 3863}

\bibitem[\protect\citeauthoryear{{Maguire} et~al.,}{{Maguire}
  et~al.}{2010}]{Maguire_etal2010_04et}
{Maguire} K.,  et~al., 2010, \mn@doi [\mnras]
  {10.1111/j.1365-2966.2010.16332.x}, \href
  {http://adsabs.harvard.edu/abs/2010MNRAS.404..981M} {404, 981}

\bibitem[\protect\citeauthoryear{{Marino} et~al.,}{{Marino}
  et~al.}{2013}]{Marino_etal2013}
{Marino} R.~A.,  et~al., 2013, \mn@doi [\aap] {10.1051/0004-6361/201321956},
  \href {https://ui.adsabs.harvard.edu/abs/2013A&A...559A.114M} {559, A114}

\bibitem[\protect\citeauthoryear{{Maza}, {Wischnjewsky}  \& {Gonzalez}}{{Maza}
  et~al.}{1983}]{Maza_etal1983_SN1983K}
{Maza} J.,  {Wischnjewsky} M.,   {Gonzalez} L.~E.,  1983, IAUC, \href
  {https://ui.adsabs.harvard.edu/\#abs/1983IAUC.3827....2M} {3827, 2}

\bibitem[\protect\citeauthoryear{{Meza} et~al.,}{{Meza}
  et~al.}{2019}]{Meza_etal2019_ASAS14jb}
{Meza} N.,  et~al., 2019, \mn@doi [\aap] {10.1051/0004-6361/201834972}, \href
  {https://ui.adsabs.harvard.edu/abs/2019A&A...629A..57M} {629, A57}

\bibitem[\protect\citeauthoryear{{Minkowski}}{{Minkowski}}{1941}]{Minkowski1941}
{Minkowski} R.,  1941, \mn@doi [\pasp] {10.1086/125315}, \href
  {https://ui.adsabs.harvard.edu/\#abs/1941PASP...53..224M} {53, 224}

\bibitem[\protect\citeauthoryear{{Moriya}, {Tominaga}, {Blinnikov}, {Baklanov}
  \& {Sorokina}}{{Moriya} et~al.}{2011}]{Moriya_etal2011}
{Moriya} T.,  {Tominaga} N.,  {Blinnikov} S.~I.,  {Baklanov} P.~V.,
  {Sorokina} E.~I.,  2011, \mn@doi [\mnras] {10.1111/j.1365-2966.2011.18689.x},
  \href {https://ui.adsabs.harvard.edu/abs/2011MNRAS.415..199M} {415, 199}

\bibitem[\protect\citeauthoryear{{Moriya}, {Yoon}, {Gr{\"a}fener}  \&
  {Blinnikov}}{{Moriya} et~al.}{2017}]{2017MNRAS.469L.108M}
{Moriya} T.~J.,  {Yoon} S.-C.,  {Gr{\"a}fener} G.,   {Blinnikov} S.~I.,  2017,
  \mn@doi [\mnras] {10.1093/mnrasl/slx056}, \href
  {https://ui.adsabs.harvard.edu/abs/2017MNRAS.469L.108M} {469, L108}

\bibitem[\protect\citeauthoryear{{Moriya}, {F{\"o}rster}, {Yoon},
  {Gr{\"a}fener}  \& {Blinnikov}}{{Moriya} et~al.}{2018}]{2018MNRAS.476.2840M}
{Moriya} T.~J.,  {F{\"o}rster} F.,  {Yoon} S.-C.,  {Gr{\"a}fener} G.,
  {Blinnikov} S.~I.,  2018, \mn@doi [\mnras] {10.1093/mnras/sty475}, \href
  {https://ui.adsabs.harvard.edu/abs/2018MNRAS.476.2840M} {476, 2840}

\bibitem[\protect\citeauthoryear{{Morozova}, {Piro}  \& {Valenti}}{{Morozova}
  et~al.}{2017}]{Morozova_etal2017}
{Morozova} V.,  {Piro} A.~L.,   {Valenti} S.,  2017, \mn@doi [\apj]
  {10.3847/1538-4357/aa6251}, \href
  {https://ui.adsabs.harvard.edu/abs/2017ApJ...838...28M} {838, 28}

\bibitem[\protect\citeauthoryear{{Morozova}, {Piro}  \& {Valenti}}{{Morozova}
  et~al.}{2018}]{Morozova_etal2018}
{Morozova} V.,  {Piro} A.~L.,   {Valenti} S.,  2018, \mn@doi [\apj]
  {10.3847/1538-4357/aab9a6}, \href
  {https://ui.adsabs.harvard.edu/abs/2018ApJ...858...15M} {858, 15}

\bibitem[\protect\citeauthoryear{{Mould} et~al.,}{{Mould}
  et~al.}{2000}]{Mould_etal2000}
{Mould} J.~R.,  et~al., 2000, \mn@doi [\apj] {10.1086/308304}, \href
  {http://adsabs.harvard.edu/abs/2000ApJ...529..786M} {529, 786}

\bibitem[\protect\citeauthoryear{{M{\"u}ller}, {Prieto}, {Pejcha}  \&
  {Clocchiatti}}{{M{\"u}ller} et~al.}{2017}]{Muller_etal2017_SNII_nickel_mass}
{M{\"u}ller} T.,  {Prieto} J.~L.,  {Pejcha} O.,   {Clocchiatti} A.,  2017,
  \mn@doi [\apj] {10.3847/1538-4357/aa72f1}, \href
  {https://ui.adsabs.harvard.edu/abs/2017ApJ...841..127M} {841, 127}

\bibitem[\protect\citeauthoryear{{Niemela}, {Ruiz}  \& {Phillips}}{{Niemela}
  et~al.}{1985}]{Niemela_etal1985_SN1983K}
{Niemela} V.~S.,  {Ruiz} M.~T.,   {Phillips} M.~M.,  1985, \mn@doi [\apj]
  {10.1086/162863}, \href {http://adsabs.harvard.edu/abs/1985ApJ...289...52N}
  {289, 52}

\bibitem[\protect\citeauthoryear{{Nugent} et~al.,}{{Nugent}
  et~al.}{2006}]{Nugent_etal2006}
{Nugent} P.,  et~al., 2006, \mn@doi [\apj] {10.1086/504413}, \href
  {http://adsabs.harvard.edu/abs/2006ApJ...645..841N} {645, 841}

\bibitem[\protect\citeauthoryear{{Olivares E.} et~al.,}{{Olivares E.}
  et~al.}{2010}]{Olivares_etal2010}
{Olivares E.} F.,  et~al., 2010, \mn@doi [\apj] {10.1088/0004-637X/715/2/833},
  \href {http://adsabs.harvard.edu/abs/2010ApJ...715..833O} {715, 833}

\bibitem[\protect\citeauthoryear{{Pastorello} et~al.,}{{Pastorello}
  et~al.}{2006}]{Pastorello_etal2006}
{Pastorello} A.,  et~al., 2006, \mn@doi [\mnras]
  {10.1111/j.1365-2966.2006.10587.x}, \href
  {http://adsabs.harvard.edu/abs/2006MNRAS.370.1752P} {370, 1752}

\bibitem[\protect\citeauthoryear{{Pastorello} et~al.,}{{Pastorello}
  et~al.}{2009}]{Pastorello_etal2009}
{Pastorello} A.,  et~al., 2009, \mn@doi [\mnras]
  {10.1111/j.1365-2966.2009.14505.x}, \href
  {http://adsabs.harvard.edu/abs/2009MNRAS.394.2266P} {394, 2266}

\bibitem[\protect\citeauthoryear{{Pejcha} \& {Prieto}}{{Pejcha} \&
  {Prieto}}{2015}]{Pejcha_Prieto2015b}
{Pejcha} O.,  {Prieto} J.~L.,  2015, \mn@doi [\apj]
  {10.1088/0004-637X/806/2/225}, \href
  {https://ui.adsabs.harvard.edu/abs/2015ApJ...806..225P} {806, 225}

\bibitem[\protect\citeauthoryear{{Penston} \& {Blades}}{{Penston} \&
  {Blades}}{1980}]{Penston_Blades1980_SN1979C_reddening}
{Penston} M.~V.,  {Blades} J.~C.,  1980, \mn@doi [\mnras]
  {10.1093/mnras/190.1.51P}, \href
  {https://ui.adsabs.harvard.edu/abs/1980MNRAS.190P..51P} {190, 51P}

\bibitem[\protect\citeauthoryear{{Persson}, {Murphy}, {Krzeminski}, {Roth}  \&
  {Rieke}}{{Persson} et~al.}{1998}]{Persson_etal1998}
{Persson} S.~E.,  {Murphy} D.~C.,  {Krzeminski} W.,  {Roth} M.,   {Rieke}
  M.~J.,  1998, \mn@doi [\aj] {10.1086/300607}, \href
  {http://adsabs.harvard.edu/abs/1998AJ....116.2475P} {116, 2475}

\bibitem[\protect\citeauthoryear{{Pettini} \& {Pagel}}{{Pettini} \&
  {Pagel}}{2004}]{Pettini_Pagel2004}
{Pettini} M.,  {Pagel} B. E.~J.,  2004, \mn@doi [\mnras]
  {10.1111/j.1365-2966.2004.07591.x}, \href
  {https://ui.adsabs.harvard.edu/abs/2004MNRAS.348L..59P} {348, L59}

\bibitem[\protect\citeauthoryear{{Phillips}, {Hamuy}, {Maza}, {Ruiz}, {Carney}
  \& {Graham}}{{Phillips} et~al.}{1990}]{Phillips_etal1990_SN1983K}
{Phillips} M.~M.,  {Hamuy} M.,  {Maza} J.,  {Ruiz} M.~T.,  {Carney} B.~W.,
  {Graham} J.~A.,  1990, \mn@doi [\pasp] {10.1086/132634}, \href
  {https://ui.adsabs.harvard.edu/\#abs/1990PASP..102..299P} {102, 299}

\bibitem[\protect\citeauthoryear{{Phillips} et~al.,}{{Phillips}
  et~al.}{2013}]{Phillips_etal2013}
{Phillips} M.~M.,  et~al., 2013, \mn@doi [\apj] {10.1088/0004-637X/779/1/38},
  \href {https://ui.adsabs.harvard.edu/abs/2013ApJ...779...38P} {779, 38}

\bibitem[\protect\citeauthoryear{{Pignata} et~al.,}{{Pignata}
  et~al.}{2009a}]{Pignata_etal2009_CHASE}
{Pignata} G.,  et~al., 2009a, in {Giobbi} G.,  {Tornambe} A.,  {Raimondo} G.,
  {Limongi} M.,  {Antonelli} L.~A.,  {Menci} N.,   {Brocato} E.,  eds,
  American Institute of Physics Conference Series Vol. 1111, American Institute
  of Physics Conference Series. pp 551--554 (\mn@eprint {arXiv} {0812.4923}),
  \mn@doi{10.1063/1.3141608}

\bibitem[\protect\citeauthoryear{{Pignata} et~al.,}{{Pignata}
  et~al.}{2009b}]{Pignata_etal2009_SN2009aj_discovery}
{Pignata} G.,  et~al., 2009b, CBET, \href
  {http://adsabs.harvard.edu/abs/2009CBET.1704....1P} {1704}

\bibitem[\protect\citeauthoryear{{Pignata} et~al.,}{{Pignata}
  et~al.}{2009c}]{Pignata_etal2009_SN2009au_discovery}
{Pignata} G.,  et~al., 2009c, CBET, \href
  {http://adsabs.harvard.edu/abs/2009CBET.1719....1P} {1719}

\bibitem[\protect\citeauthoryear{{Polshaw} et~al.,}{{Polshaw}
  et~al.}{2016}]{Polshaw_etal2016}
{Polshaw} J.,  et~al., 2016, \mn@doi [\aap] {10.1051/0004-6361/201527682},
  \href {http://adsabs.harvard.edu/abs/2016A%26A...588A...1P} {588, A1}

\bibitem[\protect\citeauthoryear{{Poon}, {Pun}, {Lam}, {Qiu}  \& {Wei}}{{Poon}
  et~al.}{2011}]{Poon_etal2011_SN1998S}
{Poon} H.,  {Pun} J.~C.~S.,  {Lam} T.~Y.,  {Qiu} Y.~L.,   {Wei} J.~Y.,  2011,
  arXiv e-prints, \href {https://ui.adsabs.harvard.edu/abs/2011arXiv1109.0899P}
  {p. arXiv:1109.0899}

\bibitem[\protect\citeauthoryear{{Poznanski}, {Prochaska}  \&
  {Bloom}}{{Poznanski} et~al.}{2012}]{Poznanski_etal2012_NaID}
{Poznanski} D.,  {Prochaska} J.~X.,   {Bloom} J.~S.,  2012, \mn@doi [\mnras]
  {10.1111/j.1365-2966.2012.21796.x}, \href
  {http://adsabs.harvard.edu/abs/2012MNRAS.426.1465P} {426, 1465}

\bibitem[\protect\citeauthoryear{{Pritchard} et~al.,}{{Pritchard}
  et~al.}{2012}]{Pritchard_etal2012_SN2007pk}
{Pritchard} T.~A.,  et~al., 2012, \mn@doi [\apj] {10.1088/0004-637X/750/2/128},
  \href {https://ui.adsabs.harvard.edu/abs/2012ApJ...750..128P} {750, 128}

\bibitem[\protect\citeauthoryear{{Quimby}, {Wheeler}, {H{\"o}flich}, {Akerlof},
  {Brown}  \& {Rykoff}}{{Quimby} et~al.}{2007}]{Quimby_etal2007}
{Quimby} R.~M.,  {Wheeler} J.~C.,  {H{\"o}flich} P.,  {Akerlof} C.~W.,  {Brown}
  P.~J.,   {Rykoff} E.~S.,  2007, \mn@doi [\apj] {10.1086/520532}, \href
  {https://ui.adsabs.harvard.edu/abs/2007ApJ...666.1093Q} {666, 1093}

\bibitem[\protect\citeauthoryear{{Reichart} et~al.,}{{Reichart}
  et~al.}{2005}]{Reichart_etal2005}
{Reichart} D.,  et~al., 2005, \mn@doi [Nuovo Cimento C Geophysics Space Physics
  C] {10.1393/ncc/i2005-10149-6}, \href
  {https://ui.adsabs.harvard.edu/abs/2005NCimC..28..767R} {28, 767}

\bibitem[\protect\citeauthoryear{{Riess} et~al.,}{{Riess}
  et~al.}{2016}]{Riess_etal2016}
{Riess} A.~G.,  et~al., 2016, \mn@doi [\apj] {10.3847/0004-637X/826/1/56},
  \href {http://adsabs.harvard.edu/abs/2016ApJ...826...56R} {826, 56}

\bibitem[\protect\citeauthoryear{{Rodr{\'{\i}}guez}, {Clocchiatti}  \&
  {Hamuy}}{{Rodr{\'{\i}}guez} et~al.}{2014}]{Rodriguez_etal2014}
{Rodr{\'{\i}}guez} {\'O}.,  {Clocchiatti} A.,   {Hamuy} M.,  2014, \mn@doi
  [\aj] {10.1088/0004-6256/148/6/107}, \href
  {http://adsabs.harvard.edu/abs/2014AJ....148..107R} {148, 107}

\bibitem[\protect\citeauthoryear{{Rodr{\'\i}guez} et~al.,}{{Rodr{\'\i}guez}
  et~al.}{2019}]{Rodriguez_etal2019}
{Rodr{\'\i}guez} {\'O}.,  et~al., 2019, \mn@doi [\mnras]
  {10.1093/mnras/sty3396}, \href
  {https://ui.adsabs.harvard.edu/\#abs/2019MNRAS.483.5459R} {483, 5459}

\bibitem[\protect\citeauthoryear{{Roy} et~al.,}{{Roy}
  et~al.}{2011}]{Roy_etal2011_SN2008gz}
{Roy} R.,  et~al., 2011, \mn@doi [\mnras] {10.1111/j.1365-2966.2011.18363.x},
  \href {https://ui.adsabs.harvard.edu/abs/2011MNRAS.414..167R} {414, 167}

\bibitem[\protect\citeauthoryear{{Saha}, {Thim}, {Tammann}, {Reindl}  \&
  {Sandage}}{{Saha} et~al.}{2006}]{Saha_etal2006}
{Saha} A.,  {Thim} F.,  {Tammann} G.~A.,  {Reindl} B.,   {Sandage} A.,  2006,
  \mn@doi [\apjs] {10.1086/503800}, \href
  {http://adsabs.harvard.edu/abs/2006ApJS..165..108S} {165, 108}

\bibitem[\protect\citeauthoryear{{Sahu}, {Anupama}, {Srividya}  \&
  {Muneer}}{{Sahu} et~al.}{2006}]{Sahu_etal2006}
{Sahu} D.~K.,  {Anupama} G.~C.,  {Srividya} S.,   {Muneer} S.,  2006, \mn@doi
  [\mnras] {10.1111/j.1365-2966.2006.10937.x}, \href
  {http://adsabs.harvard.edu/abs/2006MNRAS.372.1315S} {372, 1315}

\bibitem[\protect\citeauthoryear{{Sanders} et~al.,}{{Sanders}
  et~al.}{2015}]{Sanders_etal2015}
{Sanders} N.~E.,  et~al., 2015, \mn@doi [\apj] {10.1088/0004-637X/799/2/208},
  \href {http://adsabs.harvard.edu/abs/2015ApJ...799..208S} {799, 208}

\bibitem[\protect\citeauthoryear{{Schlafly} \& {Finkbeiner}}{{Schlafly} \&
  {Finkbeiner}}{2011}]{Schlafly_Finkbeiner2011}
{Schlafly} E.~F.,  {Finkbeiner} D.~P.,  2011, \mn@doi [\apj]
  {10.1088/0004-637X/737/2/103}, \href
  {http://adsabs.harvard.edu/abs/2011ApJ...737..103S} {737, 103}

\bibitem[\protect\citeauthoryear{{Schlegel}}{{Schlegel}}{1990}]{Schlegel1990_SNeIIn}
{Schlegel} E.~M.,  1990, \mnras, \href
  {http://adsabs.harvard.edu/abs/1990MNRAS.244..269S} {244, 269}

\bibitem[\protect\citeauthoryear{{Schlegel}, {Finkbeiner}  \&
  {Davis}}{{Schlegel} et~al.}{1998}]{Schlegel_etal1998}
{Schlegel} D.~J.,  {Finkbeiner} D.~P.,   {Davis} M.,  1998, \mn@doi [\apj]
  {10.1086/305772}, \href {http://adsabs.harvard.edu/abs/1998ApJ...500..525S}
  {500, 525}

\bibitem[\protect\citeauthoryear{{Shivvers} et~al.,}{{Shivvers}
  et~al.}{2017}]{Shivvers_etal2017}
{Shivvers} I.,  et~al., 2017, \mn@doi [\pasp] {10.1088/1538-3873/aa54a6}, \href
  {http://adsabs.harvard.edu/abs/2017PASP..129e4201S} {129, 054201}

\bibitem[\protect\citeauthoryear{{Smith} et~al.,}{{Smith}
  et~al.}{2002}]{Smith_etal2002_ugriz}
{Smith} J.~A.,  et~al., 2002, \mn@doi [\aj] {10.1086/339311}, \href
  {https://ui.adsabs.harvard.edu/\#abs/2002AJ....123.2121S} {123, 2121}

\bibitem[\protect\citeauthoryear{{Smith} et~al.,}{{Smith}
  et~al.}{2015}]{Smith_etal2015}
{Smith} N.,  et~al., 2015, \mn@doi [\mnras] {10.1093/mnras/stv354}, \href
  {https://ui.adsabs.harvard.edu/abs/2015MNRAS.449.1876S} {449, 1876}

\bibitem[\protect\citeauthoryear{{Stritzinger} \& {Morrell}}{{Stritzinger} \&
  {Morrell}}{2008}]{Stritzinger_Morrell2008_SN2008bm_classification}
{Stritzinger} M.,  {Morrell} N.,  2008, CBET, \href
  {http://adsabs.harvard.edu/abs/2008CBET.1329....1S} {1329, 1}

\bibitem[\protect\citeauthoryear{{Stritzinger}, {Morrell}, {Folatelli},
  {Covarrubias}  \& {Phillips}}{{Stritzinger}
  et~al.}{2009}]{Stritzinger_etal2009_SN2009aj_SN2009au_classification}
{Stritzinger} M.,  {Morrell} N.,  {Folatelli} G.,  {Covarrubias} R.,
  {Phillips} M.~M.,  2009, CBET, \href
  {http://adsabs.harvard.edu/abs/2009CBET.1725....1S} {1725}

\bibitem[\protect\citeauthoryear{{Szalai} et~al.,}{{Szalai}
  et~al.}{2019}]{Szalai_etal2019_SN2017eaw}
{Szalai} T.,  et~al., 2019, \mn@doi [\apj] {10.3847/1538-4357/ab12d0}, \href
  {https://ui.adsabs.harvard.edu/abs/2019ApJ...876...19S} {876, 19}

\bibitem[\protect\citeauthoryear{{Taddia} et~al.,}{{Taddia}
  et~al.}{2013}]{Taddia_etal2013_CSP_SNeIIn}
{Taddia} F.,  et~al., 2013, \mn@doi [\aap] {10.1051/0004-6361/201321180}, \href
  {http://adsabs.harvard.edu/abs/2013A%26A...555A..10T} {555, A10}

\bibitem[\protect\citeauthoryear{{Taddia} et~al.,}{{Taddia}
  et~al.}{2015}]{Taddia_etal2015}
{Taddia} F.,  et~al., 2015, \mn@doi [\aap] {10.1051/0004-6361/201525989}, \href
  {https://ui.adsabs.harvard.edu/abs/2015A&A...580A.131T} {580, A131}

\bibitem[\protect\citeauthoryear{{Tak{\'a}ts} et~al.,}{{Tak{\'a}ts}
  et~al.}{2015}]{Takats_etal2015}
{Tak{\'a}ts} K.,  et~al., 2015, \mn@doi [\mnras] {10.1093/mnras/stv857}, \href
  {http://adsabs.harvard.edu/abs/2015MNRAS.450.3137T} {450, 3137}

\bibitem[\protect\citeauthoryear{{Tartaglia} et~al.,}{{Tartaglia}
  et~al.}{2020}]{Tartaglia_etal2020_SN2015da}
{Tartaglia} L.,  et~al., 2020, \mn@doi [\aap] {10.1051/0004-6361/201936553},
  \href {https://ui.adsabs.harvard.edu/abs/2020A&A...635A..39T} {635, A39}

\bibitem[\protect\citeauthoryear{{Terreran} et~al.,}{{Terreran}
  et~al.}{2016}]{Terreran_etal2016}
{Terreran} G.,  et~al., 2016, \mn@doi [\mnras] {10.1093/mnras/stw1591}, \href
  {http://adsabs.harvard.edu/abs/2016MNRAS.462..137T} {462, 137}

\bibitem[\protect\citeauthoryear{{Th{\"o}ne} et~al.,}{{Th{\"o}ne}
  et~al.}{2017}]{Thone_etal2017_SN2015bh}
{Th{\"o}ne} C.~C.,  et~al., 2017, \mn@doi [\aap] {10.1051/0004-6361/201629968},
  \href {https://ui.adsabs.harvard.edu/abs/2017A&A...599A.129T} {599, A129}

\bibitem[\protect\citeauthoryear{{Tomasella} et~al.,}{{Tomasella}
  et~al.}{2013}]{Tomasella_etal2013}
{Tomasella} L.,  et~al., 2013, \mn@doi [\mnras] {10.1093/mnras/stt1130}, \href
  {http://adsabs.harvard.edu/abs/2013MNRAS.434.1636T} {434, 1636}

\bibitem[\protect\citeauthoryear{{Tsvetkov} \& {Pavlyuk}}{{Tsvetkov} \&
  {Pavlyuk}}{1997}]{Tsvetkov_Pavlyuk1997_SN1994Y}
{Tsvetkov} D.~Y.,  {Pavlyuk} N.~N.,  1997, Astronomy Letters, \href
  {https://ui.adsabs.harvard.edu/abs/1997AstL...23...26T} {23, 26}

\bibitem[\protect\citeauthoryear{{Tsvetkov}, {Goranskij}  \&
  {Pavlyuk}}{{Tsvetkov} et~al.}{2008}]{Tsvetkov_etal2008}
{Tsvetkov} D.~Y.,  {Goranskij} V.,   {Pavlyuk} N.,  2008, Peremennye Zvezdy,
  \href {http://adsabs.harvard.edu/abs/2008PZ.....28....8T} {28}

\bibitem[\protect\citeauthoryear{{Tsvetkov} et~al.,}{{Tsvetkov}
  et~al.}{2018}]{Tsvetkov_etal2018_SN2017eaw}
{Tsvetkov} D.~Y.,  et~al., 2018, \mn@doi [Astronomy Letters]
  {10.1134/S1063773718050043}, \href
  {https://ui.adsabs.harvard.edu/abs/2018AstL...44..315T} {44, 315}

\bibitem[\protect\citeauthoryear{{Tully}, {Courtois}  \& {Sorce}}{{Tully}
  et~al.}{2016}]{Tully_etal2016}
{Tully} R.~B.,  {Courtois} H.~M.,   {Sorce} J.~G.,  2016, \mn@doi [\aj]
  {10.3847/0004-6256/152/2/50}, \href
  {https://ui.adsabs.harvard.edu/\#abs/2016AJ....152...50T} {152, 50}

\bibitem[\protect\citeauthoryear{{Valenti} et~al.,}{{Valenti}
  et~al.}{2016}]{Valenti_etal2016}
{Valenti} S.,  et~al., 2016, \mn@doi [\mnras] {10.1093/mnras/stw870}, \href
  {http://adsabs.harvard.edu/abs/2016MNRAS.459.3939V} {459, 3939}

\bibitem[\protect\citeauthoryear{{Van Dyk}, {Li}  \& {Filippenko}}{{Van Dyk}
  et~al.}{2003}]{VanDyk_etal2003_SN2003gd}
{Van Dyk} S.~D.,  {Li} W.,   {Filippenko} A.~V.,  2003, \mn@doi [\pasp]
  {10.1086/378308}, \href
  {https://ui.adsabs.harvard.edu/abs/2003PASP..115.1289V} {115, 1289}

\bibitem[\protect\citeauthoryear{{Van Dyk} et~al.,}{{Van Dyk}
  et~al.}{2012}]{VanDyk_etal2012_SN2008bk}
{Van Dyk} S.~D.,  et~al., 2012, \mn@doi [\aj] {10.1088/0004-6256/143/1/19},
  \href {https://ui.adsabs.harvard.edu/abs/2012AJ....143...19V} {143, 19}

\bibitem[\protect\citeauthoryear{{Van Dyk} et~al.,}{{Van Dyk}
  et~al.}{2019}]{VanDyk_etal2019_SN2017eaw}
{Van Dyk} S.~D.,  et~al., 2019, \mn@doi [\apj] {10.3847/1538-4357/ab1136},
  \href {https://ui.adsabs.harvard.edu/abs/2019ApJ...875..136V} {875, 136}

\bibitem[\protect\citeauthoryear{{Vink{\'o}} et~al.,}{{Vink{\'o}}
  et~al.}{2006}]{Vinko_etal2006}
{Vink{\'o}} J.,  et~al., 2006, \mn@doi [\mnras]
  {10.1111/j.1365-2966.2006.10416.x}, \href
  {http://adsabs.harvard.edu/abs/2006MNRAS.369.1780V} {369, 1780}

\bibitem[\protect\citeauthoryear{{Wang}, {Strovink}, {Conley}, {Goldhaber},
  {Kowalski}, {Perlmutter}  \& {Siegrist}}{{Wang} et~al.}{2006}]{Wang_etal2006}
{Wang} L.,  {Strovink} M.,  {Conley} A.,  {Goldhaber} G.,  {Kowalski} M.,
  {Perlmutter} S.,   {Siegrist} J.,  2006, \mn@doi [\apj] {10.1086/500422},
  \href {http://adsabs.harvard.edu/abs/2006ApJ...641...50W} {641, 50}

\bibitem[\protect\citeauthoryear{{Woosley}, {Pinto}, {Martin}  \&
  {Weaver}}{{Woosley} et~al.}{1987}]{Woosley_etal1987}
{Woosley} S.~E.,  {Pinto} P.~A.,  {Martin} P.~G.,   {Weaver} T.~A.,  1987,
  \mn@doi [\apj] {10.1086/165402}, \href
  {https://ui.adsabs.harvard.edu/\#abs/1987ApJ...318..664W} {318, 664}

\bibitem[\protect\citeauthoryear{{Yaron} et~al.,}{{Yaron}
  et~al.}{2017}]{Yaron_etal2017}
{Yaron} O.,  et~al., 2017, \mn@doi [Nature Physics] {10.1038/nphys4025}, \href
  {http://adsabs.harvard.edu/abs/2017NatPh..13..510Y} {13, 510}

\bibitem[\protect\citeauthoryear{{Zhang} et~al.,}{{Zhang}
  et~al.}{2012}]{Zhang_etal2012_SN2010jl}
{Zhang} T.,  et~al., 2012, \mn@doi [\aj] {10.1088/0004-6256/144/5/131}, \href
  {https://ui.adsabs.harvard.edu/abs/2012AJ....144..131Z} {144, 131}

\bibitem[\protect\citeauthoryear{{de Jaeger} et~al.,}{{de Jaeger}
  et~al.}{2018}]{deJaeger_etal2018_SNII_colors}
{de Jaeger} T.,  et~al., 2018, \mn@doi [\mnras] {10.1093/mnras/sty508}, \href
  {https://ui.adsabs.harvard.edu/#abs/2018MNRAS.476.4592D} {476, 4592}

\bibitem[\protect\citeauthoryear{{de Vaucouleurs}, {de Vaucouleurs}, {Buta},
  {Ables}  \& {Hewitt}}{{de Vaucouleurs}
  et~al.}{1981}]{deVaucouleurs_etal1981_SN1979C}
{de Vaucouleurs} G.,  {de Vaucouleurs} A.,  {Buta} R.,  {Ables} H.~D.,
  {Hewitt} A.~V.,  1981, \mn@doi [\pasp] {10.1086/130772}, \href
  {https://ui.adsabs.harvard.edu/\#abs/1981PASP...93...36D} {93, 36}

\makeatother
\end{thebibliography}




\appendix

\section{Tables and figures}

\begin{table*}
\caption{Optical magnitudes of the sequence stars in the field of SN~2008bm. The full table is available online.}
\label{table:SN2008bm_CSP_griz_sequence}
\begin{tabular}{l c c c c c c c c}
\hline
Star   & $\alpha_{2000}$ & $\delta_{2000}$ & $u$ (mag)& $g$ (mag)& $r$ (mag)& $i$ (mag)& $B$ (mag)& $V$ (mag) \\
\hline
 1 & 13:03:04.20 & +10:28:43.0 &  --        &  --        &  --        & 14.570(11) & 15.606(07) & 14.907(07) \\ 
 2 & 13:02:56.57 & +10:29:47.3 & 16.667(25) & 15.587(12) & 15.120(03) & 14.938(09) & 15.918(08) & 15.311(07) \\ 
 3 & 13:02:50.77 & +10:31:10.4 & 16.569(50) & 15.497(50) & 15.498(41) & 15.567(17) & 15.763(85) & 15.490(56) \\ 
\hline
\multicolumn{9}{m{0.70\linewidth}}{\textit{Note}: Errors, in parenthesis and in units of 0.001~mag, are 1$\sigma$.}
\end{tabular}
\end{table*}

\begin{table*}
\caption{Near-IR magnitudes of the sequence stars in the field of SN~2008bm. The full table is available online.}
\label{table:SN2008bm_CSP_YJH_sequence}
\begin{tabular}{l c c c c c}
\hline
Star   & $\alpha_{2000}$ & $\delta_{2000}$ & $Y$ (mag)& $J$ (mag)& $H$ (mag)\\
\hline
1 & 13:03:16.69 & +10:33:09.6 & 13.140(18) & 12.883(29) & 12.492(09)  \\ 
2 & 13:02:55.18 & +10:28:25.5 & 13.215(14) & 12.788(34) & 12.072(32)  \\ 
3 & 13:03:05.57 & +10:26:33.2 & 13.459(14) & 13.071(34) & 12.461(35)  \\ 
\hline
\multicolumn{6}{m{0.50\linewidth}}{\textit{Note}: Errors, in parenthesis and in units of 0.001~mag, are 1$\sigma$.}
\end{tabular}
\end{table*}

\begin{table*}
\caption{Optical magnitudes of the sequence stars in the field of SN~2009aj. The full table is available online.}
\label{table:SN2009aj_CSP_griz_sequence}
\begin{tabular}{@{}l@{\,\,\,\,}c@{\,\,\,\,}c@{\,\,\,\,}c@{\,\,\,\,}c@{\,\,\,\,}c@{\,\,\,\,}c@{\,\,\,\,}c@{\,\,\,\,}c@{\,\,\,\,}c@{\,\,\,\,}c@{\,\,\,\,}c@{\,\,\,\,}c@{}}
\hline
Star   & $\alpha_{2000}$ & $\delta_{2000}$ & $u$ (mag)& $g$ (mag)& $r$ (mag)& $i$ (mag)& $z$ (mag) & $U$ (mag)& $B$ (mag)& $V$ (mag)& $R$ (mag)& $I$ (mag)\\
\hline
 1 & 13:56:26.43 & --48:31:56.8 &  --   &  --   &  --         & 13.978(15) &  --        &  --         & 14.989(32) & 14.316(31) &  --         &  --         \\ 
 2 & 13:57:02.68 & --48:30:11.4 &  --   &  --   &  --         & 13.775(15) & 13.739(40) &  --         & 15.109(25) & 14.259(27) &  --         &  --         \\ 
 3 & 13:56:55.56 & --48:33:32.8 &  --   &  --   &  --         & 14.081(03) & 14.103(42) &  --         & 15.127(25) & 14.422(29) &  --         &  --         \\ 
\hline
\multicolumn{13}{m{0.70\linewidth}}{\textit{Note}: Errors, in parenthesis and in units of 0.001~mag, are 1$\sigma$.}
\end{tabular}
\end{table*}

\begin{table*}
\caption{Near-IR magnitudes of the sequence stars in the field of SN~2009aj. The full table is available online.}
\label{table:SN2009aj_CSP_YJH_sequence}
\begin{tabular}{l c c c c c}
\hline
Star   & $\alpha_{2000}$ & $\delta_{2000}$ & $Y$ (mag)& $J$ (mag)& $H$ (mag)\\
\hline
1 & 13:56:56.40 & --48:29:39.5 & 12.653(10) & 12.400(10) & 12.092(11)  \\ 
2 & 13:56:47.70 & --48:33:36.7 & 12.501(06) & 12.189(07) & 11.754(07)  \\ 
3 & 13:56:43.43 & --48:25:06.7 & 13.101(07) & 12.896(12) & 12.648(18)  \\ 
\hline
\multicolumn{6}{m{0.50\linewidth}}{\textit{Note}: Errors, in parenthesis and in units of 0.001~mag, are 1$\sigma$.}
\end{tabular}
\end{table*}

\begin{table*}
\caption{Optical magnitudes of the sequence stars in the field of SN~2009au. The full table is available online.}
\label{table:SN2009au_CSP_griz_sequence}
\begin{tabular}{@{}l@{\,}c@{\,\,\,}c@{\,\,\,\,}c@{\,\,\,\,}c@{\,\,\,\,}c@{\,\,\,\,}c@{\,\,\,\,}c@{\,\,\,\,}c@{\,\,\,\,}c@{\,\,\,\,}c@{\,\,\,\,}c@{}}
\hline
Star   & $\alpha_{2000}$ & $\delta_{2000}$ & $u$ (mag)& $g$ (mag)& $r$ (mag)& $i$ (mag)& $z$ (mag)& $B$ (mag)& $V$ (mag) & $R$ (mag)& $I$ (mag) \\
\hline
 1 & 12:59:31.60 & --29:32:03.1 & 16.470(35)  & 15.097(22) & 14.621(11) & 14.400(21) & 14.327(20) & 15.513(30) & 14.801(26) & 14.414(22) & 13.983(22) \\ 
 2 & 12:59:51.12 & --29:37:51.5 & 17.176(107) & 15.316(26) & 14.694(11) & 14.437(22) & 14.329(16) & 15.805(25) & 14.931(22) & 14.467(26) & 14.000(26) \\ 
 3 & 12:59:38.00 & --29:32:02.3 & 16.256(48)  & 14.940(24) & 14.483(11) & 14.236(09) & 14.155(18) & 15.349(27) & 14.654(25) & 14.279(24) & 13.822(24) \\ 
\hline
\multicolumn{12}{m{0.70\linewidth}}{\textit{Note}: Errors, in parenthesis and in units of 0.001~mag, are 1$\sigma$.}
\end{tabular}
\end{table*}

\begin{table*}
\caption{Near-IR magnitudes of the sequence stars in the field of SN~2009au. The full table is available online.}
\label{table:SN2009au_CSP_YJH_sequence}
\begin{tabular}{l c c c c c}
\hline
Star   & $\alpha_{2000}$ & $\delta_{2000}$ & $Y$ (mag)& $J$ (mag)& $H$ (mag) \\
\hline
1 & 13:00:04.82 & --29:38:23.8 & 12.453(21) & 11.976(15) & 11.317(09)  \\ 
2 & 13:00:10.53 & --29:35:09.2 & 12.587(22) & 12.206(18) & 11.675(12)  \\ 
3 & 12:59:47.94 & --29:39:32.1 & 12.863(17) & 12.596(14) & 12.309(09)  \\ 
\hline
\multicolumn{6}{m{0.50\linewidth}}{\textit{Note}: Errors, in parenthesis and in units of 0.001~mag, are 1$\sigma$.}
\end{tabular}
\end{table*}

\begin{table*}
\caption{CSP-I photometry of SN~2008bm. The full table is available online.}
\label{table:SN2008bm_photometry}
\begin{tabular}{@{}l@{\,\,\,\,}c@{\,\,\,\,}c@{\,\,\,\,}c@{\,\,\,\,}c@{\,\,\,\,}c@{\,\,\,\,}c@{\,\,\,\,}c@{\,\,\,\,}c@{\,\,\,\,}c@{\quad}}
\hline
MJD         & $u$ (mag)& $g$ (mag)& $r$ (mag)& $i$ (mag)& $B$ (mag)& $V$ (mag)& $Y$ (mag)& $J$ (mag)& $H$ (mag)\\
\hline
54558.23 & 18.548(24) & 17.701(09) & 17.536(08) & 17.579(12) & 17.945(10) & 17.763(10) &     --     &     --     &     --     \\ 
54559.34 &     --     &     --     &     --     &     --     &     --     &     --     & 17.257(15) & 17.125(21) & 17.096(59) \\ 
54560.25 & 18.755(32) & 17.757(08) & 17.584(08) & 17.602(09) & 18.023(11) & 17.764(09) &     --     &     --     &     --     \\
\hline
\multicolumn{10}{l}{\textit{Note}: Errors, in parenthesis and in units of 0.001~mag, are 1$\sigma$.}
\end{tabular}
\end{table*}

\begin{table*}
\caption{Photometry of SN~2009aj. The full table is available online.}
\label{table:SN2009aj_photometry}
\begin{tabular}{@{}l@{\,\,\,\,}c@{\,\,\,\,}c@{\,\,\,\,}c@{\,\,\,\,}c@{\,\,\,\,}c@{\,\,\,\,}c@{\,\,\,\,}c@{\,\,\,\,}c@{\,\,\,\,}c@{\,\,\,\,}c@{\,\,\,\,}c@{\,\,\,\,}c@{\,\,\,\,}c@{\,\,\,\,}c@{\quad}}
\hline
MJD        & $u$ & $g$ & $r$ & $i$ & $z$ & $U$ & $B$ & $V$ & $R$ & $I$ & $Y$ & $J$ & $H$ & Telescope\\
           &(mag)&(mag)&(mag)&(mag)&(mag)&(mag)&(mag)&(mag)&(mag)&(mag)&(mag)&(mag)&(mag)&          \\
\hline
54873.20 & -- & -- & -- & -- & -- & -- &  -- & >18.412$^a$    & -- & -- & -- & -- & -- & PROMPT\\ 
54886.26 & -- & -- & -- & -- & -- & -- &  -- & 15.631(29)$^a$ & -- & -- & -- & -- & -- & PROMPT\\ 
54888.18 & -- & -- & -- & -- & -- & -- &  -- & 15.417(31)$^a$ & -- & -- & -- & -- & -- & PROMPT\\ 
\hline
\multicolumn{14}{l}{\textit{Note}: Errors, in parenthesis and in units of 0.001~mag, are 1$\sigma$.}\\
\multicolumn{14}{l}{$^a$Unfiltered photometry calibrated to the $V$-band.}
\end{tabular}
\end{table*}

\begin{table*}
\caption{Photometry of SN~2009au. The full table is available online.}
\label{table:SN2009au_photometry}
\begin{tabular}{@{}l@{\,\,\,\,}c@{\,\,\,\,}c@{\,\,\,\,}c@{\,\,\,\,}c@{\,\,\,\,}c@{\,\,\,\,}c@{\,\,\,\,}c@{\,\,\,\,}c@{\,\,\,\,}c@{\,\,\,\,}c@{\,\,\,\,}c@{\,\,\,\,}c@{\,\,\,\,}c@{\quad}}
\hline
MJD      &$u$  &$g$  &$r$  &$i$  & $z$ & $B$ & $V$ & $R$ & $I$ & $Y$ & $J$ & $H$ & Telescope\\
         &(mag)&(mag)&(mag)&(mag)&(mag)&(mag)&(mag)&(mag)&(mag)&(mag)&(mag)&(mag)&          \\
\hline
54886.18 & -- & -- & -- & -- & -- & -- & >19.042$^a$    & -- & -- & -- & -- & -- & PROMPT   \\ 
54893.15 & -- & -- & -- & -- & -- & -- & >17.656$^a$    & -- & -- & -- & -- & -- & PROMPT   \\ 
54901.17 & -- & -- & -- & -- & -- & -- & 17.629(77)$^a$ & -- & -- & -- & -- & -- & PROMPT   \\ 
\hline
\multicolumn{13}{l}{\textit{Note}: Errors, in parenthesis and in units of 0.001~mag, are 1$\sigma$.}\\
\multicolumn{13}{l}{$^a$Unfiltered photometry calibrated to the $V$-band.}
\end{tabular}
\end{table*}

\begin{table*}
\caption{Optical spectroscopy of SN~2009au.}
\label{table:SN09au_spectra}
\begin{tabular}{l c c c c c l}
\hline
UT date    & MJD      & Phase$^a$ & Instrument setup$^b$  & Wavelength range & Exposure time & Slit \\
           &          &  (d)  &                & (\AA)            & (s)           & (") \\
\hline
2009-03-22 & 54912.11 & 14.8  & TNG+LRS (+LR-B, LR-R) & 3320--10360 & 1200, 1200    & 1.5 \\
2009-03-30 & 54920.03 & 22.6  & TNG+LRS (+LR-B, LR-R) & 3800--9190  & 1800, 1800    & 1.5 \\
2009-04-01 & 54922.05 & 24.6  & NOT+ALFOSC (+gm4)     & 3420--9110  &  900          & 1.0 \\
2009-04-14 & 54935.04 & 37.5  & NOT+ALFOSC (+gm4)     & 3490--8000  & 1200          & 1.0 \\
2009-04-15 & 54936.05 & 38.5  & NOT+ALFOSC (+gm4)     & 3640--9070  & 2700          & 1.3 \\
2009-05-05 & 54956.01 & 58.3  & NOT+ALFOSC (+gm4)     & 3700--8000  & 1800          & 1.0 \\
2009-05-09 & 54960.99 & 63.2  & NOT+ALFOSC (+gm4)     & 3640--9080  &  900          & 1.3 \\
2009-05-18 & 54969.13 & 71.2  & NTT+EFOSC  (+gr11)    & 3340--10290 & 1200          & 1.0 \\
2009-07-16 & 55028.08 & 129.6 & NTT+EFOSC  (+gr16)    & 6000--9400  & 2400          & 1.5 \\
2009-08-01 & 55044.98 & 146.4 & Gemini-S+GMOS (+R400) & 4700--8120  & 1800          & 1.0 \\
\hline
\multicolumn{7}{m{0.88\linewidth}}{$^a$Since explosion (MJD~54897.2).}\\
\multicolumn{7}{m{0.88\linewidth}}{$^b$TNG+LRS: 3.58~m Telescopio Nazionale Galileo (La Palma, Spain) + Low Resolution Spectrograph; NOT+ALFOSC: 2.56~m Nordic Optical Telescope (La Palma, Spain) + Alhambra Faint Object Spectrograph and Camera; NTT+EFOSC: 3.58~m New Technology Telescope (La Silla, Chile) + ESO Faint Object Spectrograph and Camera; Gemini-S+GMOS: 8.1~m Gemini South Telescope (Cerro Pach\'on, Chile) + Gemini Multi-Object Spectrograph.}
\end{tabular}
\end{table*}

\begin{table*}
\caption{Near-IR spectroscopy of SN~2009aj.}
\label{table:SN09aj_spectra}
\begin{tabular}{l c c c c c c c l}
\hline
UT date    & MJD      & Phase$^a$ & Instrument setup$^b$       & Wavelength range & Exposure time & Slit \\
           &          &  (d)     &               &           (\AA)            & (s)           & (") & &\\
\hline
2009-03-04 & 54895.38 & 15.4  & VLT+ISAAC (+SWS1-LR)   &  9700--25000     & 600         &  1.0   \\
2009-03-16 & 54907.36 & 27.3  & NTT+SOFI (+GB, GR)     &  9400--24400     & 2700, 5400  &  0.6   \\
2009-04-26 & 54948.17 & 67.7  & SOAR+OSIRIS (+LR)      &  11700--19400           & 2400, 2400  &  1.0   \\
\hline
\multicolumn{9}{m{0.83\linewidth}}{$^a$Since explosion (MJD~54879.8).}\\
\multicolumn{9}{m{0.83\linewidth}}{$^b$VLT+ISAAC: 8.2~m Very Large Telescope UT1 (Paranal, Chile) + Infrared Spectrometer And Array Camera; NTT+SOFI: 3.58~m New Technology Telescope (La Silla, Chile) + Son of ISAAC; SOAR+OSIRIS: 4.1~m Southern Astrophysical Research Telescope (Cerro Pach\'on, Chile) + Ohio State Infrared Imager/Spectrometer.}
\end{tabular}
\end{table*}

\begin{figure*}
\includegraphics[width=1.0\columnwidth]{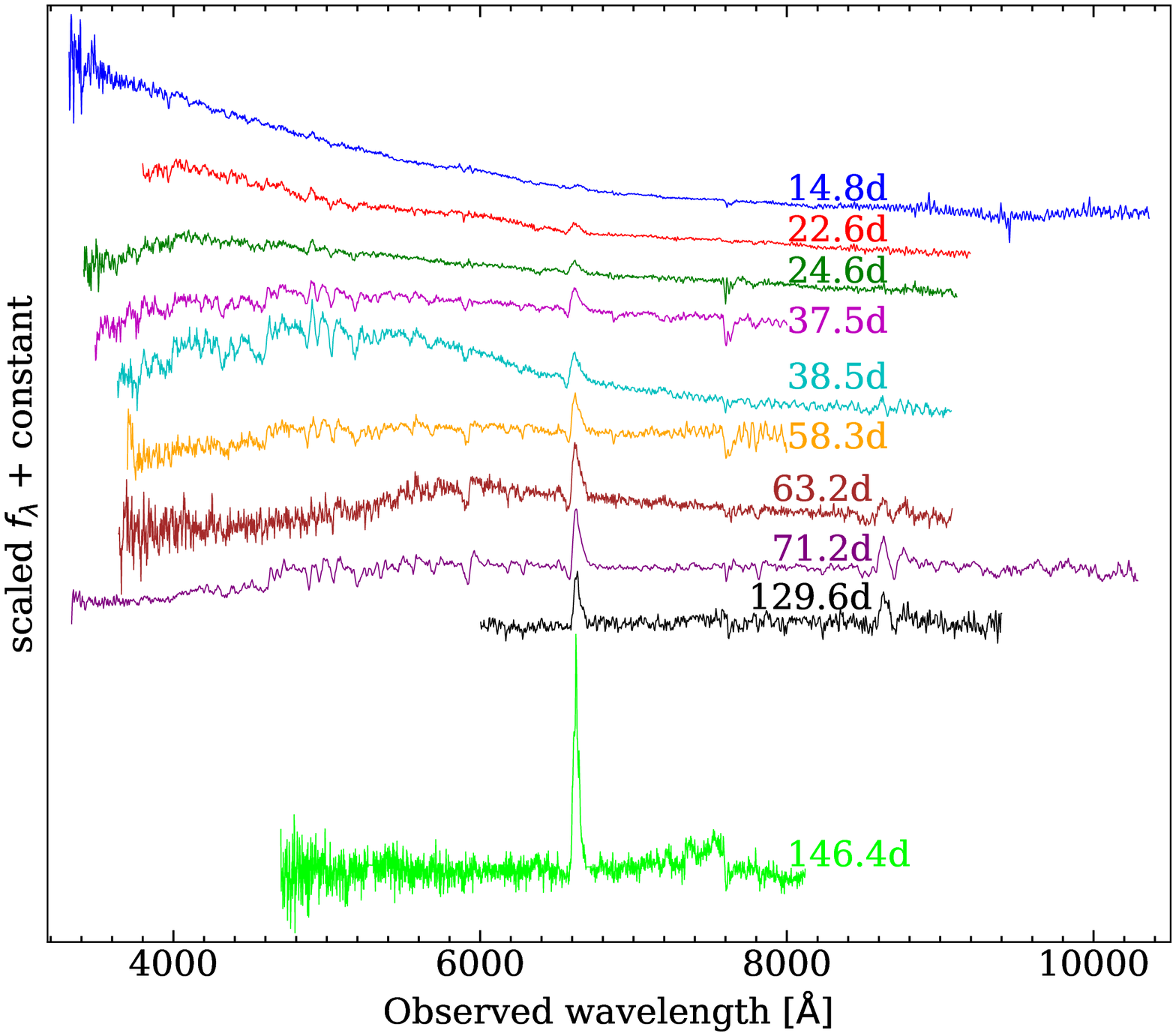}
\includegraphics[width=1.0\columnwidth]{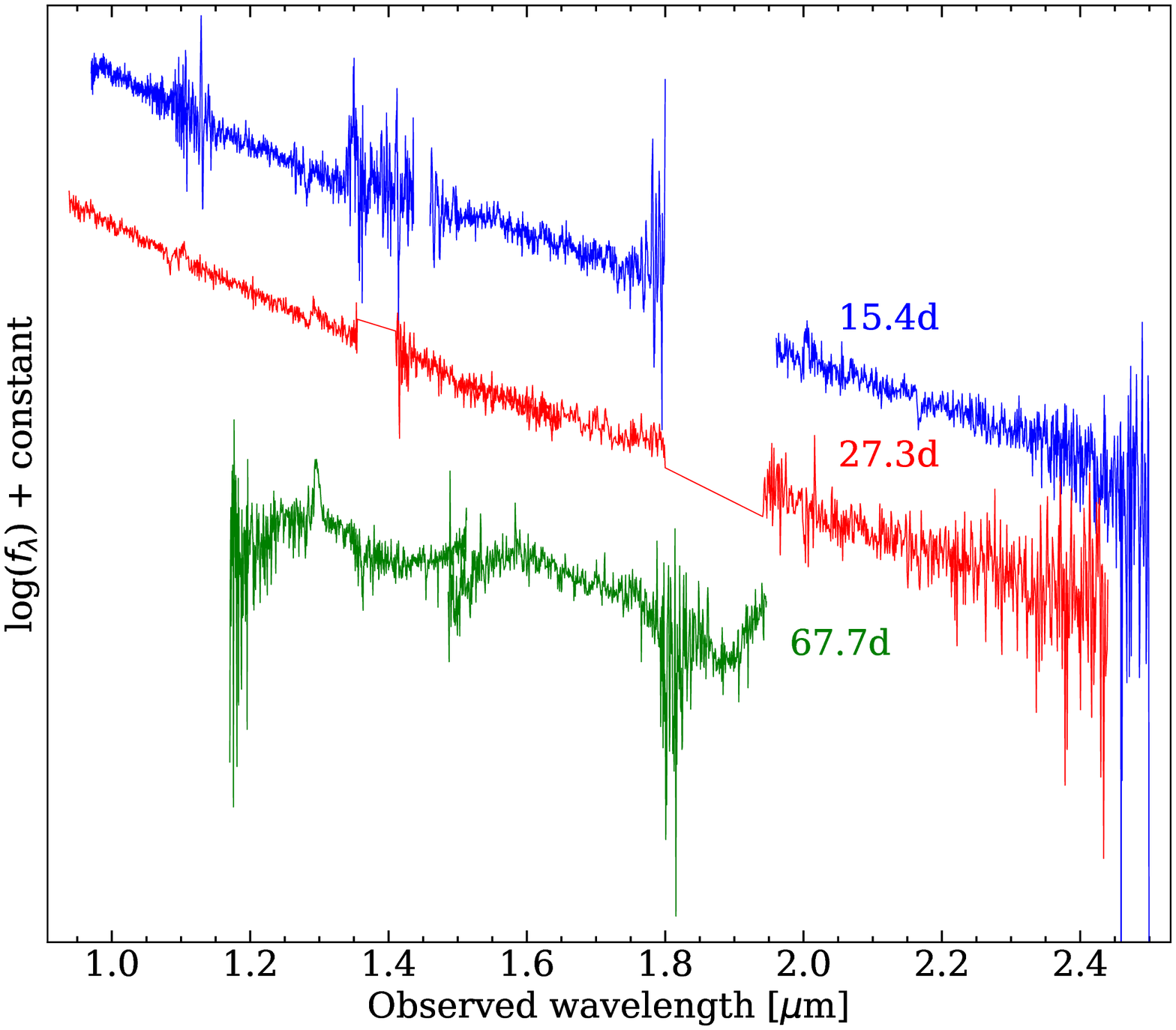}
\caption{Left: optical spectra of SN~2009au. Right: near-IR spectra of SN~2009aj.}
\label{fig:SN09au_spectra}
\end{figure*}

\begin{table*}
\caption{SN and host galaxy parameters of the SN~IIn/II sample.}
\label{table:SN_IInII_sample}
\begin{tabular}{l c c c c c c}
\hline
 SN data & 1979C & 1998S & 2007pk & 2008fq & PTF11iqb & 2013fc\\
\hline
 Host galaxy                             &M100                  &NGC~3877              &NGC~579        &NGC~6907         &NGC~151        & ESO~154-G10     \\
 Host type$^a$                           &SAB(s)bc              &SA(s)c?               &Scd?           &SB(s)bc          &SB(r)bc        &(R')SB(r)a?      \\
 \EGBV\ (mag)$^b$                        &$0.023$               &$0.020$               &$0.045$        &$0.055$          &$0.028$        &$0.026$          \\
 \EhBV\ (mag)                            &$0.13_{-0.02}^{+0.06}$&$0.20_{-0.08}^{+0.11}$&$0.09\pm0.04^c$&$0.99\pm0.14^d$  &$0.0$          &$0.91\pm0.06$    \\
 Explosion epoch (MJD)                   &$43970.4\pm8.4$       &$50871.2\pm3.5$ &$54411.8\pm2.5$ &$54719.8\pm4.5$       &$55762.2\pm2.7$&$56516.2\pm2.0$  \\
 $cz_\mathrm{helio}$ (\kms)$^{a,e}$      &$1571$                &$895$           &$4993$          &$3182$                &$3747$         &$5586$           \\
 $\mu_z$ (mag)$^{f}$                     &$30.74\pm0.81$        &$31.24\pm0.64$  &$34.14\pm0.17$  &$33.37\pm0.24$        &$33.51\pm0.23$ &$34.43\pm0.15$   \\
 $\mu_{z\mathrm{i}}$ (mag)$^g$           &$31.18\pm0.05^\ddagger$&$30.92\pm0.45^\dagger$ &--      &$32.80\pm0.40^\dagger$&$33.89\pm0.40^\dagger$ &--   \\
 Adopted $\mu$ (mag)$^h$                 &$31.18\pm0.05$        &$31.03\pm0.37$  &$34.14\pm0.17$  &$33.22\pm0.21$        &$33.60\pm0.21$ &$34.43\pm0.15$   \\
 12+log(O/H) (dex)$^i$                   &--&$8.56\pm0.10^\star$&$8.42\pm0.04^\diamond$&$8.57\pm0.04^\otimes$&$8.61\pm0.03^\star$& $8.55\pm0.04^\oplus$   \\
 $s_2$ (\sunit)                          &$3.05\pm0.08$         &$3.62\pm0.09$   &$2.28\pm0.05$   &$1.77\pm0.08$         &$1.40\pm0.10$  &$3.59\pm0.12$    \\
 $M_V^{\mathrm{max}}$ (mag)              &$-19.55\pm0.14$       &$-19.46\pm0.49$ &$-18.54\pm0.21$ &$-20.82\pm0.50$       &$-18.34\pm0.23$& $-20.13\pm0.24$ \\
 $M_V^{50\mathrm{d}}$ (mag)              &$-18.26\pm0.31$       &$-18.39\pm0.49$ &$-17.64\pm0.21$ &$-19.78\pm0.50$       &$-17.39\pm0.25$&$-18.85\pm0.24$  \\
 \isoni\ mass (\msun)                    &$>0.069\pm0.010^j$      &$>0.155\pm0.069^j$&--              &--                    &$0.029\pm0.007$&$>0.279\pm0.070^j$ \\
 (\bv)$_{50\mathrm{d}}$ (mag)            &$0.53\pm0.05$         & $0.40\pm0.11$  &$0.69\pm0.09$   & $0.51\pm0.14$        &$0.46\pm0.07$  &$0.42\pm0.16$    \\
 pEW$_{\mathrm{FeII}\lambda5018}^{50\mathrm{d}}$ (\AA) &$8.03\pm0.40$ & $9.68\pm0.31$ &$11.49\pm0.14$ & $14.09\pm0.46$   &--             &  $14.78\pm0.98$ \\
 $v_\mathrm{FeII}^{50\mathrm{d}}$ (\kms) &$6086\pm315$          &$4355\pm183$    &$5086\pm185$    &$5365\pm228$          &--             & $5682\pm196$    \\
 References$^k$                          & 1, 2, 3, 4, 5        & 6, 7, 8        & 9, 10          & 11, 12               & 13            & 14              \\
\hline
\multicolumn{7}{l}{$^a$From NED.}\\
\multicolumn{7}{l}{$^b$Galactic colour excesses from \citet{Schlafly_Finkbeiner2011}, with a statistical uncertainty of 16\% \citep{Schlegel_etal1998}.}\\
\multicolumn{7}{l}{$^c$Average of the \EhBV\ values reported in \citet{Pritchard_etal2012_SN2007pk} and \citet{Inserra_etal2013}.}\\
\multicolumn{7}{m{0.99\linewidth}}{$^d$SN~2008fq has a \naid\ pEW of $2.9\pm0.2$~\AA\ \citep{Taddia_etal2013_CSP_SNeIIn}, which is unhelpful to estimate \EhBV\ (see Section~\ref{sec:2009au_info}). As for SN~2009au, we compute \EhBV\ matching the \bv\ colour curve of SN~2008fq to the rest of SNe~IIn/II.}\\
\multicolumn{7}{l}{$^e$Heliocentric velocities, with an error of 162~\kms\ to take into account the host galaxy rotational velocity \citep{Anderson_etal2014_blueshifted_emission}.}\\
\multicolumn{7}{m{0.99\linewidth}}{$^f$Distance moduli computed from recessional velocities, corrected for the infall of the Local Group towards the Virgo cluster and the Great Attractor \citep{Mould_etal2000}, and assuming $H_0=73$~\hnotunits. We include an error of 382~\kms\ due to peculiar velocities \citep{Wang_etal2006}.}\\
\multicolumn{7}{m{0.99\linewidth}}{$^g$Redshift-independent distance moduli (when available): ($\ddagger$) Cepheids distance from \citet{Saha_etal2006}; ($\dagger$) Tully-Fisher distance from the Extragalactic Distance Database (EDD, \url{http://edd.ifa.hawaii.edu/}).}\\
\multicolumn{7}{m{0.99\linewidth}}{$^h$Weighted average of $\mu_z$ and $\mu_{zi}$ (if it is available). For SN~1979C we adopt the Cepheid distance modulus.}\\
\multicolumn{7}{m{0.99\linewidth}}{$^i$Oxygen abundances, in the N2 calibration of \citet{Marino_etal2013}, measured by \citet{Anderson_etal2016} ($\otimes$), \citet{Taddia_etal2015} ($\star$), \citet{Inserra_etal2013} ($\diamond$), and \citet{Kangas_etal2016_SN2013fc} ($\oplus$). The latter three are recalibrations of the original values reported in the N2 calibration of \citet{Pettini_Pagel2004}.}\\
\multicolumn{7}{m{0.99\linewidth}}{$^i$\isoni\ masses measured with the \citet{Hamuy2003} relation and $\mathrm{BC}=0.26\pm0.06$ \citep{Hamuy2001}. The upper limits are because the slope during the radioactive tail is $>1.5$~\sunit.}\\
\multicolumn{7}{m{0.99\linewidth}}{$^j$ Since $s_3>1.5$ \sunit, \isoni\ mass estimations are lower limits.}\\
\multicolumn{7}{m{0.99\linewidth}}{$^k$(1) \citet{Balinskaia_etal1980_SN1979C};
(2) \citet{deVaucouleurs_etal1981_SN1979C};
(3) \citet{Branch_etal1981_SN1979C};
(4) \citet{Barbon_etal1982};
(5) \citet{Penston_Blades1980_SN1979C_reddening};
(6) \citet{Fassia_etal2000_SN1998S};
(7) \citet{Poon_etal2011_SN1998S};
(8) \citet{Fassia_etal2001_SN1998S};
(9) \citet{Inserra_etal2013};
(10) \citet{Hicken_etal2017};
(11) \citet{Taddia_etal2013_CSP_SNeIIn};
(12) \citet{Faran_etal2014_IIL};
(13) \citet{Smith_etal2015};
(14) \citet{Kangas_etal2016_SN2013fc}.
}\\
\end{tabular}
\end{table*}

\begin{table*}
\caption{Properties of the normal SNe~II with \bv\ colours during the radioactive tail.}
\label{table:normal_SNII_sample}
\begin{tabular}{@{}l@{\quad}c@{\quad}c@{\quad}c@{\,\,}c@{\,\,\,}l@{\quad}l@{\quad}c@{\quad}c@{\quad}c@{\,\,}c@{\,\,\,}l@{}}
\hline
SN     & $t_0$ & \EGBV$^\dagger$ &\EhBV\ &  $cz_\mathrm{helio}^\ddagger$ & Ref$^\ast$ & SN     & $t_0$ & \EGBV$^\dagger$ &\EhBV\ &  $cz_\mathrm{helio}^\ddagger$ & Ref$^\ast$\\
       & (MJD) & (mag) & (mag) & (\kms)  &  & & (MJD) & (mag) & (mag) & (\kms)  & \\
\hline
   1969L & $40550.0\pm5.0$  & $0.053$ & $0.000\pm0.097$ & $518$  & 1, 2      &   2007it & $54348.0\pm1.0$ & $0.099$ & $0.019\pm0.013$ & $1193$ & 15, 16   \\ 
   1992H & $48660.5\pm10.0$ & $0.015$ & $0.000\pm0.097$ & $1793$ & 2, 3      &   2008bk & $54540.9\pm8.0$ & $0.017$ & $0.000\pm0.016$ & $230$  & 7, 16, 17 \\
   1996W & $50179.5\pm3.0$  & $0.036$ & $0.187$         & $1617$ & 4         &   2008gz & $54693.5\pm5.0$ & $0.036$ & $0.030\pm0.040$ & $1862$ & 18 \\
  1999ca & $51277.5\pm7.0$  & $0.094$ & $0.039\pm0.100$ & $2791$ & 5, 6, 7   &   2009bw & $54916.0\pm3.0$ & $0.197$ & $0.080$         & $1155$ & 19 \\
  1999em & $51475.2\pm3.8$  & $0.035$ & $0.100\pm0.052$ & $800$  & 5, 6, 8   &    2012A & $55928.7\pm4.7$ & $0.027$ & $0.009$         & $753$  & 20 \\
  2003B  & $52621.7\pm4.3$  & $0.023$ & $0.000\pm0.081$ & $1141$ & 5, 6      &   2012aw & $56002.1\pm0.8$ & $0.024$ & $0.046\pm0.008$ & $778$  & 21, 22 \\
  2003gd & $52715.0\pm3.0$  & $0.060$ & $0.130\pm0.100$ & $657$  & 5, 6, 9   &   2014cx & $56901.9\pm0.5$ & $0.096$ & $0.0$           & $1646$ & 23 \\
  2004dj & $53186.5\pm3.0$  & $0.034$ & $0.161\pm0.081$ & $221$  & 6, 10, 11 & ASAS14jb & $56945.6\pm3.0$ & $0.015$ & $0.0$           & $1808$ & 24 \\
  2004et & $53270.5\pm0.5$  & $0.293$ & $0.000\pm0.081$ & $40$   & 6, 12     & ASAS15oz & $57261.1\pm4.0$ & $0.078$ & $0.0$           & $2078$ & 25 \\
  2005cs & $53548.4\pm0.5$  & $0.032$ & $0.040\pm0.050$ & $463$  & 13, 14    &  2017eaw & $57886.2\pm1.0$ & $0.293$ & $0.0$           & $40$   & 26, 27, 28 \\
\hline
\multicolumn{12}{m{1.0\linewidth}}{$^\dagger$Galactic colour excesses from \citet{Schlafly_Finkbeiner2011}, with a statistical error of 16 per cent \citep{Schlegel_etal1998}.}\\
\multicolumn{12}{m{1.0\linewidth}}{$^\ddagger$Heliocentric redshifts from NED.}\\
\multicolumn{12}{m{1.0\linewidth}}{$^\ast$References: (1) \citet{Ciatti_etal1971_SN1969L};
(2) \citet{Hamuy2003};
(3) \citet{Clocchiatti_etal1996_SN1992H};
(4) \citet{Inserra_etal2013};
(5) \citet{Galbany_etal2016};
(6) \citet{Olivares_etal2010};
(7) \citet{Gutierrez_etal2017_I};
(8) \citet{Elmhamdi_etal2003};
(9) \citet{VanDyk_etal2003_SN2003gd};
(10) \citet{Vinko_etal2006};
(11) \citet{Tsvetkov_etal2008};
(12) \citet{Maguire_etal2010_04et};
(13) \citet{Pastorello_etal2009};
(14) \citet{Dessart_etal2008};
(15) \citet{Andrews_etal2011_SN2007it};
(16) \citet{Anderson_etal2014_V_LC};
(17) \citet{VanDyk_etal2012_SN2008bk};
(18) \citet{Roy_etal2011_SN2008gz};
(19) \citet{Inserra_etal2012_SN2009bw};
(20) \citet{Tomasella_etal2013};
(21) \citet{Bose_etal2013};
(22) \citet{DallOra_etal2014};
(23) \citet{Huang_etal2016_SN2014cx};
(24) \citet{Meza_etal2019_ASAS14jb};
(25) \citet{Bostroem_etal2019_ASAS15oz};
(26) \citet{Tsvetkov_etal2018_SN2017eaw};
(27) \citet{Szalai_etal2019_SN2017eaw};
(28) \citet{VanDyk_etal2019_SN2017eaw}.
}
\end{tabular}
\end{table*}

\begin{table*}
\caption{Properties and references of the SNe~IIn used in this work.}
\label{table:SN_IIn_sample}
\begin{tabular}{l c c c c l}
\hline
SN     & $t_0$ & \EGBV$^\dagger$ &\EhBV\ &  $cz_\mathrm{helio}^\ddagger$ & References$^\ast$\\
       & (MJD) & (mag)                         & (mag)                 & (\kms)                        &  \\
\hline
1994Y  & $49570.1\pm5.9$  & $0.008$& $0.0$         & $2558$ & 1, 2 \\
1996al & $50255.0\pm2.0$  & $0.010$& $0.10\pm0.05$ & $1970$ & 3    \\
2010jl & $55473.5\pm5.0$  & $0.023$& $0.02$        & $3214$ & 4, 5 \\
2011ht & $55833.2$        & $0.072$& $0.0$         & $1093$ & 6    \\
2015bh & $57154.8\pm3.9$  & $0.019$& $0.21\pm0.07$ & $1947$ & 7, 8 \\
2015da & $57030.4\pm1.5$  & $0.012$& $0.97\pm0.27$ & $2165$ & 9    \\
\hline
\multicolumn{6}{m{0.55\linewidth}}{$^\dagger$Galactic colour excesses from \citet{Schlafly_Finkbeiner2011}, with a statistical error of 16 per cent \citep{Schlegel_etal1998}.}\\
\multicolumn{6}{m{0.55\linewidth}}{$^\ddagger$Heliocentric redshifts from NED.}\\
\multicolumn{6}{m{0.55\linewidth}}{$^\ast$(1) \citet{Ho_etal2001_SN1994Y};
(2) \citet{Tsvetkov_Pavlyuk1997_SN1994Y};
(3) \citet{Benetti_etal2016_SN1996al};
(4) \citet{Fransson_etal2014_SN2010jl};
(5) \citet{Chandra_etal2015_SN2010jl};
(6) \citet{Humphreys_etal2012_SN2011ht};
(7) \citet{EliasRosa_etal2016_SN2015bh};
(8) \citet{Thone_etal2017_SN2015bh};
(9) \citet{Tartaglia_etal2020_SN2015da}.
}
\end{tabular}
\end{table*}

\begin{table*}
\caption{New estimation of the nickel masses for the SNe~II in the Hamuy sample.}
\label{table:new_Ni_masses}
\begin{tabular}{l c c c c c c c}
\hline
SN$^\star$ & \EGBV$^\dagger$ &\EhBV$^\ast$&$\mu_z^\diamond$&$\mu_{z\mathrm{i}}^\oslash$&$\mu^\ddagger$&$M_V^{50\mathrm{d}}$&$M$(\isoni) \\
       & (mag)                         & (mag)                 & (mag)      & (mag)  & (mag)       & (mag) & (\msun) \\
\hline
 1969L  & $0.053$ & $0.000\pm0.097^a$ & $29.78\pm1.26$ & $29.84\pm0.04^c$ & $29.84\pm0.04$ & $-16.65\pm0.31$ & $0.068\pm0.021$ \\
 1970G  & $0.008$ & $0.000\pm0.097^a$ & $28.96\pm1.83$ & $29.14\pm0.05^d$ & $29.14\pm0.05$ & $-17.06\pm0.34$ & $0.030\pm0.010$ \\
 1973R  & $0.028$ & $0.452\pm0.097^a$ & $29.07\pm1.74$ & $30.18\pm0.04^e$ & $30.18\pm0.04$ & $-17.11\pm0.31$ & $0.093\pm0.036$ \\
 1986I  & $0.034$ & $0.065\pm0.097^a$ & $33.00\pm0.29$ & $30.71\pm0.40^f$ & $30.71\pm0.40$ & $-16.47\pm0.54$ & $0.076\pm0.036$ \\
 1988A  & $0.035$ & $0.000\pm0.097^a$ & $30.71\pm0.82$ & $31.11\pm0.40^f$ & $31.03\pm0.36$ & $-16.14\pm0.47$ & $0.054\pm0.027$ \\
 1990E  & $0.022$ & $0.468\pm0.097^a$ & $31.07\pm0.69$ & $30.89\pm0.45^f$ & $30.94\pm0.38$ & $-16.56\pm0.52$ & $0.044\pm0.022$ \\
 1990K  & $0.012$ & $0.065\pm0.097^a$ & $31.66\pm0.53$ & $31.82\pm0.45^f$ & $31.75\pm0.34$ & $-17.49\pm0.50$ & $0.036\pm0.017$ \\
 1991G  & $0.017$ & $0.000\pm0.097^a$ & $30.90\pm0.75$ & $30.64\pm0.45^f$ & $30.71\pm0.39$ & $-15.23\pm0.50$ & $0.020\pm0.009$ \\
 1992H  & $0.015$ & $0.000\pm0.097^a$ & $32.53\pm0.35$ & $32.23\pm0.45^f$ & $32.42\pm0.28$ & $-17.48\pm0.41$ & $0.139\pm0.054$ \\
1992ba  & $0.050$ & $0.139\pm0.052^b$ & $30.76\pm0.80$ & --               & $30.76\pm0.80$ & $-15.92\pm0.82$ & $0.024\pm0.018$ \\
1999gi  & $0.014$ & $0.181\pm0.052^b$ & $30.01\pm1.13$ & $30.77\pm0.06^c$ & $30.77\pm0.06$ & $-16.47\pm0.18$ & $0.041\pm0.007$ \\
\hline
\multicolumn{8}{m{0.90\linewidth}}{\textit{Note}. To recompute $M_V^{50\mathrm{\,d}}$ and $M$(\isoni) we use the same observables listed in Table~2 of \citet{Hamuy2003}.}\\
\multicolumn{8}{m{0.90\linewidth}}{$^\star$We recompute $M_V^{50\mathrm{\,d}}$ and $M$(\isoni) values only for the SNe that do not have $M$(\isoni) estimations in the A14 sample, with three or more $V$-band photometric points during the nebular phase, and with $s_3<1.3$ \sunit.}\\
\multicolumn{8}{m{0.90\linewidth}}{$^\dagger$Galactic colour excesses from \citet{Schlafly_Finkbeiner2011}, with a statistical error of 16 per cent \citep{Schlegel_etal1998}.}\\
\multicolumn{8}{m{0.90\linewidth}}{$^\ast$Host galaxy reddening values from \citet{Hamuy2003} (a) or \citet{Olivares_etal2010} (b).}\\
\multicolumn{8}{m{0.90\linewidth}}{$^\diamond$Distance moduli computed from recessional velocities, corrected for the infall of the Local Group towards the Virgo cluster and the Great Attractor \citep{Mould_etal2000}, and assuming $H_0=73$~\hnotunits. We include an error of 382~\kms\ due to peculiar velocities \citep{Wang_etal2006}.}\\
\multicolumn{8}{m{0.90\linewidth}}{$^\oslash$Redshift-independent distance moduli: Cepheid distance from \citet{Saha_etal2006} (c) and \citet{Riess_etal2016} (d); TRGB distance from \citet{Jang_Lee2017_V} (e); Tully-Fisher distance from EDD (f). We do not consider distances computed with the SN itself.}\\
\multicolumn{8}{m{0.90\linewidth}}{$^\ddagger$Weighted average of $\mu_z$ and $\mu_{z\mathrm{i}}$ (if it is available). For SN~1969L, SN~1970G, SN~1973R, SN~1986I, and SN~1999gi we adopt the redshift-independent distance modulus.} 
\end{tabular}
\end{table*}


\bsp	
\label{lastpage}
\end{document}